\documentclass[twocolumn]{aastex631}

\accepted{to ApJ September 7th, 2023}

\shorttitle{GRAVITY HD79246B}
\shortauthors{Balmer et al.}
\graphicspath{{./}{figures/}}

\definecolor{mylinkcolor}{RGB}{16, 37, 110}
\hypersetup{linkcolor=mylinkcolor,citecolor=mylinkcolor,urlcolor=mylinkcolor}

\makeatletter
\def\blfootnote{\xdef\@thefnmark{}\@footnotetext}
\makeatother

\usepackage{flafter} 

\begin{document}

\title{VLTI/GRAVITY Observations and Characterization of the Brown Dwarf Companion HD~72946~B}

\collaboration{100}{The ExoGRAVITY Collaboration}
\author[0000-0001-6396-8439]{William O. Balmer}
\altaffiliation{Johns Hopkins University George Owen Fellow}
\correspondingauthor{William O. Balmer}
\email{wbalmer@stsci.edu}
\affiliation{ Department of Physics \& Astronomy, Johns Hopkins University, 3400 N. Charles Street, Baltimore, MD 21218, USA}
\affiliation{Space Telescope Science Institute, 3700 San Martin Dr, Baltimore, MD 21218, USA}
\author[0000-0003-3818-408X]{ Laurent Pueyo}
\affiliation{Space Telescope Science Institute, 3700 San Martin Dr, Baltimore, MD 21218, USA}
\author[0000-0002-5823-3072]{ Tomas Stolker}
\affiliation{ Leiden Observatory, Leiden University, P.O. Box 9513, 2300 RA Leiden, The Netherlands}
\author[0000-0001-6533-6179]{Henrique Reggiani}
\altaffiliation{Carnegie Fellow}
\affiliation{The Observatories of the Carnegie Institution
for Science, 813 Santa Barbara St, Pasadena, CA 91101, USA}
\author{ A.-L.~Maire}
\affiliation{ STAR Institute/Universit\'e de Li\`ege, Belgium}
\affiliation{ Max Planck Institute for Astronomy, K\"onigstuhl 17, 69117 Heidelberg, Germany}
\author[0000-0002-6948-0263]{ S.~Lacour}
\affiliation{ LESIA, Observatoire de Paris, PSL, CNRS, Sorbonne Universit\'e, Universit\'e de Paris, 5 place Janssen, 92195 Meudon, France}
\affiliation{ European Southern Observatory, Karl-Schwarzschild-Stra\ss e 2, 85748 Garching, Germany}
\author[0000-0003-4096-7067]{ P.~Molli\`ere}
\affiliation{ Max Planck Institute for Astronomy, K\"onigstuhl 17, 69117 Heidelberg, Germany}
\author{ M.~Nowak}
\affiliation{ Institute of Astronomy, University of Cambridge, Madingley Road, Cambridge CB3 0HA, United Kingdom}
\author[0000-0001-6050-7645]{ D.~Sing }
\affiliation{ Department of Physics \& Astronomy, Johns Hopkins University, 3400 N. Charles Street, Baltimore, MD 21218, USA}
\affiliation{ Department of Earth \& Planetary Sciences, Johns Hopkins University, Baltimore, MD, USA}
\author[0000-0001-9431-5756]{ N.~Pourr\'e}
\affiliation{ Universit\'e Grenoble Alpes, CNRS, IPAG, 38000 Grenoble, France}
% ...



\begin{abstract}
Tension remains between the observed and modeled properties of substellar objects, but objects in binary orbits, with known dynamical masses can provide a way forward. HD~72946~B is a recently imaged brown dwarf companion to the nearby, solar type star. We achieve $\sim100$ \textmu as relative astrometry of HD~72946~B in the K-band using VLTI/GRAVITY, unprecedented for a benchmark brown dwarf. We fit an ensemble of measurements of the orbit using \texttt{orbitize!} and derive a strong dynamical mass constraint $\mathrm{M_B}=69.5\pm0.5 \mathrm{M_{Jup}}$ assuming a strong prior on the host star mass $\mathrm{M_A}=0.97\pm0.01 \mathrm{M_\odot}$ from an updated stellar analysis. We fit the spectrum of the companion to a grid of self-consistent \texttt{BT-Settl-CIFIST} model atmospheres, and perform atmospheric retrievals using \texttt{petitRADTRANS}. A dynamical mass prior only marginally influences the sampled distribution on effective temperature, but has a large influence on the surface gravity and radius, as expected. The dynamical mass alone does not strongly influence retrieved pressure-temperature or cloud parameters within our current retrieval setup. Independent of cloud prescription and prior assumptions, we find agreement within $\pm2\,\sigma$ between the C/O ratio of the host ($0.52\pm0.05)$ and brown dwarf ($0.43$ to $0.63$), as expected from a molecular cloud collapse formation scenario, but our retrieved metallicities are implausibly high (0.6-0.8) in light of an excellent agreement of the data with the solar abundance model grid. Future work on our retrieval framework will seek to resolve this tension. Additional study of low surface-gravity objects is necessary to assess the influence of a dynamical mass prior on atmospheric analysis.
\end{abstract}

\section{Introduction} \label{sec:intro}
\par Brown Dwarfs (BDs) are substellar objects unable to fuse hydrogen \citep[$\mathrm{M}\lesssim75-80\mathrm{M_{Jup}}$][]{Saumon2008, Baraffe2015, Dupuy2017, Fernandes2019}. Due to their insufficient mass, their cores do not reach the temperatures required for nuclear fusion to balance radiative losses, and they become supported by electron-degeneracy pressure \citep{Chabrier2000}. BDs burn away deuterium \citep[and lithium, for $\mathrm{M}\gtrsim65\mathrm{M_{Jup}}$,][]{Dupuy2017, Zhang2019}, soon exhausting these relatively scarce fuels and cool ``inexorably like dying embers plucked
from a fire" \citep{Burrows2001}. The known population of BDs are rich in spectral diversity \citep{Cushing2005} because their lack of nuclear heating leads to low effective temperature atmospheres with dense, overlapping molecular opacities that evolve over time as the BD continues to cool. This complexity necessitates precise luminosity, age, and mass measurements in order to properly test models of BD evolution and composition. This is much easier said than done.
\par Theories of star formation suggest that a significant number of higher-mass BDs form via molecular cloud collapse, during the process of star formation, either alone, in binary pairs, or near stellar mass hosts but undergoing subsequent ejection \citep[e.g.,][]{Padoan2004, Bate2009, Umbreit2005}. It is still vigorously debated whether or not (or, more realistically, what proportion of) low mass BDs might arise from the fragmentation of the circumstellar disk around more massive stars \citep{Boss1997, Stamatellos2007, Stamatellos2009, Kratter2010, Li2016, Squicciarini2022}, and in what ways these processes are related to planet formation. Around main sequence FGK stars, a ``brown dwarf desert" exists \citep{Grether2006}, where few BD companions can be found at solar system scale separations \citep{Ma2014}. New work appears to reveal similar trends, scaled up or down in mass, around earlier type stars \citep{Duchene2023} and late type (even substellar) objects \citep{2018MNRAS.479.2702F}. This makes the known companions of this nature, the inhabitants of the desert, interesting in the broader context of BD studies.
\par Moreover, the similarity (in physics, chemistry, composition) between BDs and giant planets has motivated careful study of these objects, particularly their atmospheres, as more readily accessible laboratories for studying the physics of giant planets. Giant planets span the L-T-Y spectral sequence of BDs, and in many ways our current modeling of directly detected super-jovian exoplanet atmospheres depend on models and observations of BDs \citep{Bowler2016}. Many evolutionary and spectral models exist \citep[e.g.,][]{Burrows1997, Allard2003, Saumon2008, Allard2013, Baraffe2015, Phillips2020}, but the degeneracy between BD age and mass (as younger, less massive BDs can appear as hot and luminous as older, more massive BDs) has complicated the process of testing these models, as masses cannot be independently determined for isolated field BDs. 
\par BDs in binary orbits around main sequence stars are important to study for two key reasons. Firstly, because a combination of radial velocities and astrometry (relative and/or absolute) can yield their dynamical mass \citep[e.g.,][]{Dupuy2017, Brandt2019, Fontanive2019, Rickman2020, mBrandt2021, Rickman2022, Bonavita2022, Franson2022, Li2023, Franson2023}, a model independent mass determination derived from orbital motion. Second, because we expect binaries that form via molecular cloud collapse to exhibit similar chemical compositions, and because brown dwarf interiors are fully convective \citep{Chabrier2000}, we can test our ability to retrieve atmospheric abundances for substellar companions on these objects of approximately known composition (that is, provided disequilibrium effects in the atmosphere do not impede our ability to infer the composition, the composition throughout the BD interior is not distinct from the composition of the atmosphere). This is not the case, for example, for planets formed via core accretion, where a rocky core might enrich the atmosphere with metals \citep[e.g.][]{Thorngren2019}. For these ``benchmark" companions to stars, we can measure the abundances of the host and compare to the abundances derived for the more opaque (physically and theoretically) atmosphere of the substellar object. Many spectral modeling frameworks are benchmarked against one another, but not against these benchmark BDs, whose dynamical mass is known after measuring their motion in a binary orbit, and whose composition in assumed to be approximately stellar. There is a relatively new and emerging body of work, that this paper contributes to, attempting to detect benchmark candidates, measure their dynamical masses, observe their atmospheres, apply existing evolutionary models to the benchmarks to check for consistency, and finally test the vast array of atmospheric model codes against objects of known mass \citep{Line2015, Peretti2019, Wang2022, Xuan2022}. 
\subsection{A sequence of L's: difficulty modeling clouds?}
\par The observed population of directly imaged, young giant planets follow the L-type sequence of spectral types \citep[][see their Figure 7]{Bowler2016}, and so the study of L-type BDs is well motivated by those seeking to understand the atmospheres of directly imaged giant planets. This sequence is distinguished by the presence of carbon monoxide CO dominated atmospheres whose near infrared colors redden as they become fainter. The challenge of accurately modeling atmospheres in the L-type regime is the apparent presence of condensate clouds. The influence of these clouds was first observed in the near-infrared color-magnitude relation for brown dwarfs, as later type L dwarfs become increasingly red, before the L-T transition where the condensate clouds no longer dominate NIR colors for field BDs \citep[e.g.][]{Knapp2004, Dupuy2012}. Additionally, the presence of these clouds can be inferred directly, by measuring the absorption due to cloud grains in the mid-infrared \citep[e.g.][]{Suarez2022, Suarez2023}, or indirectly, via the impact of cloud opacity on the shape the spectral slope of shorter wavelengths, or via variability studies that indicate the rotation of patchy clouds in and out of view \citep[e.g.][]{Vos2022}.
\par The carbon-to-oxygen ratio (C/O) of an object is theorized to encode information about the formation location or history of giant exoplanets, assuming the form, composition, and evolution of the circumstellar disk \citep{Oberg2011}. As demonstrated in, e.g. \citet{Molliere2022} the actual practice of linking C/O to the formation history of a given planet is challenging because these planet formation model assumptions can strongly influence the interpretation of a given measured C/O ratio. For young, massive directly imaged planets inhabiting the L dwarf sequence, even obtaining accurate measurements of this quantity is confounded by the presence of clouds. Indeed, in conducting an analysis of two late-T dwarf companions (whose atmospheres may be less strongly affected by the same kinds of clouds as L dwarfs), \citet{Line2015} find good agreement between the retrieved abundances of the BDs and their hosts. This has been more difficult to reproduce for L dwarfs.
\par \citet{Burningham2017} analyse the L4 spectral template 2MASS~J05002100+0330501, and the L4 dwarf 2MASSW~J2224438-015852, both field objects. They identify a major disagreement between their CO abundance and a solar abundances, and they note that future work focusing on benchmark L dwarfs may be necessary to rigorously test their ability to retrieve gas abundances. Interestingly, \citet{Peretti2019} find a similar disagreement when studying SPHERE YJH spectrum and K-band photometry of the benchmark L9 BD HD~4747~B. They derive a dynamical mass of $65.3\pm4.4\mathrm{M_{Jup}}$ from radial velocity and direct imaging, and use a retrieval analysis to identify C and O abundances that are discrepant with their measured abundances for the host star HD~4747~A. They note, however, that spectroscopic measurements of the object in K-band, at the 2.29~\textmu m CO absorption band, are needed to truly constrain these abundances.
\par \citet{Gonzales2020} conducted a retrieval analysis of the L7+T7 binary SDSS J1416+1348AB where they recovered an equivalent C/O ratio between the binary pair. \citet{Wang2022} analysed a high spectral resolution observation of the benchmark L-type BD HR~7672~B, a close-in companion to the solar type star HR~7672~A, with the Keck Planet Imager and Characterizer (KPIC). They find elemental abundances consistent with the primary star using an extension of the \texttt{petitRADTRANS} code (see our description in \S\ref{sec:petit} for references). \citet{Xuan2022}, also using KPIC in conjunction with \texttt{petitRADTRANS} find the C and O
abundances derived for HD~4747~B in exceptional agreement, and the C/O ratio only discrepant at the $2\,\sigma$ level with its host star. They cite the uncertainty in stellar abundances (due to non-LTE effects) and/or sequestration of O in cloud condensates making up the difference.
\subsection{HD~72946: a high-contrast L5 benchmark}
\par HD~72946~A is a bright, nearby star \citep[G=7.02 mag, d=$25.87\pm0.08$ pc,][]{GaiaCollaboration2021}. It is comoving with the spectroscopic binary HD~72945~AB, at a separation of $\sim\!10"$ \citep{GaiaCollaboration2021}. From high resolution optical spectroscopy, \citet[][]{Bouchy2016} measured a $\mathrm{T_{eff}}=5686\pm40\mathrm{K}$, $\log{g}=4.50\pm0.06~\mathrm{dex}$, and $\mathrm{[Fe/H]}=0.11\pm0.03~\mathrm{dex}$ stellar atmosphere. Various studies corroborate a super-solar metallicity, \citep[e.g.,][]{Luck2006,Casagrande2011,Aguilera-Gomez2018}. \citet{Luck2017} find a stellar $\mathrm{C/O}\sim\!0.5$. \citet{Maire2020} conducted an analysis of available data in order to determine the age of the system, and found a range of 0.8-3 Gyr, with a most probably value of 1-2 Gyr from a combination of Lithium data, stellar kinematics, isochronal analysis, and a rough gyrochronological estimate. \citet{mBrandt2021} conduct a similar age analysis, a ``Bayesian activity age method," from which they derive a posterior distribution of $1.9_{-0.5}^{+0.6}$ Gyr, in agreement with \citet{Maire2020}.
\par In 2016, \citeauthor{Bouchy2016} reported a radial-velocity (RV) signal, measured using the ELODIE and SOPHIE instruments, that was best fit by a low mass companion with minimum mass $M\sin(i)=60.4\pm2.2~\mathrm{M_{Jup}}$ and a $\sim\!16~\mathrm{yr}$ period. The RV data covered the full phase of the companion's orbit. Subsequently, the companion was directly imaged using the VLT/SPHERE instrument \citep{Maire2020}. \citeauthor{Maire2020} jointly fit the RV measurements with both their SPHERE relative astrometry and absolute astrometry from the Hipparcos-Gaia Catalog of Accelerations \citep[HGCA,][]{Brandt2018}, assuming a stellar mass prior of $0.986\pm0.027\mathrm{M_\odot}$ based on their isochronal analysis. They derive a dynamical mass of $72.4\pm1.6\mathrm{M_{Jup}}$ for the companion, confirming its substellar nature. They compute $\log{L_{bol}/L_\odot}=-4.11\pm0.10~\mathrm{dex}$ by converting their SPHERE/IFS J-band magnitude into a $\mathrm{J_s}$ magnitude and applying the empirical relation in \citet{Filippazzo2015}.
\par With updated absolute astrometry from \textit{Gaia} eDR3 \citep{GaiaCollaboration2021}, adopting the same stellar mass prior as \citet{Maire2020}, \citet{mBrandt2021} compute an orbital solution that gives $\mathrm{M_B} = 72.5\pm1.3\mathrm{M_{jup}}$. They use the computed dynamical mass to benchmark three evolutionary models: Burrows97 \citep{Burrows1997}, SM08-hybrid \citep{Saumon2008}, and \texttt{ATMO2020} \citep{Phillips2020}. They compute $\log{L_{bol}/L_\odot}=-4.133\pm0.023~\mathrm{dex}$ by computing the MKO and 2MASS photometry of the best fit template found in \citet{Maire2020} and using the relation in \citet{Dupuy2017}. They use measured age and bolometric luminosity to derive a model mass, and measured mass and bolometric luminosity to derive model ages. They find the age for HD~72946~B consistent with predictions from SM08-hybrid and \texttt{ATMO2020}, but discrepant by $1.4\,\sigma$ with the \citet{Burrows1997} models. This discrepancy is expected, as those models are cloud-free. They interpret their results as evidence that HD~72946~B's substellar cooling age is in agreement with it's host age.
\par Despite its relatively recent discovery and close separation, HD~72946~B is becoming an important benchmark for brown dwarfs near the hydrogen-burning limit. This work presents new observations of HD~72946~B in the K-band using the VLTI/GRAVITY instrument, and seeks to demonstrate GRAVITY's exceptional ability to improve studies of benchmark objects. We then seek to prove HD~72946~B's utility as a spectral benchmark by interrogating the atmospheric retrieval codes employed in the study of L-type exoplanets by the ExoGRAVITY Large Program \citep{Lacour2019}.





\section{Observations and Data Reduction} \label{sec:observations}
\subsection{VLTI/GRAVITY}

\begin{deluxetable*}{ccccccccccc}
\tablewidth{\textwidth}
\tablecaption{Observing log. NEXP, NDIT, and DIT denote the number of exposures, the number of detector integrations per exposure, and the detector integration time, respectively, and $\tau_0$ denotes the atmospheric coherence time. The fiber pointing is the placement of the science fiber relative to the fringe tracking fiber (which is placed on the central star), $\gamma$ is the coupling efficiency at the position of the companion (see Table \ref{tab:astrometry}). \label{tab:obslog}}
\tablehead{Date & \multicolumn{2}{c}{UT time} & \multicolumn{2}{c}{NEXP/NDIT/DIT} & Airmass & $\tau_0$ & Seeing & Fiber pointing & $\gamma$ \\
& Start & End & HD~72946~B & HD~72946~A & & & & $\Delta$RA/$\Delta$DEC &}
\startdata
\hline\hline
2020-02-09 & 05:11:28 & 05:36:55 & 2/8/60~s & 3/64/1~s & 1.22--1.28 & 5.2--6.0~ms & 0.68--0.95$^{\prime\prime}$ & 151.9,153.4 & 0.968 \\
2021-01-05 & 05:37:04 & 06:12:42 & 2/8/100~s & 3/64/1~s & 1.17--1.19 & 2.5--4.2~ms & 0.96--1.36$^{\prime\prime}$ & 157.0,101.0 & 0.999 \\
2021-01-30 & 06:09:31 & 07:03:12 & 3/8/100~s & 4/64/1~s & 1.27--1.46 & 2.1--2.9~ms & 1.09--1.45$^{\prime\prime}$ & 155.6,100.3 & 0.991 \\
2022-01-25 & 05:01:25 & 05:26:05 & 2/16/30~s & 3/64/1~s & 1.17--1.18 & 6.2--8.1~ms & 0.57--0.74$^{\prime\prime}$ & 156.5,38.0 & 0.994 \\
\hline
\enddata
\end{deluxetable*}

\begin{figure*}
    \centering
    \includegraphics[width=\textwidth]{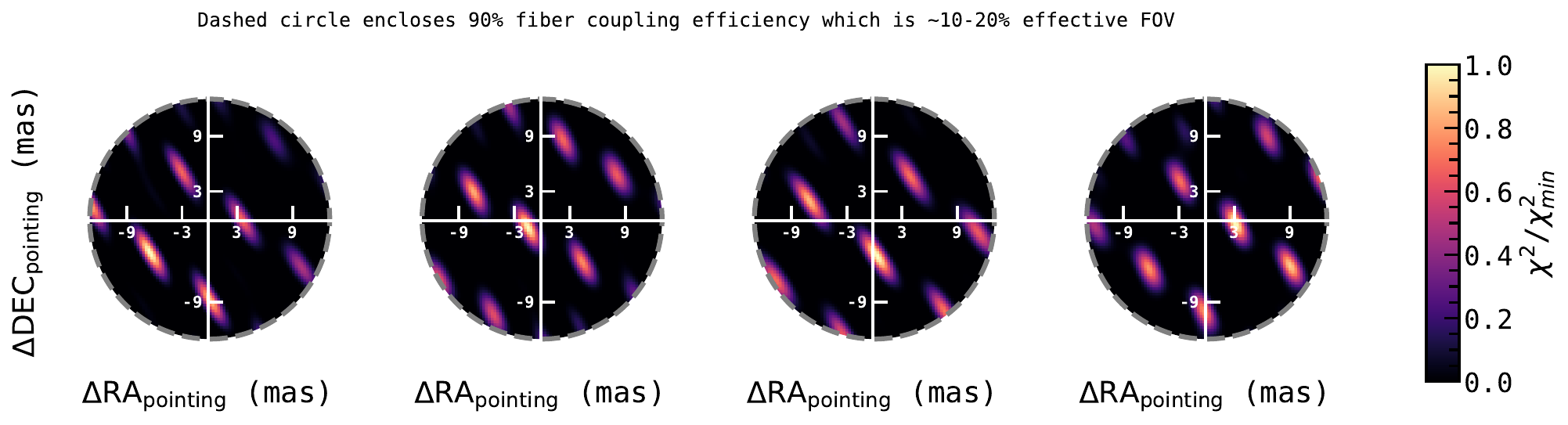}
    \caption{Detections of HD~72946~B with VLTI/GRAVITY. Each panel visualizes the periodogram $\chi^2$ map calculated after subtracting the stellar residuals. Each epoch in Table \ref{tab:obslog} is presented chronologically, left to right. The dashed grey circle indicates the radius beyond which the coupling efficiency into the instrument fiber is $<90\%$ (this is much smaller than the complete fiber FOV, which is $\sim\!60\mathrm{mas}$). The origin is the placement of the science fiber on-sky for a given observation, a prediction based on the previous available orbit fit. The strongest peak in the $\chi^2$ map indicates the position of the companion, with characteristic interferometric side-lobes whose shape and distribution depend on the u-v plane coverage.}
    \label{fig:chi2map}
\end{figure*}

\par We observed HD~72946~B on the 9th of February, 2020, again on the 1st and 30th of January, 2021, and on the 25th of January, 2022 (UTC) using the European Southern Observatory (ESO) Very Large Telescope Interferometer (VLTI)'s four 8.2m Unit Telescopes (UTs) and the GRAVITY instrument \citep{GravityCollaboration2017} in fringe tracking mode \citep{Lacour2019}. The observations were carried out as target visibility and bad weather backups to the ExoGRAVITY large program \citep{Lacour2020}, via programs 1104.C-0651(B) and 1103.B-0626(D). The observing log, presented in Table \ref{tab:obslog}, records the length of the observations and number of files recorded. The atmospheric conditions were rather good during most of the observations. 
The placement of the science fiber was based on preliminary orbit fits to the available relative astrometry and radial velocities of the system \citep{Maire2020} and is reported in Table \ref{tab:obslog}, along with the analytical coupling efficiency at the location of the companion \citep[a function of the distance between the companion location and the fiber pointing, see Appendix A in][]{Wang2021}. The coupling efficiency was $>95\%$ for all observations, so we do not correct for the effect of coupling efficiency on the observed spectra.

\par We extracted the complex visibilites on the host and the companion, which were phase-referenced with the metrology system, for each observation using the Public Release 1.5.0 (1 July 2021\footnote{\url{https://www.eso.org/sci/software/pipelines/gravity/}}) of the ESO GRAVITY pipeline \citep{Lapeyrere2014}. We then decontaminated the flux on the companion due to the host using a custom python pipeline developed by our team
This pipeline is described in detail in Appendix A of \citet{GravityCollaboration2020}. 
\par We obtained astrometry for each epoch by analysing the phase of the ratio of coherent fluxes. In short, our pipeline generates a $\chi^2$ periodogram power map over the fiber's field-of-view (Figure \ref{fig:chi2map}). The astrometry is taken to be the minimum of the $\chi^2$ map. We estimated the uncertainty on each astrometric point from the RMS of astrometric values fit to each individual exposure\footnote{The typical precision is on the order of $\sim\!100$~\textmu as $\gg 16.5$~\textmu as (the theoretical limit of VLTI/GRAVIY), due to high and low frequency phase errors induced by instrumental systematics.}
Then, the pipeline extracts the ratio of the coherent flux between the two sources, i.e. the ``contrast spectrum" of the companion, which is robust to variations in atmospheric quality and instrument stability \citep{Nowak2020}, at the location of the companion.

\begin{deluxetable*}{cccccc}
\tablewidth{\textwidth}
\tablecaption{New relative astrometry of HD~72946~B around HD~72946~A. \label{tab:astrometry}}
\tablehead{
\multicolumn{6}{c}{GRAVITY} \\ 
\colhead{Epoch [MJD]} & \colhead{$\Delta$RA [mas]} & \colhead{$\sigma_{\Delta\mathrm{RA}}$ [mas]} & \colhead{$\Delta$Dec [mas]} & \colhead{$\sigma_{\Delta\mathrm{Dec}}$ [mas]} & \colhead{$\rho$}
}
\startdata
58888.22    & 145.54   & 0.08           & 149.92    & 0.12       & -0.8165            \\
59219.24    & 155.64   & 0.02           & 100.24    & 0.04       &  0.0253               \\
59244.27    & 155.95   & 0.06           & 96.41     & 0.06       & -0.500               \\
59604.22    & 159.67   & 0.04           & 37.78     & 0.0       & -0.8825  \\   
\hline\hline
\enddata
\tablecomments{The co-variance matrix can be reconstructed using $\sigma_{\Delta\mathrm{RA}}^2$ and $\sigma_{\Delta\mathrm{Dec}}^2$ on the diagonal, and $\rho\times\sigma_{\Delta\mathrm{RA}}\times\sigma_{\Delta\mathrm{Dec}}$ on the off-diagonal.}
\end{deluxetable*}

\begin{figure}
    \centering
    \includegraphics[width=0.45\textwidth]{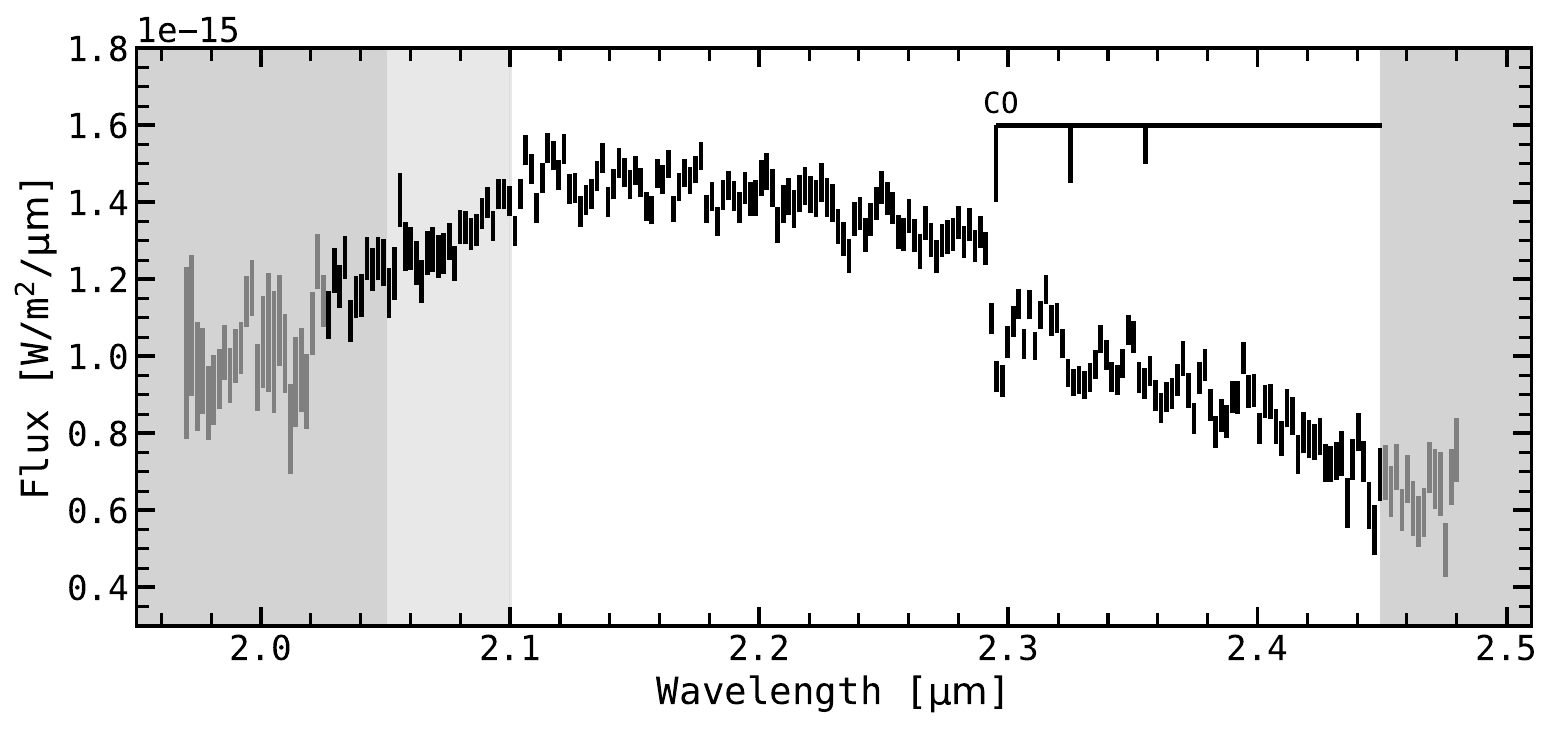}
    \caption{The K-band spectrum of HD~72946~B. The flux calibrated companion spectrum is a weighted combination of the 4 observed contrast spectra, multiplied by the flux calibrated host spectrum (Figure \ref{fig:hostspec}). The errorbars represent the diagonal of the full co-variance matrix. The 2.29~\textmu m carbon monoxide bandhead is indicated, and regions of poor atmospheric transmission are shaded grey \citep{Vacca2003}. The grey data points are excluded from our atmospheric model fits.
    }
    \label{fig:spectrum}
\end{figure}

\par The relative astrometry is listed in Table \ref{tab:astrometry}. For GRAVITY observations, we give the RA and Dec positions and their uncertainties as measured, in addition to the correlation coefficient that describes the elliptical confidence interval. The extracted spectrum, the contrast spectrum multiplied by a synthetic host spectrum (see \S\ref{sec:host}), is shown in Figure \ref{fig:spectrum}.

\subsection{Previous observations of HD~72946}
\par We obtained the optical stellar spectrum used in \S\ref{sec:host} and the radial velocities of HD~72946A published in \citet{Bouchy2016} through the ELDOIE and SOPHIE archives\footnote{\href{http://atlas.obs-hp.fr/sophie/}{atlas.obs-hp.fr/sophie/}} \citep{Moultaka2004}. The spectrum used to determine HD~72946~A's stellar parameters was observed on 2008-03-12, and has a signal-to-noise ratio (S/N) $\approx 150$ at $5000$ \AA.
\par We obtained the HGCA eDR3 edition \citep{tBrandt2021} within \texttt{orbitize!} \citep{Blunt2020} on a branch of the \texttt{orbitize} github that automates retrieval of the HGCA\footnote{Our orbit fit including proper motion anomaly used a development version of \texttt{orbitize!} recorded in \href{https://github.com/sblunt/orbitize/commit/5ffc5c01cd5318bce2b17392bec12578f8f5dcb4}{commit 5ffc5c0} from May 2022}.
\par We also make use of the previous direct detection of the companion with the SPHERE instrument in the Y-, J-, and H-bands \citep{Maire2020}. We include their epoch of relative astrometry in our orbital analysis and the spectro-photometry in our spectral analysis. We do not re-reduce the data, and adopt the values as recorded in \citet{Maire2020}, except that we use their spectrophotometric contrast measurements (as opposed to their absolute flux measurements) and transform them into absolute flux measurements using a synthetic host spectrum based on our analysis in \S\ref{sec:host}, in order to be consistent with the absolute flux we determine for the VLTI/GRAVITY spectrum. We propagated uncertainties in quadrature, and decided to inflate the uncertainty estimates of the H-band photometry, and the last 7 blue spectral channels in the Y-J spectrophotometry. We justify this on the basis that the blue channels in the SPHERE data suffer from significantly decreased throughput compared to the red channels, and that the H-band photometry likely included correlated noise that, unlike the spectroscopic datasets, we could not account for via a correlation matrix. For the last 7 blue spectral channels in the Y-J spectrophotometry, we added the median uncertainty in quadrature with the median uncertainty estimate from the remaining channels. We similarly inflated the uncertainty on the H1 and H2 photometry by a factor of 3.5 (from $\pm1.36\cdot10^{-17}~\mathrm{W/m^2/\mu m}$ to $\pm4.88\cdot10^{-17}~\mathrm{W/m^2/\mu m}$), adding the median uncertainty from reliable Y-J channels in quadrature to the median H-band photometric uncertainties.

\section{Host Analysis} \label{sec:host}

\par In order to properly assess the properties of HD~72946~B, we conducted an up-to-date analysis of the host star. We sought to determine the most precise mass for the host possible. While our orbit fits that include the proper motion anomaly will technically weigh the host as well as the companion, our knowledge of the host's properties from isochronal and spectroscopic modeling can place a much stronger constraint on the host mass (and therefore aid in a more precise mass determination for the companion). Since we are also interested in the abundances of the brown dwarf, we sought to measure the abundances of the host, primarily the stellar carbon-to-oxygen (C/O) ratio. Although HD~72946~A is a solar neighborhood, solar type star, assuming solar abundances without verification can lead to wrong conclusions \citep[e.g.][for details on one such cautionary tale]{Reggiani2022}.

\subsection{Stellar Parameters}
\par Using the algorithm outlined in \cite{Reggiani2022} we obtained the host star fundamental and photospheric parameters. Our analysis makes use of both the classical spectroscopy-only approach\footnote{The classical spectroscopy-only approach to photospheric stellar parameter estimation involves simultaneously minimizing for individual line-based iron abundance, inferences the difference between \ion{Fe}{1} \& \ion{Fe}{2}-based abundances, as well as their dependencies on transition excitation potential and measured reduced equivalent width.} and isochrones to infer accurate, precise, and self-consistent photospheric and fundamental stellar parameters. The method improves the composition determination by leveraging the isochrones to help determine the effective temperature and surface gravity using archival photometry and \textit{Gaia} parallax, while the  spectrum determines the abundances and microturbulence parameters. 
\par The inputs to our photospheric and fundamental stellar parameter inference include the equivalent widths of \ion{Fe}{1} and \ion{Fe}{2} atomic absorption lines. The absorption lines data are from \cite{galarza2019} for lines from \citet{melendez2014} found to be insensitive to stellar activity. We measured the equivalent widths by fitting Gaussian profiles with the \texttt{splot} task in \texttt{IRAF} \citep{Tody1986, Tody1993} to our continuum-normalized spectrum.  Whenever necessary, we use the \texttt{deblend} task to disentangle absorption lines from adjacent spectral features. We included multiwavelength photometry (Gaia Data Release 3 G, Two Micron All Sky Survey (2MASS) J, H, and Ks, and Tycho B, and V), and Gaia DR3 parallax \citep{gaia2016,gaia2018,fouesneau2022,skrutskie2006,hogg2000}. We assume \citet{asplund2021} solar abundances and follow the steps described in \cite{Reggiani2022} to obtain the fundamental and photospheric stellar parameters of the host star from a combination of spectral information and a fit to MESA Isochrones and Stellar Tracks \cite[MIST;][]{dotter2016,choi2016,paxton2011,paxton2013,paxton2015,paxton2018,paxton2019}. We do the fitting through the \texttt{isochrones} package\footnote{\href{https://github.com/timothydmorton/isochrones}{github.com/timothydmorton/isochrones}} \citep{morton2015}, which uses \texttt{MultiNest}\footnote{\href{https://ccpforge.cse.rl.ac.uk/gf/project/multinest/}{ccpforge.cse.rl.ac.uk/gf/project/multinest/}} \citep{Feroz2008,Feroz2009,Feroz2019} via \texttt{PyMultinest} \citep{Buchner2014}. In particular, the parameters are computed via efficient interpolations across the MIST grid space and precomputed synthetic photometry for each isochrone is compared to the observed photometry of the star. The updated \textit{Gaia} parallax provides a well constrained surface gravity, and the broad wavelength photometric coverage constrains the effective temperature. These lead to a precise mass estimate. 
As a consistency check we also employed the 
\textit{colte}\footnote{\href{https://github.com/casaluca/colte}{github.com/casaluca/colte}} code \citep{casagrande2021} 
to estimate the stellar effective temperature via the infrared flux method (IRFM). 
The IRFM effective temperature is $\mathrm{T_{eff}}=5592\pm71$, fully consistent 
with our adopted effective temperature.
Our adopted fundamental and photosperic parameters are displayed in Table \ref{tab:host}. 

\begin{deluxetable}{lcc}
\tablecaption{Adopted Stellar Parameters}\label{tab:host} 
\tablewidth{0pt}
\tablehead{
\colhead{Property} & \colhead{Value} & \colhead{Unit}}
\startdata
Gaia DR3 $G$   & $7.024\pm0.002$ & Vega mag \\
2MASS $J$ & $5.882\pm0.024$ & Vega mag\\
2MASS $H$ & $5.609\pm0.027$ & Vega mag\\
2MASS $Ks$ & $5.497\pm0.021$ & Vega mag\\
Tycho $B$ & $7.933\pm0.02$ & Vega mag\\
Tycho $V$ & $7.159\pm0.02$ & Vega mag\\
Gaia DR3 parallax & $9.459\pm0.056$ & mas\\
\hline
\multicolumn{3}{l}{\textbf{Isochrone-inferred parameters}} \\
Effective temperature $T_{\mathrm{eff}}$ & $5638\pm14$ & K \\
Surface gravity $\log{g}$ & $4.51 \pm 0.01$ & cm s$^{-2}$ \\
Stellar mass $M_{\ast}$ & $0.97 \pm 0.01$ & $M_{\odot}$ \\
Stellar radius $R_{\ast}$ & $0.91\pm0.01$ & $R_{\odot}$ \\
Luminosity $L_{\ast}$ & $0.77 \pm 0.01$ & $L_{\odot}$ \\
\hline
\multicolumn{3}{l}{\textbf{Spectroscopically inferred parameters}} \\
$[\mathrm{Fe/H}]_\mathrm{1D LTE}$ &  $0.036\pm0.023$ & \\
$[\mathrm{Fe/H}]_{\mathrm{1D non-LTE}}$ & $0.069\pm0.080$ \\
$\mathrm{C/O}_{\mathrm{1D LTE}}$ & $0.512^{+0.047}_{-0.043}$ & \\
$\mathrm{Mg/Si}_{\mathrm{1D LTE}}$ & $1.047\pm0.124$ & \\
$\xi$ & $0.98 \pm 0.09$ & km s$^{-1}$ \\
\hline
\multicolumn{3}{l}{\textbf{Age Estimates}}\\
Isochrone-based age $\tau_{\mathrm{iso}}$ & $2.67^{+0.25}_{-0.49}$ & Gyr \\
$\mathrm{[Y/Mg]}$-based Age & $1.9\pm1.5$ & Gyr \\
$\mathrm{[Y/Al]}$-based Age & $0.5^{+1.5}_{-0.5}$ & Gyr
\enddata
\end{deluxetable}

\subsection{Chemical composition of HD~72946~A}
\par We inferred the elemental abundances of \ion{C}{1}, \ion{Na}{1}, \ion{Mg}{1}, \ion{Al}{1}, \ion{Si}{1}, \ion{Ca}{1}, \ion{Sc}{1}, \ion{Sc}{2}, \ion{Ti}{1}, \ion{Ti}{2}, \ion{V}{1}, \ion{Cr}{1}, \ion{Cr}{2}, \ion{Fe}{1}, \ion{Fe}{2}, \ion{Ni}{1}, \ion{Cu}{1}, \ion{Zn}{1} , \ion{Sr}{1}, \ion{Sr}{2}, \ion{Y}{2}, \ion{Zr}{2}, \ion{Ba}{2}, \ion{Ce}{2}, and \ion{Dy}{2}, from the equivalent widths (EWs) of absorption lines. The EWs were measured from our continuum-normalized spectrum by fitting Gaussian profiles with the \texttt{splot} task in \texttt{IRAF}. We use the \texttt{deblend} task to disentangle absorption lines from adjacent spectral features whenever necessary. We assume \citet{asplund2021} solar abundances and local thermodynamic equilibrium (LTE) and use the 1D plane-parallel solar-composition ATLAS9 model atmospheres and the 2019 version of \texttt{MOOG} \citep{Sneden1973, Sneden2012} to infer elemental abundances based on each equivalent width measurement.
\par SOPHIE spectra have a wavelength range of $3900$-$6940$ \AA, which does not include the oxygen triplet at $7771$-$7775$ \AA. Therefore, our oxygen abundance comes from the forbidden oxygen transition at $6300.3$ \AA. This transition is blended with a nickel line, and to obtain the abundance one must synthesize the spectral region \citep[e.g.][]{teske2014}. We used the \textit{linemake}\footnote{\href{https://github.com/vmplacco/linemake}{github.com/vmplacco/linemake}} code \citep{sneden2009,sneden2016,placco2021} to create a linelist for our analysis, and we updated the atomic data of the oxygen and nickel transitions according to \citet{johansson2003} and \citet{magg2022}. As this transition is not a strong transition, and due to the spectral quality at the region, our quoted oxygen abundance should be viewed as a lower limit. Optical spectrum of HD~72946~A covering the oxygen triplet region would allow us to determine a definitive oxygen abundance.
\par Whenever possible we also apply line-by-line abundance corrections for non-local thermodynamical equilibrium (NLTE) effects and three dimensional effects. We applied 1D non-LTE corrections for aluminum \citep{amarsi2020}, calcium \citep{amarsi2020}, and iron \citep{amarsi2016}.  We also make use of 3D non-LTE corrections for carbon \citep{amarsi2019}. We present our full set of chemical abundances, including the non-LTE corrected abundances, in Table \ref{host_chemical_abundances}. 

\subsection{Stellar Age}
The isochrones-based age of the host star is $\tau = 2.67^{+0.25}_{-0.49}$ Gyr. As HD~72946~A 
is part of the thin disk, it also follows the same chemistry-age relations, determined 
via different chemical clocks, that were derived using large samples of solar-twins 
in the solar neighborhood. We apply the chemical-clocks from \cite{Spina2016} and 
find $\tau_{\mathrm{[Y/Mg]}}=1.9\pm1.5$ Gyr and $\tau_{\mathrm{[Y/Al]}}=0.5^{+1.5}_{-0.5}$ Gyr. 

We note the agreement between this age and the 0.8-3 Gyr estimate made by \citet[][]{Maire2020}. Our [Y/Mg] age agrees well with the $1.9_{-0.5}^{+0.6}$ Gyr estimate from \citet{mBrandt2021}, but our isochrone-based age is slightly older. For this work we adopt the isochrone-based age of $\tau = 2.67^{+0.25}_{-0.49}$ Gyr, because it is consistent with the other fundamental stellar parameters determined here.

\subsection{Synthetic Stellar Spectrum}
\par In order to transform the contrast spectra measured for HD~72946~B with GRAVITY and SPHERE, we used \texttt{species} \citep{Stolker2020} to scale a \texttt{BT-NextGen} \citep{Allard2012} synthetic stellar spectrum with the inferred parameters we derived above to the archival photometry (see Table \ref{tab:host}). In order to scale the spectrum, we let the radius float while fixing all other parameters to their inferred values, except for parallax, where we placed a Gaussian prior around the \textit{Gaia} parallax. The resulting radius is consistent with our inferred radius to within $1\,\sigma$ uncertainties, and the residuals to the fit are within $2.5\,\sigma$ for each photometric measurement (see Appendix \ref{hostappendix}, Figure \ref{fig:hostspec}), which we take as validation that the BT-NextGen model is an appropriate synthetic spectrum to use in order to generate our absolute flux calibrated companion spectra. We sampled the spectrum at the resolution of the GRAVITY and SPHERE spectra using \texttt{spectres} \citep{Carnall2017}, and at over the bandpass of the SPHERE H-band photometry using \texttt{synphot} \citep{STScIDevelopmentTeam2018}. We found that at the resolution of the R=500 GRAVITY data, reasonable changes in the stellar properties produce no noticeable change in the absolute fluxed spectrum of the companion, and any line depth variations due to the (non-solar) abundances of the host are negligible at this resolution. To test the impact systematic errors in our assumed stellar properties would have on our flux calibrated spectrum of HD~72946~B, we tested the difference using a synthetic stellar spectrum with $\mathrm{T_{eff}}$ set to the IRFM effective temperature derived in \S3.1; we found that the two spectra based on independently derived $\mathrm{T_{eff}}$ produced a negligible difference compared to the uncertainties on the contrast spectrum itself.

\section{Orbit Analysis} \label{sec:orbit}
\par We now have four new astrometric observations of HD~72946~B at a precision unprecedented for a ``benchmark" BD (Table \ref{tab:astrometry}). This exquisite precision ($\sim\!30\times$ more precise than previous imaging) allows us to determine the orbital parameters for the object very finely. In both analyses we conduct, we include the SPHERE and GRAVITY relative astrometry, as well as the SOPHIE and ELODIE radial velocities. In one analysis, we exclude the proper motion anomaly and parallax data, fitting for only RVs and relative astrometry. In another analysis, we include the proper motion anomaly measurements from the HGCA, and the parallax from \textit{Gaia} in the fit.
\par We use the parallel-tempered \citep{Vousden2016} Affine-invariant \citep{Foreman-Mackey2013} MCMC algorithm packaged within \texttt{orbitize!}\footnote{\href{https://orbitize.readthedocs.io/en/latest/}{orbitize.readthedocs.io}.} that fits for the 6 parameter visual orbit \citep{Green1985}, the system parallax, offset and jitter terms for each RV instrument, and the masses of the star and companion. We place a physically motivated normally distributed prior $\mathcal{N}(0.97~\mathrm{M_\odot},\,0.03~\mathrm{M_\odot})$ on the mass of the primary based on the isochronal analysis conducted in \S\ref{sec:host}, and otherwise implement default priors on all orbital elements as described in \citet[][]{Blunt2020}. In the analysis excluding the proper motion anomaly, we set a uniform prior ($\pm1\,\mathrm{mas}$) around the system parallax and allow the MCMC to converge on the parallax naturally. In the analysis including the proper motion anomaly, we set a tight prior on the \textit{Gaia} eDR3 parallax recorded for the system $\mathcal{N}(38.981~\mathrm{mas}, 0.041~\mathrm{mas})$ following \citet{mBrandt2021} to facilitate comparison with their work.

\begin{figure}
    \centering
    \includegraphics[width=0.45\textwidth]{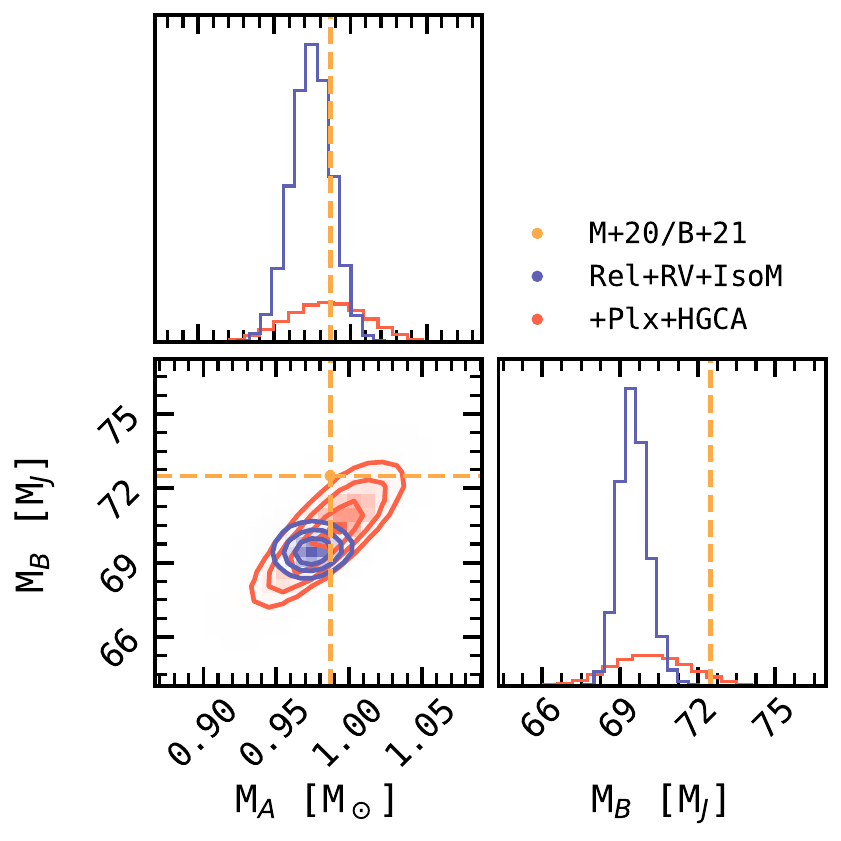}
    \caption{The posterior distribution of the masses of HD~72946~A and B from the orbit fits conducted in this work, compared to literature values. The fit excluding the proper motion anomaly and parallax (red) and including the proper motion anomaly (blue) are plotted against the median values reported in \citeauthor{Maire2020} and \citeauthor{mBrandt2021} (the single orange line, as the central values in both are effectively equivalent). We find a best fit mass $\mathrm{M_B}=69.5^{+0.5}_{-0.5}~\mathrm{M_{Jup}}$, still consistent with previous results within 1-2$\sigma$, but about $\sim\!2\,\mathrm{M_{Jup}}$ lower. In all works, $\mathrm{M_A}$ is effectively fixed by a strong prior based on an isochronal analysis of the primary star. Notably, our revised primary mass is slightly lower than that used in previous works, which explains the subsequent offset in secondary mass.}
    \label{fig:masspost}
\end{figure}

\begin{figure*}
    \centering
    \includegraphics[width=\textwidth]{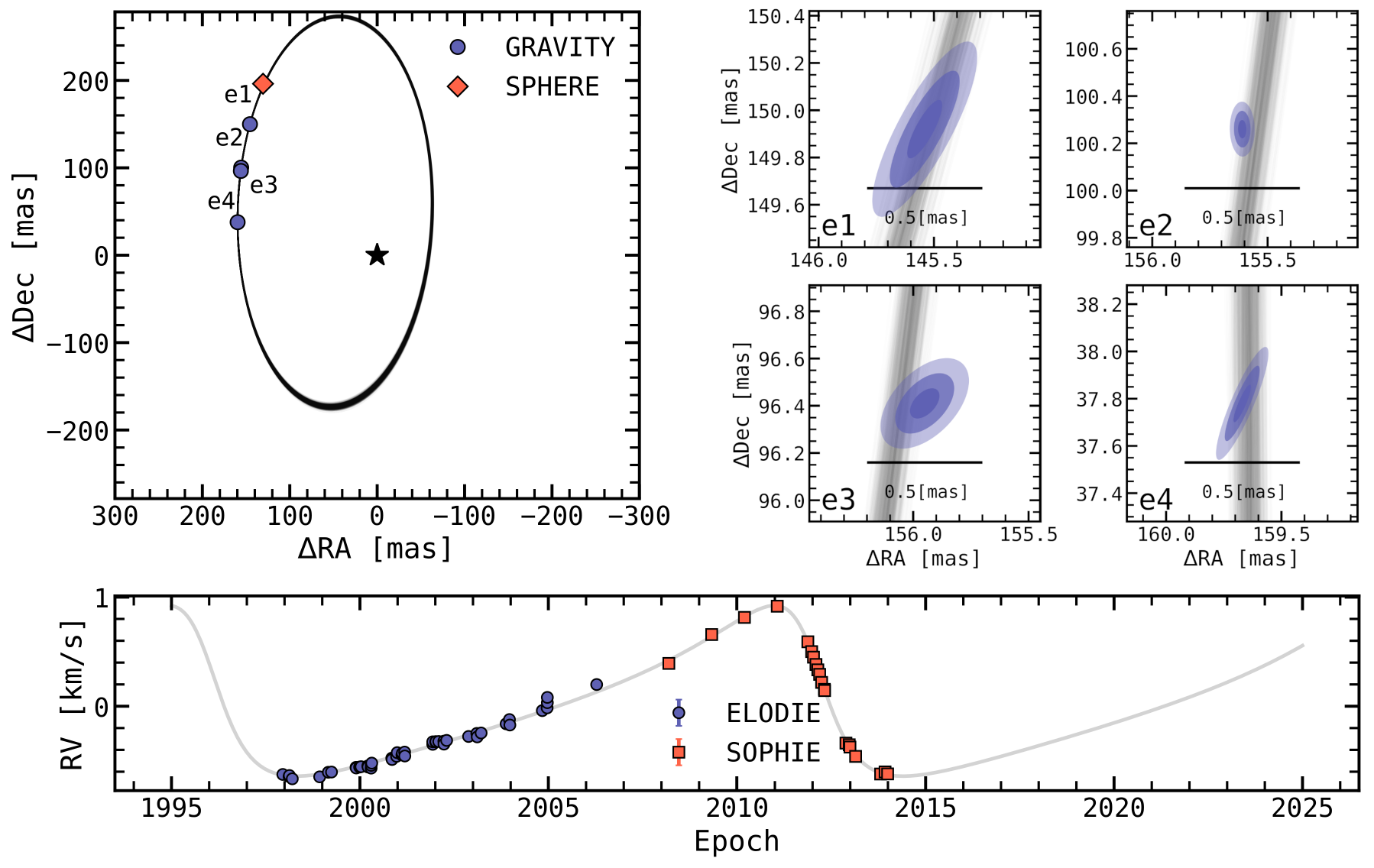}
    \caption{The orbit of HD~72946~B. Top left: 100 randomly drawn orbits from the posterior of orbital fits made using \texttt{orbitize!} are plotted in RA and Dec relative to HD~72946~A (black star); overplotted are the direct detections of HD~72946~B from VLT/SPHERE and VLTI/GRAVITY, whose errorbars are too small to visualize at this scale. Each GRAVITY detection is labeled a-d in chronological order. Top right: GRAVITY astrometry on HD~72946~B are overplotted against 100 randomly drawn orbits. The solid crosses represent the 1$\,\sigma$ errors on the astrometry in RA/Dec reported in Table \ref{tab:astrometry}, while the semi-transparent ellipses represent the 1, 2, and 3$\,\sigma$ confidence ellipses determined by the correlation coefficient (also in Table \ref{tab:astrometry}) that fully describes the confidence on the astrometry. Lower panel: Radial velocities of HD~72946~A from ELODIE and SOPHIE \cite{Bouchy2016}, overplotted against the 50 randomly drawn orbits in the above panel.}
    \label{fig:orbitize}
\end{figure*}

\par We initialize \texttt{orbitize!} with 50 walkers and 20 temperatures, and run 200,000 steps that are discarded as ``burn-in" before 20,000 are accepted to estimate the posterior distribution, for a total of 11,000,000 orbits fit. We visually inspected the chains to check for convergence. The full posterior distribution of orbital elements for both cases is visualized in Appendix \ref{sec:posteriors}, Figure \ref{fig:orbpost}. The median and 1$\,\sigma$ uncertainty on the orbital parameters derived from the posterior distribution of fits are reported in Appendix \ref{sec:posteriors}, Table \ref{tab:conservativeorbit} for the fit without the proper motion anomaly and in Table \ref{tab:optimisticorbit} for the fit with the proper motion anomaly. 
\par Figure \ref{fig:orbitize} plots the relative astrometry in Table \ref{tab:astrometry} with 50 randomly drawn orbits from the posterior distribution of orbital fits, illustrating the refined orbit achieved by combining a full phase of radial velocity measurements and incredibly precise on-sky astrometry. 

\par The orbital period of the system ($\sim\!16~\mathrm{yr}$) is low enough that there is significant non-linear motion of the primary and secondary components over the \textit{Gaia} observing baseline. There may be errors in the parallax/proper motion fit for the eDR3 data release for a host star like HD~72946~A because of this orbital motion. This can be solved with the addition of the per-scan epoch astrometry, or a parallax fit assuming a binary-solution (as was done for the system in DR3). For this reason, because we conducted the analysis using the eDR3 data, we fit two cases of orbits, one with and one without including the proper motion anomaly. When fitting including the proper motion anomaly, we used measurements of the proper motion anomaly from the eDR3 HGCA, which strategically inflates the \textit{Gaia} errors to avoid biasing the orbit fit. This appears to work well, leading to an increased precision on our mass, that still agrees with our fit that excludes absolute astrometry, but we present the fit without absolute astrometry for completeness.
\par We find primary and secondary masses slightly lower than in \citet{mBrandt2021}, but still in agreement at $2\,\sigma$. This is attributable to our revised host analysis, which slightly lowered the primary star mass estimate. Our semi-major axis and eccentricity also largely agree with their values. The precise GRAIVTY astrometry effectively fixes semi-major axis in our fits, and the radial velocities provide a significant period constraint, leading to a refinement in the remaining free parameter in Kepler's equation, the mass, as well as the on-sky orientation of the ellipse. 
\par Excluding absolute astrometry we find a secondary mass error of $3.134\%$, while including it we find a secondary mass error of $1.129\%$. In either case, this strong dynamical mass constraint makes HD~72946~B a standout benchmark candidate for the L5 class and for high mass BDs in general. Fitting with epoch astrometry (to be released in \textit{Gaia} DR4) could work to push the uncertainty on the secondary mass below $1\%$ for this object.


\par Using the isochronal age from \S\ref{sec:host} and our updated dynamical mass, we compared the age and mass of HD~72946~B to two evolutionary models (one cloudless, one cloudy), primarily to determine the expected radius for the BD. Our updated dynamical mass and isochronal age are not entirely inconsistent with previous findings \citep{Maire2020, mBrandt2021}, and so we do not repeat their benchmarking analysis (for instance, comparing marginalized combinations of the bolometric luminosity, age, and dynamical mass samples to a variety of evolutionary models to check for consistency) in this paper. Based on our dynamical mass and age estimate we compute the effective temperature, surface gravity, radius, and bolometric luminosity of HD~72946~B using the cloudless Sonora-Bobcat model \citep{Marley2021} and the cloudy ($f_{sed}=2$) SM08 model \citep{Saumon2008}. Both models assume solar metallicity. We drew 10,000 random samples from the posterior distributions on the dynamical mass and the isochronal age, and computed the corresponding parameters from each evolutionary model grid by interpolating linearly between grid points using the \texttt{species} package \citep{Stolker2020}.
\par We find an estimate of the radius of HD~72946~B of $0.89\pm0.01\,\mathrm{R_{J}}$ for the cloudy SM08 model and of $0.84\pm0.02\,\mathrm{R_{J}}$ for the cloudless Sonora Bobcat model. The estimated effective temperatures are $1567\pm70$K and $1470\pm52$K respectively. 

\section{Spectral Analysis} \label{sec:spectrum}

\par Having nearly doubled the wavelength coverage for HD~72946~B with the addition of our GRAVITY observations, we conducted an initial atmospheric analysis using both a self-consistent grid of model atmospheres and an atmospheric inversion (a ``retrieval") of a model atmosphere. Throughout, we were predominately interested in leveraging the dynamical mass to more accurately fit for atmospheric parameters. We generally avoided constructing models that would require intensive computational resources (e.g. free temperature or pressure node P-T profiles), which we postpone to future work. Here, we present a qualitative exploration of the new data and an overview of the atmospheric properties we computed for HD~72946~B.

\subsection{Self-consistent \texttt{BT-Settl} model grid} \label{subsec:specgrid}

\begin{figure*}
    \centering
    \includegraphics[width=\textwidth]{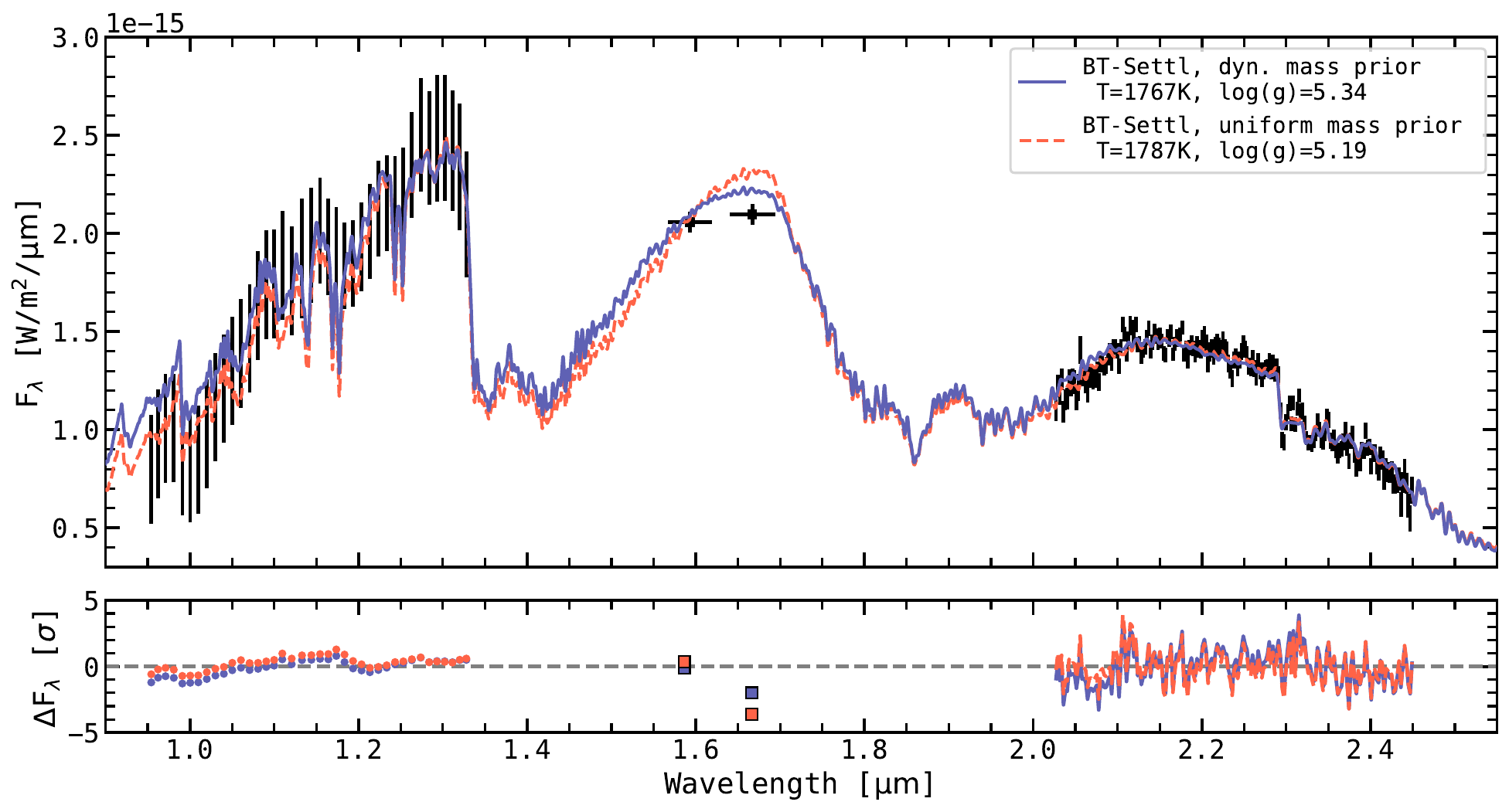}
    \caption{The observed spectrum of HD~72946~B fit to the \texttt{BT-Settl} grid. The best fit model spectrum assuming the dynamical mass is plotted as the solid blue curve, while the fit with mass as a free parameter is plotted as a dashed red curve; both models have been smoothed to R=500. The SPHERE/IFU spectrum, SPHERE/IRDIS photometry, and the GRAVITY spectrum are plotted in black, from left to right, respectively. Residuals to each fit are plotted below, and show some non-Gaussian structure. In particular, the SPHERE Y-J spectral slope appears to. be in tension with the model. Aside from a noticable underluminosity near 2.05\textmu m, the residuals to the GRAVITY spectrum are relatively normally distributed.}
    \label{fig:btspec}
\end{figure*}

\par In order to investigate the atmosphere of HD~72946~B we began by comparing the observed spectrum to a self-consistently produced grid of model atmospheres. The \texttt{BT-Settl-CIFIST} grid was chosen because of its microphysical cloud treatment for L-type BDs \citep{Allard2003, Allard2011, Allard2013, Baraffe2015}. We expected the model grid to provide good results for an old, high surface gravity brown dwarf based on the excellent match the model has demonstrated to the observed moderate (R=3000) spectra of the L5.5 field brown dwarf 2MASS 1507-16 \citep[see Figure 1,][]{Allard2014}. The model performs admirably for young, low surface gravity but like its contemporaries, it struggles to reproduce the H-band slope and J-band flux of L dwarfs \citep{Bonnefoy2010, Patience2012, Manjavacas2014}, overpredicts the effective temperature and underpredicts the radius compared to evolutionary models.
\par We used the \texttt{species} package to fit the \texttt{BT-Settl-CIFIST} grid, linearly interpolating spectra between grid points. We initialized UltraNest\footnote{\href{https://johannesbuchner.github.io/UltraNest/}{johannesbuchner.github.io/UltraNest/}} \citep{Buchner2017, Buchner2021} via \texttt{species} to sample the interpolated grid with 500 live points. We measured the posterior distribution on the grid parameters, namely effective temperature ($\mathrm{T_{eff}}$), log(g), radius, parallax, and included a Gaussian process parameterized by a squared-exponential kernel to account for correlated noise between wavelength channels in the SPHERE data \citep[see \S4.1 in][]{Wang2020}. When fitting the GRAVITY data, \texttt{species} accounts for the correlation matrix of the spectrum in the fit. 
\par To investigate its discerning power, we fit the \texttt{BT-SETTL-CIFIST} grid to the data twice, first without a prior on the mass of the object, and then assuming a Gaussian prior on the mass equivalent to the dynamical mass we derived in \S\ref{sec:orbit}. Effectively, this experiment let the sampler determine the log(g) and radius based only on the spectral measurements, and then constrained log(g) and radius based on the dynamical mass. Figure \ref{fig:btspec} plots the best fit \texttt{BT-Settl-CIFIST} model fits.
\par Even using Ultranest, which implements a ``safer" uncertainty estimation than other nested sampling packages \citep{Nelson2020}, the uncertainties on physical parameters derived here are likely underestimated due to systematic errors between various grids of self-consistent models and unaccounted-for sources of error in the spectra (for instance, uncorrected higher-order telluric or instrumental effects). That being said, the experiment yielded two distinct solutions for the atmospheric parameters; the constrained mass favoring a slightly cooler, higher surface gravity solution, while the free mass estimated a slightly hotter, lower surface gravity. The free mass maximum a posteriori spectrum yields a higher log evidence, with a $\Delta\ln{Z}=50\pm0.5$ compared to the mass prior fit. The difference between the two cases appears small, especially considering the inability of the implemented sampling methods to estimate systematic errors. Nevertheless, the difference in log(g) is significant when given the statistical uncertainties ($\Delta\mathrm{log(g)}=0.22$, $\delta\mathrm{log(g)\simeq0.02}$), and results in a mass discrepant from our dynamical mass by $32\,\mathrm{M_J}$. The posterior distribution of effective temperatures are discrepant with those derived from both a cloudless and cloudy evolutionary model (see Table \ref{tab:atmos}). Interestingly, with the inclusion of the dynamical mass prior, the spectroscopically inferred radius is shifted towards agreement with the evolutionary models.
\par The posteriors for both fits (Figure \ref{fig:btsettlpost}) are well behaved and effectively Gaussian. We report median and $1\,\sigma$ confidence intervals on each parameter in Table \ref{tab:atmos} for both fits. The best fit parameters from the dynamical mass constrained fit give a total luminosity of $\log{\mathrm{L/L_\odot}}=-4.15$.
\par The \texttt{BT-Settl} self-consistent models provide an excellent fit to the data. While the model appears a marginally poor fit to the 0.95-1.05~\textmu m slope of the SPHERE data, it otherwise captures the SPHERE and especially GRAVITY spectrum remarkably well, and with solar abundances. Small deviations could be due to data deficiencies (residual telluric contamination or poorly corrected instrumental throughput) or model deficiencies. Notably, there is systematic uncertainty involved when linearly interpolating spectra between widely spaced grid points, because line or molecular feature depths do not vary linearly with temperature or surface gravity \citep{Czekala2015}. Our experiment proves the usefulness of the dynamical mass as an independent physical prior, as it helped the sampler avoid over-fitting less reliable regions of the observation (or the model), and constrained the log(g) in the absence of a strong spectral constraint on the H-band slope, which is sensitive to variations in surface gravity.

\movetabledown=2.1in
\begin{table*}
\footnotesize
\centering
\begin{rotatetable}
\caption{Subset of parameters for HD~72946~B derived from evolutionary and atmospheric analysis.}
\label{tab:atmos}
\begin{tabular}{ccccccccccc}
\hline\hline
Model & \multicolumn{9}{c}{Notable Model Parameters}  & $\chi_\mathrm{red.}^2$/d.o.f. \\
 & Mass [$\mathrm{M_{J}}$] & $\mathrm{T_{eff}}$ [$\mathrm{K}$] & log(g) & Radius [$\mathrm{R_J}$] & [Fe/H] & C/O & \multicolumn{2}{c}{Cloud parameters} & log(L/L$_\odot$) & \\
\hline
Evolutionary & & & & & & & & & \\
Sonora Bobcat & $69.5\pm0.5$ & $1567\pm70$ &  $5.38\pm0.02$ & $0.84\pm0.02$ & solar & solar & \multicolumn{2}{c}{cloudless} & $-4.39\pm0.09$ & \nodata \\
SM08 (cloudy) & $69.5\pm0.5$ & $1470\pm52$ & $5.33\pm0.01$ & $0.89\pm0.01$ & solar & solar & \multicolumn{2}{c}{$\mathrm{f_{sed}}=2$} & $-4.45\pm0.07$ & \nodata \\
\hline
Self-consistent (forward model) & & & & & & & \\
\texttt{BT-Settl-CIFIST}, $\mathcal{U}(\mathrm{M_B})$ & $37.52\pm2.52$ & $1790\pm5$ & $5.17\pm0.02$ & $0.79\pm0.01$ & solar & solar & \multicolumn{2}{c}{microphysical} & $-4.15\pm0.01$ & 1.36/228 \\
\texttt{BT-Settl-CIFIST}, $\mathcal{N}(\mathrm{M_B})$ & $69.5\pm0.5$ & $1770\pm4$ & $5.39\pm0.01$ & $0.84\pm0.01$ & solar & solar & \multicolumn{2}{c}{microphysical} & $-4.140\pm0.005$ & 1.58/228 \\
\hline
Atmospheric Inversion (retrieval) & & & & & & & & & \\
\texttt{petitRADTRANS} Cloudless & & & & & & & & & \\
$\mathcal{N}(\mathrm{M_B})$ & $69.5\pm0.5$ & $1656\pm18$ & $5.29\pm0.02$ & $0.94\pm0.02$ & $0.86\pm0.05$ & $0.63\pm0.01$ & & & $-4.20\pm0.01$ & 1.51/221  \\
\texttt{petitRADTRANS} Grey Cloud & & & & & & & log($\kappa_\mathrm{grey}$) & log(P$_\mathrm{top}$) \\
$\mathcal{N}(\mathrm{M_B})$ & $69.5\pm0.5$ & $1720\pm20$ & $5.39\pm0.02$ & $0.86\pm0.01$ & $0.69\pm0.05$ & $0.58\pm0.01$ & $2.20\pm1.60$ & $0.52\pm0.03$ & $-4.21\pm0.01$ & 1.50/219 \\
\texttt{petitRADTRANS} EddySed & & & & & & & $\mathrm{f_{sed}}$ & log($\mathrm{K_{zz}}$) \\
$\mathcal{U}(\mathrm{M_B})$ & $119^{+50}_{-40}$ & $1868\pm21$ & $5.70^{+0.17}_{-0.20}$ & $0.77\pm0.02$ & $0.79\pm0.14$ & $0.58\pm0.02$ & $14.2\pm3.4$ & $3.2^{+0.61}_{-0.55}$ & $-4.17\pm0.01$ & 1.42/214 \\

$\mathcal{N}(\mathrm{M_B})$ & $69.5\pm0.5$ & $1846\pm23$ & $5.45\pm0.02$ & $0.78\pm0.02$ & $0.72\pm0.14$ & $0.59\pm0.02$ & $15.5^{+2.6}_{-2.9}$ & $3.07^{+0.77}_{-0.62}$ & $-4.16\pm0.01$ & 1.45/214 \\

$\mathcal{U}(\mathrm{M_B})$, $\mathcal{N}(\mathrm{[Fe/H]_A, C/O_A})$& $38.0^{+13.0}_{-8.1}$ & $1765\pm16$ & $5.12\pm0.13$ & $0.85\pm0.02$ & $0.03\pm0.01$ & $0.48^{+0.02}_{-0.03}$ & $15.8^{+2.8}_{-3.57}$ & $2.63^{+0.43}_{-0.38}$ & $-4.17\pm0.01$ & 1.97/214 \\

$\mathcal{N}(\mathrm{M_B})$, $\mathcal{N}(\mathrm{[Fe/H]_A, C/O_A})$ & $69.5\pm0.5$ & $1785\pm24$ & $5.39\pm0.02$ & $0.83\pm0.02$ & $0.03\pm0.01$ & $0.43\pm0.01$ & $14.0^{+3.4}_{-3.6}$ & $2.58^{+0.40}_{-0.36}$ & $-4.17\pm0.01$ & 1.58/214 \\
\hline
\end{tabular}
\tablecomments{For each type of model considered (evolutionary, self-consistent atmospheric, atmospheric inversion) we record the mean and $1\,\sigma$ standard deviation for parameters of interest. $\mathcal{U(P)}$ or $\mathcal{N(P)}$ denotes a uniform or normally distributed prior on the parameter P.  There are 39+193+2-$\Sigma P_i$=234-$\Sigma P_i$ degrees of freedom for each spectral fit.}
\end{rotatetable}
\end{table*}




\subsection{\texttt{petitRADTRANS} Retrievals} \label{sec:petit}
\par While pre-computed grids of model atmospheres can treat the interrelation between pressure-temperature (P-T) structure of the atmosphere, particle clouds, and chemistry self-consistently from first principles, in order to determine the precise atmospheric abundances of the atmosphere it is necessary to make a number of assumptions and instead conduct an atmospheric inversion, or ``retrieval," where rapidly computed model atmospheres based on simplifying assumptions are sampled from a posterior space and compared to the data.
\par We selected the open source \texttt{petitRADTRANS} \citep{Molliere2019} radiative transfer code to perform our atmospheric retrievals. The code has been used to conduct retrievals of young, low surface gravity objects \citep[e.g. HR~8799~e, $\beta$~Pic~b,][]{Molliere2020, Nowak2020} successfully, capturing the slopes and absorption features of spectra well, as well as retrieving reasonable cloud properties, with the usual caveats concerning effective temperature and radii, which are incongruous compared to expectations from evolutionary models. Despite its success with directly imaged planets, and having been benchmarked against self-consistent models, the code remains relatively untested for older, higher surface gravity objects in the literature. We used \texttt{species} as a wrapper for \texttt{petitRADTRANS}, and the scripts used to generate the retrievals presented herein are available upon request \footnote{A generalized retrieval example can be found online at \href{https://species.readthedocs.io/en/latest/tutorials/atmospheric_retrieval.html}{species.readthedocs.io}}. The \texttt{species} retrieval implementation uses \texttt{pyMultiNest} \citep{Feroz2009, Buchner2014} to sample model parameters before passing these parameters to \texttt{petitRADTRANS} to generate a model spectrum by solving the radiative transfer equation, and then compares the model spectrum to the data. 
\par To limit computational expense, in this initial retrieval reconnaissance we run \texttt{petitRADTRANS} in constant sampling efficiency mode. Unfortunately, this choice limits our ability to inter-compare the likelihood and posterior probability of each retrieval. We intended instead to qualtiatively explore a handful of questions informed by the retrievals.
\par We defer the reader to \citet{Molliere2020} for many details, but attempt to describe the most essential elements of the retrieval here. In general, our \texttt{petitRADTRANS} retrievals assume abundances are in chemical equilibrium.
We include CO, H$_2$O, CH$_4$, NH$_3$, CO$_2$, Na, K, TiO, VO, FeH, H$_2$S line species and collision induced absorption (CIA) of H$_2$ and He. The line opacities are taken from the ExoMolOP database \citep{Chubb2021}, in the ``petitRADTRANS" format\footnote{\href{https://www.exomol.com/data/data-types/opacity/}{www.exomol.com}}, while the CIA opacities are those from \citet{Borysow1988, Borysow1989, Borysow2001, Richard2012}. The abundances are parameterized as functions of the carbon-to-oxygen ratio C/O and metallicity [Fe/H], as well as the pressure-temperature structures discussed below. As in \S\ref{subsec:specgrid}, we fit for the correlated noise in the SPHERE spectrum, and the GRAVITY spectrum correlation matrix is accounted for in the fit. 

\subsubsection{P-T structure}
\par We implemented the ``3-part" or ``Molli\`ere" P-T parameterization that is described in \S2.2 of \citet{Molliere2020}. This P-T profile is motivated by the desire to strike a balance between a purely parametric and purely free P-T structure. 
This 3-part P-T profile splits pressure space into three regions, where at high temperatures there are free pressure nodes, in the photospheric region (between $\tau = 0.1$ and the radiative-convective boundary) we apply the Eddington approximation, and below the radiative-convective boundary (high pressures), we apply a moist adiabat. The moist adiabatic gradient is found by interpolating in the T-P-[Fe/H]-C/O space of the chemistry table, and the atmosphere is forced onto the moist adiabat once it becomes Schwarzschild-unstable. This P-T profile avoids enforcing the isothermal upper atmosphere required by the Eddington approximation, while enjoying the benefits of the approximation, namely the physical motivation, speed of analytic computation, and the reduced number of free parameters. We did not test fully flexible retrievals using free-pressure or free-temperature nodes \citep[e.g.][]{Wang2022}, even though it has been suggested that the rigid analytical retrievals could bias retrieved abundances. We decided not to test fully flexible P-T profile retrievals largely because of their computational intensity. Future work could indeed investigate this assumption using the existing data or data of higher spectral resolution, which would prevent overfitting. 

\subsubsection{Cloud parameterization}
\par We also investigated the affect of clouds and cloud parameterization on our retrievals. \citet[][]{Suarez2022} find that the MIR \textit{Spitzer} spectra of mid-L-type field brown dwarfs had the strongest absorption signatures due to silicate grains at 8-11~\textmu m. These silicates are thought to nucleate the clouds whose opacity dominates the NIR spectra of substellar objects \citep[e.g.][]{Cushing2006}. As HD~72646~B is an L5-type, it is squarely within the peak of the silicate cloud sequence \citep[][see their Figure 6]{Suarez2022} and thus could reasonably be expected to host silicate clouds. Alternatively, \citet{Tremblin2015, Tremblin2016} note that the NIR spectra of substellar objects can be explained by a reduced temperature gradient generated by diabatic convection driven by the thermal and compositional gradients in the atmosphere \citep{Tremblin2019}. 
\par The first and more simplistic cloud model treats the cloud deck as a grey -- meaning wavelength independent -- non-scattering absorber, parameterized by the opacity of the cloud, $\log{\kappa_\mathrm{grey}}$, and the pressure of the cloud top, $\mathrm{P_{top}}$. The physical implications of such a simple cloud model might be overstated, and a more honest interpretation of this simplistic cloud model is that, marginalized over both parameters, it acts to damp the spectrum uniformly across wavelength. 
\par The second cloud model follows the the \citet{Ackerman2001} parameterization \citep[see \S2.4 in][for the implementation of this cloud model in \texttt{petitRADTRANS}]{Molliere2020}, which we dub the EddySed model. We include the opacities from \citet[Fe]{Henning1996} and \citet[MgSiO$_3$]{Scott1996, Jaeger1998}. This physically motivated model implements absorption and scattering of MgSiO$_3$ and Fe clouds, expected for a BD of these temperatures. We effectively vary the sedementation efficiency, vertical mixing of cloud particles, and particle size distribution by freely sampling three parameters, $\mathrm{f_{sed}}$, $\mathrm{K_{zz}}$, and $\sigma_\mathrm{g}$, respectively.
In order to enforce a cloudy atmosphere, we use a parameter $\mathrm{\log\tau_{cloud}}$ to scale the cloud opacity at the photosphere (where $\tau=1$). When this parameter is used, instead of retrieving for the logarithmic abundance of enstatite ($\mathrm{\log{\tilde{X}_{MgSiO_3}}}$) and iron ($\mathrm{\log{\tilde{X}_{Fe}}}$) separately, we retrieved $\mathrm{\log\tau_{cloud}}$ and the logarithmic ratio of abundances ($\mathrm{\log{\tilde{X}_{Fe}}/\tilde{X}_{MgSiO_3}}$). An example and further discussion of the implementation of this $\mathrm{\log\tau_{cloud}}$ parameter can be found in \citet{Brown-Sevilla2022}. When running retrievals with clouds, we implement Adaptive Mesh Refinement (AMR), increasing the vertical resolution of the atmosphere about the cloud base in order to resolve the abrupt onset of the cloud deck \citep{Molliere2020}.

\subsection{Retrieval Experiments}
\par Table \ref{tab:atmos} records the median and $1\,\sigma$ CI of a subset of the parameters of interest, derived from the retrievals fit to the observed spectrum of HD~72946~B.
\par We investigated the role of clouds in the atmosphere, leveraging our dynamical mass prior. Our first retrieval is cloudless, and relies on our flexible P-T profile to reproduce qualitatively the reduced temperature gradient to fit the data. The second implements the ``grey" cloud model, and the third the EddySed model. Figure \ref{fig:3ptcompcloud} plots the pressure-temperature profiles for each retrieval.
\par
\begin{figure*}
    \centering
    \includegraphics[width=0.95\textwidth]{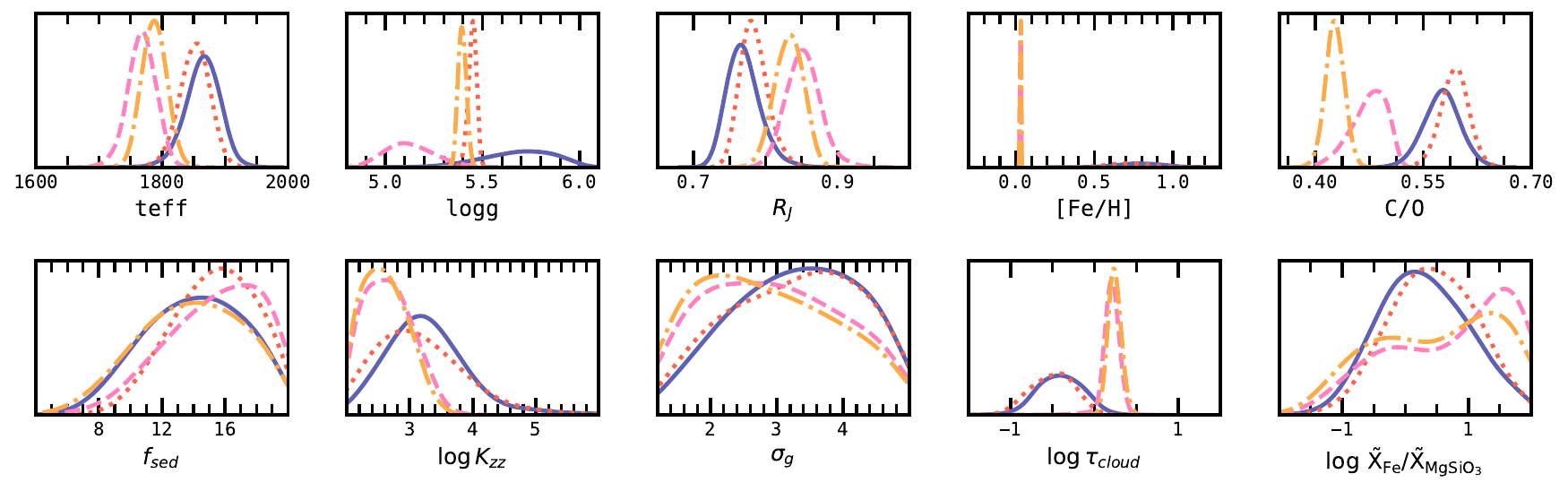}
    \caption{Comparing the 1-D marginalized posterior distributions of notable retrieval parameters, for retrievals recorded in \ref{tab:atmos} and illustrated in \ref{fig:retrievalspec}. The uniform mass and abundance prior retrieval is shown as a dotted red curve; the dynamical mass, uniform abundance prior retreival is shown as a solid blue curve; the uniform mass, stellar abundance prior retrieval is shown as a pink dashed curve; the dynamical mass and stellar abundance prior retrieval is shown as a dash-dotted yellow curve. Assuming a dynamical mass does not appear to affect the outcome of the retrieval as dramatically as setting a prior on the abundances. Without an abundance prior, the retrieval estimates C/O ratios slightly higher, but still consistent with stellar values, but [Fe/H] much higher than stellar (0.8 compared to 0.03). A stellar abundance prior strongly shapes the location of the cloud ($\log{\tau_{\mathrm{cloud}}}$ is more sharply peaked, and the ratio of cloud particles is shifted towards the prior bounds), whereas for a uniform abundance prior, cloud parameters are more widely distributed.}
    \label{fig:1dretrievalpost}
\end{figure*}

 We sought to test the influence a prior on the dynamical mass of the object has on an atmospheric free retrieval. In principle, the dynamical mass prior indirectly constrains the surface gravity and radius of the atmosphere, but we wanted to test whether this would lead to more accurately retrieved abundances (e.g. abundances similar to the host star) and whether the dynamical mass prior might help constrain cloud properties. We also sought to test whether placing priors from the stellar abundances ([Fe/H] and C/O) might inform our retrieval of the BD properties. Model spectra from these four retrievals are shown in Figure \ref{fig:retrievalspec}, illustrating the influence different prior assumptions have on the resulting spectrum. Figure \ref{fig:1dretrievalpost} shows the 1-D marginalized posterior for notable retrieval parameters in this experiment, while Appendix \ref{sec:posteriors}, Figures \ref{fig:retrievalpost} plots the full comparative posterior distributions. 

\section{Discussion} \label{sec:discuss}



\begin{figure*}
    \centering
    \includegraphics[width=\textwidth]{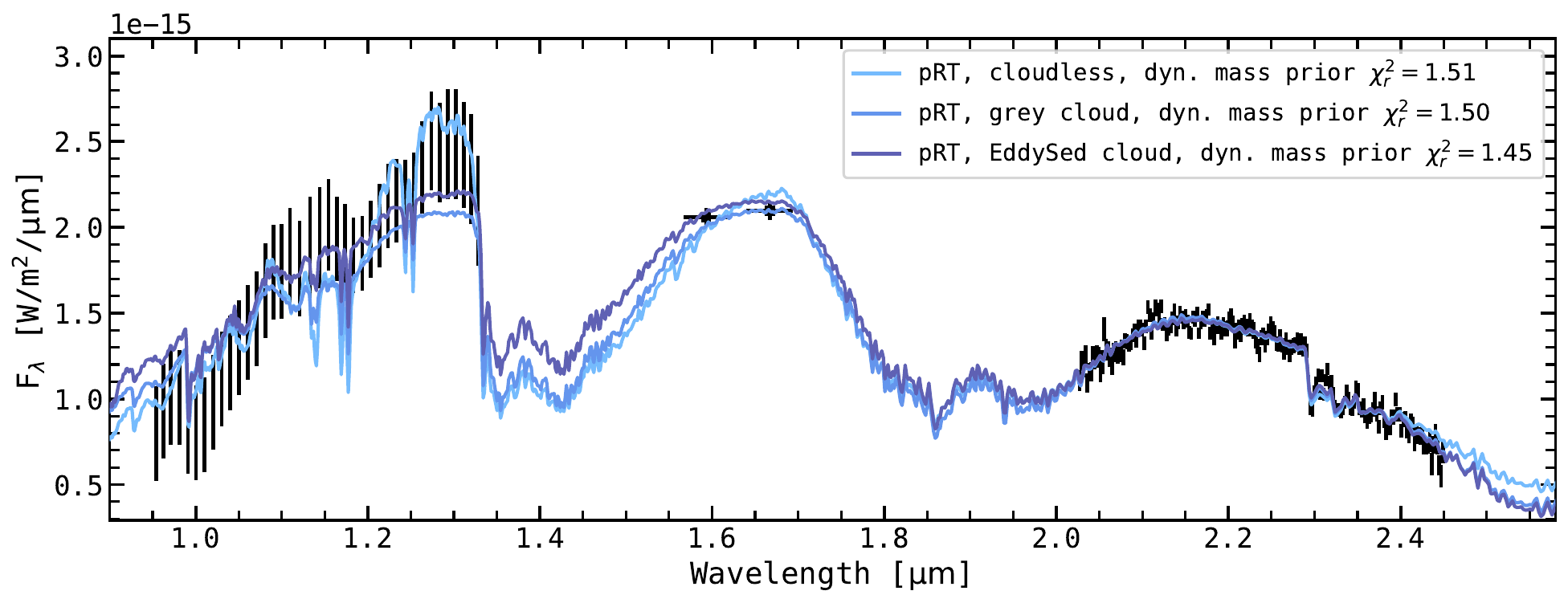}
    \caption{The maximum a posteriori spectra for the three retrievals that vary cloud model, but apply a dynamical mass prior. Observations of HD~72946~B are plotted as in \ref{fig:btspec}. The best fit retrievals fit the GRAVITY data well in each case, but the choice of cloud model appears to strongly influence the goodness of the fit to the SPHERE Y-J spectrophotometry.}
    \label{fig:retrievalcloudspec}
\end{figure*}

\begin{figure*}
    \centering
    \includegraphics[width=\textwidth]{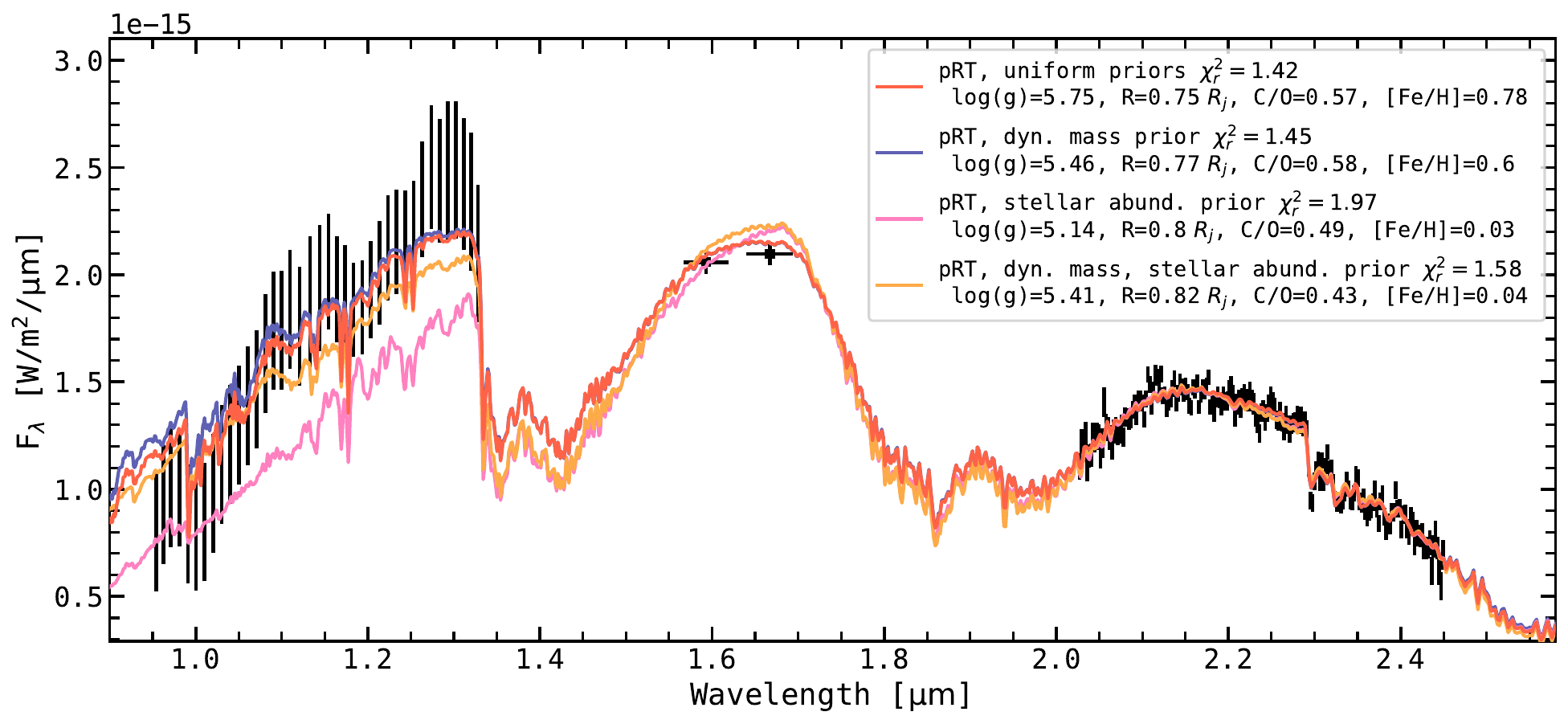}
    \caption{The maximum a posteriori spectra for the four EddySed retrievals, varying prior assumptions of the mass and abundances of the atmosphere. Observations of HD~72946~B are plotted as in \ref{fig:btspec}. The best fit retrievals fit the GRAVITY data well in each case, but struggle to capture the slope of the SPHERE Y-J spectrophotometry.}
    \label{fig:retrievalspec}
\end{figure*}


\par Our atmospheric analysis is summarized in Table \ref{tab:atmos}. In general, we expected to be able to construct some model atmosphere that can fit the data arbitrarily well given enough independent free parameters, with or without clouds \citep[see, for instance, the results of][]{Lueber2022}. Instead of comparing the Bayesian evidences for our retrievals (which, in order to do properly, would require running the retrievals with a non-constant sampling efficiency, which would dramatically increase the computation time required), we sought to conduct a preliminary qualitative assessment of the BD atmosphere.
\par We compared our dynamical mass and age estimates to two example evolutionary models, the cloudy SM08 \citep{Saumon2008} and the cloudless Sonora Bobcat \citep{Marley2021}, deriving effective temperatures, surface gravities, radii, and luminosities predicted by the dynamical mass (as opposed to inferring a mass via a luminosity and age estimate as is common for non-benchmark objects). We find that in general the effective temperatures from the evolutionary models disagree by about 200-300\,K with the effective temperatures from our spectroscopic modeling.
\par Our observations of HD~72946~B proved an excellent match to a solar composition, cloudy \texttt{BT-Settl-CIFIST} model atmosphere (\S5.1) with $\mathrm{T_{eff}}\sim1780\,\mathrm{K}$, but the derived surface gravity and radius dependent on our inclusion of our dynamical mass as a prior in our fit. Without the dynamical mass, the fit to the model grid derived a lower surface gravity and radius (5.2 and 0.79$\mathrm{R_J}$, respectively) than the fit with the dynamical mass (5.4 and 0.84$\mathrm{R_J}$), which lead to a wildly incorrect mass determination (37.5$\mathrm{M_J}$ compared to the 69.5$\mathrm{M_J}$ dynamical mass). Interestingly, while neither spectroscopic effective temperature agrees with the results from the evolutionary models we explored, our dynamical mass prior run sampled surface gravities consistent with those derived from the cloudy SM08 evolutionary models. This indicates the importance of a dynamical mass prior in assessing model deficiencies for evolutionary and self-consistent atmospheric models.

\subsection{Presence of Clouds}

\par We verified the claim made by \citet{Tremblin2015}, that the shape of the spectrum can be reproduced with either a cloud opacity or a reduced temperature gradient, by constructing a cloudless, simple cloud, and EddySed cloud retrieval comparison (see Figure \ref{fig:retrievalcloudspec}). As seen in Figure \ref{fig:3ptcompcloud}, the simple one layer cloud model still relies on the flexibility of the P-T profile to reproduce the data with an ``isothermal knee", whereas the the EddySed model with Fe and MgSiO$_3$ cloud decks show an approximately Eddington profile even below the photosphere. Each retrieval within this cloud experiment provided an acceptable fit to the data, with $\chi^2_\mathrm{red}<2$ and 220-230 degrees of freedom, but because the posteriors were sampled with a constant efficiency, and have differing numbers of free parameters, a rigorous Bayesian inter-comparison is not possible.

\par In general, the cloud parameters converged on similar results, producing a compact cloud layer between 10 and 1 bar. The scattering clouds, parameterized using the EddySed model \citep{Ackerman2001, Molliere2020} produce low vertical diffusion ($\mathrm{K_{zz}}$) and high sedimentation parameter ($\mathrm{f_{sed}}$). The values for $\mathrm{K_{zz}}$ are below 5, the ``minimum" value for the baseline model in \citet{Ackerman2001}. The very high $\mathrm{f_{sed}}\simeq12-16$ contrasts results for the red, low surface gravity planet HR~8799~e \citep[$\mathrm{f_{sed}}\simeq2$][]{Molliere2020}. These $\mathrm{f_{sed}}$ values are much higher than those typically invoked to model brown dwarfs of similar spectral types \citep[1-4,][]{Saumon2008, Stephens2009}, and, similar to the derived abundances, appear to indicate a deficiency in the retrieval framework, and are likely related to the difficulties in retrieving the correct abundances (see \S6.3). These cloud parameters merely encode the opacity necessary to fit the data within the retrieval framework, but having both $\mathrm{f_{sed}}$ and $\mathrm{K_{zz}}$ as free parameters appears to lead to an unphysical cloud layer that then requires a high metallicity to compensate for in attempting to reproduce the spectral slope. From a data-standpoint alone, the SPHERE Y-J data appears to have the strongest influence on the derived cloud and abundance parameters, but has a higher uncertainty than our GRAVITY observations.

A quick back of the envelope calculation, using a $T_{\mathrm{eff}}=1800\,\mathrm{K}$, $\mathrm{log(g)}=5.4$, and mean molecular weight of 2.33 amu yields an atmospheric scale height of about $\mathrm{H}=25\,\mathrm{m}$, and following Equation 5 in \citet{Ackerman2001}, assuming $\mathrm{L}=\mathrm{H}$ and $\Gamma/\Gamma_{ad}=1$, at the photospheric pressure of a few bars gives a $\log(\mathrm{K_{zz}})$ estimate of about 15, 5 times larger than our sampled values. Future work will seek to constrain the sampled $\mathrm{K_{zz}}$ so that effectively, only $\mathrm{f_{sed}}$ and $\sigma_\mathrm{g}$ are sampled as free parameters during the retrieval. 

\par We have seen the effect the choice of cloud model has on the retrieved P-T structure. In general, the EddySed cloud model arrives at a solution where a deeper iron cloud deck (near the altitude of the isothermal knee in the gray slab retrieval), and then an silicate cloud (near the altitude of the gray slab cloud) appear. This overcompensation from the grey cloud retrieval, forcing a change in the P-T structure to mimic opacity that could as easily be described by a physical cloud deck, might indicate that the grey cloud assumption oversimplifies the structure of the clouds on HD~72946~B, if these clouds do exist. This is supported by the findings of other retrieval studies which indicate a qualitatively similar set of clouds (a deep iron cloud and higher silicate cloud) best fit the 1-20~\textmu m spectrum of isolated brown dwarfs \citep{Burningham2017, Burningham2021}. However, as we've noted, the assumption of radiative-convective equilibrium (which results in a large temperature gradient) precludes the alternative solution: the BD is relatively cloudless (or, the clouds do not play a strong role in producing NIR opacity) and the shape near-infrared spectra is produced by a lower temperature gradient generated by fingering convection arising from a chemical gradient \citep{Tremblin2015, Tremblin2016, Tremblin2019}. Since our data does not probe these deeper pressure layers, the results of this retrieval experiment are somewhat ambiguous.  

\par The EddySed retrievals arrived at similar abundances of iron and enstatite (a log abundance ratio near zero) when the abundances were unconstrained. When constrained to the stellar values, the posterior distribution of the abundance ratio appears multimodal, with a mode near zero, but a more dominant mode favoring iron grains by nearly a factor of 100, pushing against the prior boundary. This can be seen in the poor fit of these spectra to the SPHERE Y-J spectrophotometry, as an abundant iron cloud layer and low metallicity results in a flattened Y-J slope.

\subsection{Retrieved abundances}
\par
\begin{figure}
    \centering
    \includegraphics[width=0.45\textwidth]{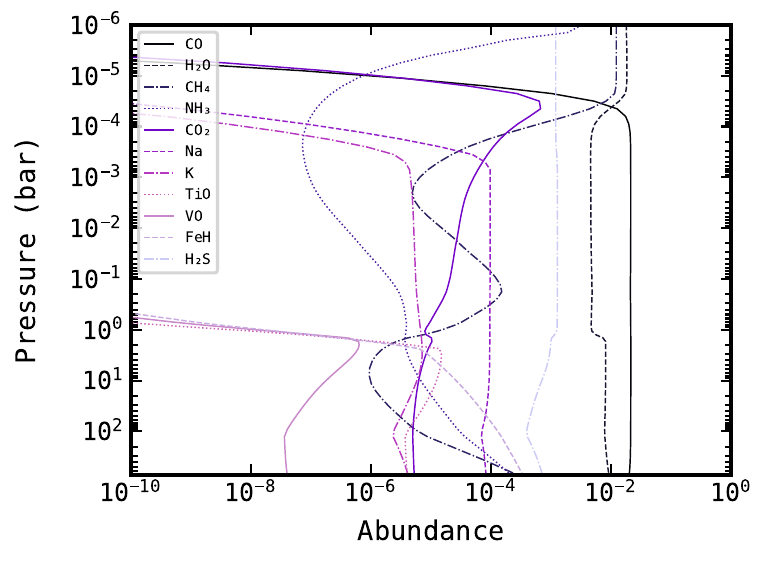}
    \caption{The mass fraction ``abundance" of each molecule (directly proportional to the volume mixing ratio of said molecule) included in our \texttt{petitRADTRANS} retrieval, under assumptions of chemical equilibrium, using a dynamical mass prior, EddySed cloud with freely sampled parameters, and no abundance prior. Chemical equilibrium results in the rapid decrease in abundance of FeH, TiO, and VO near the location of the photosphere, but the exact abundances of these metal bearing species is likely confounded by the unphysical sampled cloud parameters.}
    \label{fig:vmr_vs_p}
\end{figure}

\begin{figure*}
    \centering
    \includegraphics[width=0.85\textwidth]{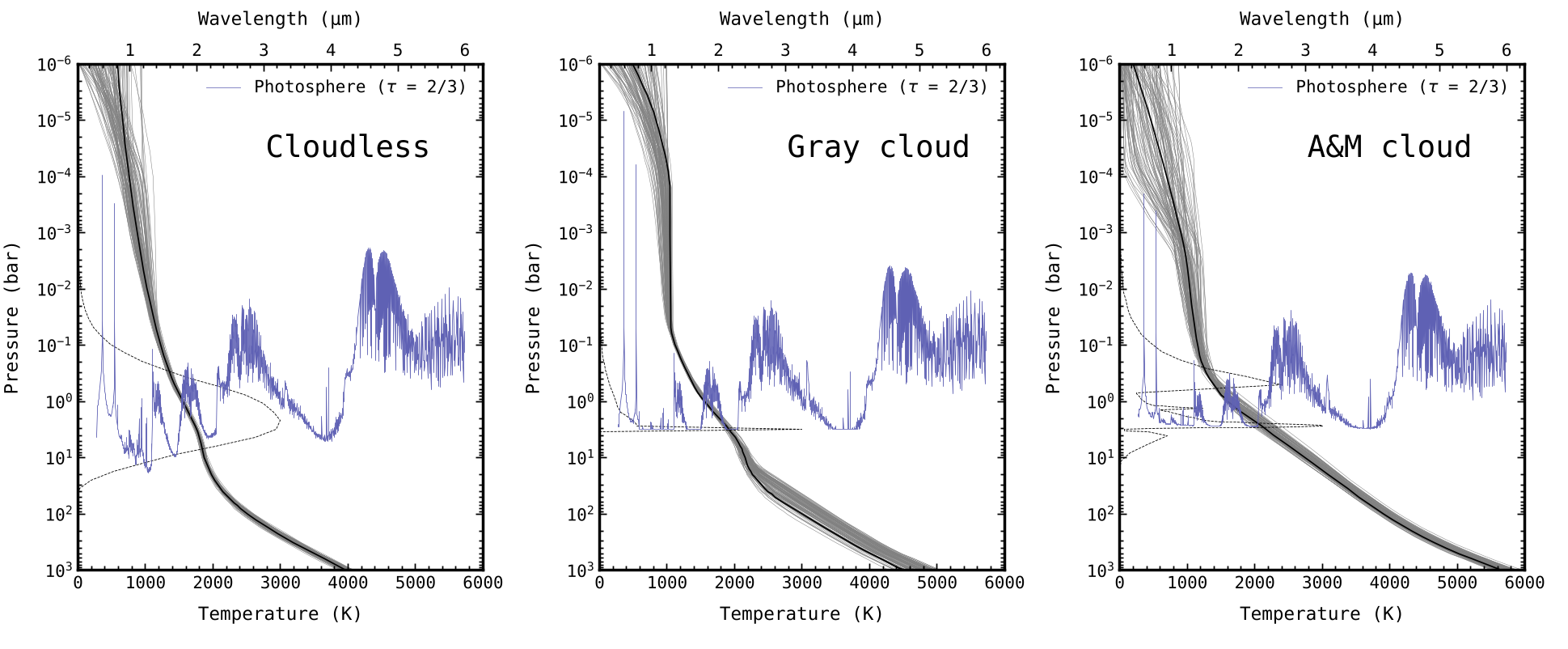}
    \caption{A comparison between the retrieved P-T structure of HD~72946~B using the 3-part P-T parameterization, fixing the mass to the dynamical mass, but varying the cloud prescription between cloudless (left), a grey slab cloud (middle), and a scattering cloud of iron and enstatite grains parameterized by the sedimentation efficiency (right). The ordinate plots pressure in units of bar. The temperature is shown along the bottom abscissa and corresponds to the black and grey solid curves, while wavelength is shown along the top abscissa and corresponds to the blue solid curve, which shows the pressure at which the optical depth is 2/3 (e.g. the location in altitude of the photosphere). A 1-D histogram of the contribution function is plotted along the left axis as a dotted curve, showing the contribution of a given pressure to the opacity of the atmosphere averaged over all computed wavelengths. Notably, the P-T profile for each retrieval deviates below the photosphere: where the iron condensate cloud appears in the EddySed retrieval, the P-T grows more isothermal in the grey cloud retrieval, and the isothermal region becomes larger in the cloudless retrieval.}
    \label{fig:3ptcompcloud}
\end{figure*}

 It is expected that a high-mass brown dwarf like HD~72946~B would have formed from the same molecular cloud as its host, and therefore have effectively the same abundances as HD~72946~A ([Fe/H]$=0.036\pm0.023$, C/O$=0.51\pm0.05$). We note that, as discussed in \S3, our stellar abundance determination should be viewed as a lower limit on O, indicating that the C/O ratio for the star could be lower than 0.51. With the dynamical mass prior, we find that in the absence of a source of cloud opacity, the retrieval is driven towards very high [Fe/H]$=0.86\pm0.05$ compared to the stellar value. Cloudy retrievals (either with, or without a dynamical mass prior), but with uniform abundance priors estimate C/O$\simeq0.58\pm0.02$, and high [Fe/H]$\simeq0.7\pm0.1$. The uniform abundance, dynamical mass prior, EddySed retrieval  abundances are visualized in Figure \ref{fig:vmr_vs_p}, which plots the mass fractions of each molecule versus pressure throughout the atmosphere. The interplay between cloud opacity and metallicity necessarily indicate a bias in the exact values in Figure \ref{fig:vmr_vs_p}, which is included for illustrative purposes. The chemical equilibrium treatment results in rainout of FeH, TiO, and VO in the photosphere as expected, which have been shown to strongly affect the retrieved bulk properties of L-type brown dwarfs \citep{Rowland2023}. This C/O is still systematically higher than the stellar value, by $1-2\,\sigma$, but consistent with the stellar value. 
 Placing a prior on the abundances (both [Fe/H] and C/O) based on the stellar values results in a systematically lower retrieved C/O ratio and an overall a worse fit to the SPHERE Y-J data. In light of our uncertain stellar oxygen abundance determination, it could be feasible that these retrieved C/O ratios could still be consistent with the stellar value, in which case the quality of the SPHERE Y-J spectrum (in particular the overall slope) would be called into question. These inconclusive retrieval abundance results are necessarily overshadowed by the well fit, solar abundance BT-Settl model, and appear therefore to indicate a deficiency in the retrieval framework's cloud treatment. 

\par Previous studies have indicated that parameterized P-T profiles can bias the retrieval of most bulk properties \citep{Wang2022, Rowland2023}, and so future work should explore freely parameterized P-T profiles. Our retrievals assume chemical equilibrium, which could bias our retrieved C/O and [Fe/H] if there is significant mixing in the atmosphere. It seems likely that disequilibrium could affect the derived abundances of the major oxygen and carbon bearing species, which can currently be explored with our retrieval framework. Future work will detail this. Changes in the optical depth of the cloud could in principle reduce the opacity required to reproduce the shorter wavelength data, and resolve the anomalously high metallicity we retrieved here; for instance, retrieving on only $\mathrm{f_{sed}}$ and parameterizing $\mathrm{K_{zz}}$ based on equations 5-6 in \citet{Ackerman2001}. In Figure \ref{fig:retrievalspec}, we see that varying the prior assumptions of our retrievals has the strongest impact on the quality of the fit to the SPHERE spectrophotometry, where water and FeH (iron hydride) has a large role in shaping the slope of the low resolution spectrum, which could also be impacted by biases induced by the data reduction or instrumental throughput treatment. Future work should investigate whether implementing the changes to retrievals mentioned above on the spectrum of HD~72946~B result in more reasonable [Fe/H], and whether a dynamical mass might improve the efficacy of these improved retrievals.

\subsection{Dynamical mass prior for atmospheric retrieval}
\par We investigated leveraging the dynamical mass (\S\ref{sec:orbit}) and well determined properties of the host star (\S\ref{sec:host}) when comparing retrieval fits. We find an informative, thought not necessarily satisfying set of results. As with the self-consistent model grid, without the dynamical mass prior, the derived surface gravities, radii, and therefore derived masses from the spectral fits produce wildly inconsistent masses, as the retrieval struggles to reconcile the interplay between surface gravity, clouds, and molecular opacities with a large number of free parameters. 
\par Comparing the uniform abundance prior retrievals, only examining varying our prior assumption on the dynamical mass, we find that the dynamical mass prior retrieval finds a lower log(g), radius, and metallicity than the uninformative mass prior retrieval, but a marginally higher C/O. Notably, the posterior distributions on P-T profile, cloud, and other nuisance parameters show no effective differences between the two retrievals. This is evidence that the inclusion of the dynamical mass prior does not effectively constrain the P-T structure or cloud properties of this retrieval. 

It is not unprecedented, from a physical standpoint, that the inclusion of a dynamical mass prior in our retrieval does not effectively constrain the A\&M cloud model parameters when they are all are left free. Based on a microphysics model \citet{Gao2018} find that $\mathrm{f_{sed}}$ is dependent on $\mathrm{K_{zz}}$ but not on gravity, for a constant $\mathrm{K_{zz}}$; of course, outside of the EddySed model, $\mathrm{K_{zz}}$ is not necessarily constant, and could even be constructed to depend on gravity via mixing-length-theory arguments \citep{Mukherjee2022}.

\par The systematic shift in log(g), radius, [Fe/H], and C/O is noticable in the full posterior distribution plot (Figure \ref{fig:retrievalpost}), and is related to the scale height of the atmosphere, which the dynamical mass prior does indirectly constrain. Without this prior, in order to fit the data, the scale height of the atmosphere decreases, descreasing the radius, and the number density in the photosphere increases, which affects the cross sections implied by the abundances. That is, left free, the retrieval appears to probe too deep, resulting in a smaller radii and higher log(g), which affects the determination of the abundances. It is evident that, currently, the dynamical mass prior is useful insofar as it helps isolate or determine retrieval deficiencies, but it is not apparent whether future improvements to the retrieval framework (or atmospheric modeling in general) will necessitate the inclusion of a dynamical mass prior.

\subsection{In Context}
\par In a similar retrieval analysis, \citeauthor{Xuan2022} show that their high ($R\sim\!35,000$) resolution Keck Planet Imager and Characterizer (KPIC) spectra of a cloudy brown dwarf is not sensitive to clouds, because the clouds contribute the most opacity at lower altitudes. While the high-res constrains C/O by probing the K-band CO bandhead at higher pressures, their retrieved abundances appear independent of a variety of cloud assumptions. This is not the case for their retrieval only to their low resolution data. We find a qualitatively similar result here. From our analysis it appears that the VLTI/GRAVITY spectrum ($R\sim\!500$) of HD~72946~B is able to probe a wide enough range of pressures above the cloud deck to enable a retrieval of the C/O independent from the cloud parameterization. 
\par It still remains unclear whether a dynamical mass prior dramatically improves the retrieval of cloud properties for this brown dwarf. We can safely argue that, assuming the EddySed cloud model, a parameterized P-T profile, and chemical equilibrium, a dynamical mass prior does not appear to strongly constrain the cloud or P-T profile parameters for a high surface gravity L-type object. This is in part also due to the lack of MIR data on HD~72946~B, where the spectral signatures of silicate grains exhibit strong wavelength dependence, and are able to constrain the cloud properties directly \citet{Burningham2021}. This motivates future studies of more widely separated benchmark brown dwarfs (with strong dynamical mass constraints) with JWST/MIRI. Coupling direct observations of silicate absorption in benchmark BDs with retrievals, or improving retrievable parameteric cloud models might yet constrain the composition, shape, and size distribution of cloud particles in these L-type substellar companions. In the short term, implementing a free P-T profile and non-uniform abundance profiles for key molecules might improve the accuracy of the physical parameters derived from the retrieved spectrum of HD~72946~B. Observations from GRAVITY can quickly refine the orbital parameters of detected companions, even those on longer period orbits, and can therefore shape the landscape of viable targets for future benchmark/atmosphere studies. 

\par Regardless of the exact cloud composition or parameterization, or the structure of the P-T profile, the C/O of the object appears in agreement with the stellar value to within $2\,\sigma$, and each retrieved C/O value is precise ($\pm0.02\,$dex) indicating the efficacy of the GRAVITY spectrum in constraining C/O at close separations. 

\section{Summary} \label{sec:conclusion}

\par In this paper we presented new observations of the brown dwarf companion HD~72946~B from VLTI/GRAVITY. The observations yielded 4 astrometric points with impressive precision and a moderate resolution R=500 K-band spectrum of the companion. The precise astrometry enabled us to refine the orbit of the companion and improve the dynamical mass measurement, when fit jointly with radial velocities and absolute astrometry. The enhanced resolution in the K-band allows for a precise observation of the carbon monoxide bandhead at 2.3\textmu m, even at a separation of only $\sim\!125~\mathrm{mas}$.
\par We also present an updated joint isochronal and spectral analysis of the host using archival photometry and visible light spectroscopy. We determined the mass ($0.97\pm0.01~\mathrm{M_\odot}$), age ($2.67^{+0.25}_{-0.49}~\mathrm{Gyr}$), and elemental abundances (Tables \ref{tab:host}, \ref{host_chemical_abundances}) for the star self-consistently. These parameters helped to inform our understanding of the brown dwarf companion's evolution, orbit, and composition.
\par We compared two orbit fitting schemes, where we considered fits with and without absolute astrometry from the Hipparchos-Gaia Catalogue of Accelerations and the reported \textit{Gaia} eDR3 system parallax, and found that the inclusion of absolute astrometry brought the error on the dynamical mass to $1.1\%$. With this precise a mass, HD~72946~B will be a crucial system to test evolutionary models for substellar objects in the near future. Future work should investigate whether the existing GRAVITY astrometry has the precision to constrain the influence of the widely separated (10") co-moving stellar binary HD~72945~AB on the orbit of HD~72946~AB, as was attempted without the GRAVITY dataset in \citet{mBrandt2021}.
\par We compared the observed spectrum of HD~72946~B to the \texttt{BT-Settl-CIFIST} grid of self-consistent, microphysical cloudy model atmospheres. Based on our forward model fits, the object has an effective temperature of $\sim\!1750~\mathrm{K}$, a log(g) of $\sim\!5.4$, and a radius of $\sim\!0.86~\mathrm{R_{Jup}}$. We then constructed free retrievals with \texttt{petitRADTRANS}. We investigated the effect of a dynamical mass prior on the retrieved atmosphere. We briefly compared the interdependence of the retrieved pressure-temperature profiles on our cloud model, verifying that the observed spectrum could be fit with either a reduced temperature gradient or a cloud deck. 
\par The retrieval analysis presented here is only a sample of what can be explored using the now well determined properties of the HD~72946 system. Our investigation indicated the need to include rainout of FeH in our retrieval setup in the future. Future work could benefit from measuring the H-band and MIR silicate features of the companion. Future work utilizing different retrieval codes under different assumptions (for instance, chemical disequilibrium) could also benefit from the dynamical mass prior, as could the application of additional self-consistent forward models.
\par VLTI/GRAVITY observations of exoplanets have already proven the instrument's unique capability to observe multi-planet interactions \citep{Lacour2021}, precisely determine substellar C/O ratios \citep[]{Molliere2020, Nowak2020, Kammerer2021}, and dramatically refine a companion's orbital parameters \citep[][]{Wang2020, Kammerer2021}. With our observations of HD~72946~B we have demonstrated its capacity to conduct the same studies of benchmark brown dwarfs at even higher signal-to-noise. Atmospheric studies of these objects can inform our understanding of the best practices for determining the elemental abundances for exoplanets, which are key quantities for constraining planetary formation pathways and histories.

\section*{Acknowledgements}

\par W.O.B. would like to thank Jacob Hamer for his input regarding the age estimates of the HD~72946 system, and Sagnick Mukherjee for discussions regarding clouds. The authors would like to thank the Paranal Observatory astronomers and local staff for their tremendous support in completing these observations.
\par TS acknowledges the support from the Netherlands Organisation for Scientific Research (NWO) through grant VI.Veni.202.230.
\par This work used the Dutch national e-infrastructure with the support of the SURF Cooperative using grant no. EINF-1620.
\par SL acknowledges the support of the French Agence Nationale de la Recherche (ANR), under grant ANR-21-CE31-0017 (project ExoVLTI)
\par Based on observations collected at the European Southern Observatory under ESO programme(s) 1104.C-0651(B), 1103.B-0626(D).
\par Based on spectral data retrieved from the ELODIE archive at Observatoire de Haute-Provence (OHP), available at atlas.obs-hp.fr/elodie. Based on data retrieved from the SOPHIE archive at Observatoire de Haute-Provence (OHP), available at atlas.obs-hp.fr/sophie.
\par This research has made use of the VizieR catalogue access tool, CDS, Strasbourg, France (DOI: 10.26093/cds/vizier).
\par This research has made use of the Jean-Marie Mariotti Center \texttt{Aspro} service.
\par This publication makes use of data products from the Two Micron All Sky Survey, which is a joint project of the University of Massachusetts and the Infrared Processing and Analysis Center/California Institute of Technology, funded by the National Aeronautics and Space Administration and the National Science Foundation. 
\par This work has made use of data from the European Space Agency (ESA) mission {\it Gaia} (\url{https://www.cosmos.esa.int/gaia}), processed by the {\it Gaia} Data Processing and Analysis Consortium (DPAC, \url{https://www.cosmos.esa.int/web/gaia/dpac/consortium}). Funding for the DPAC has been provided by national institutions, in particular the institutions participating in the {\it Gaia} Multilateral Agreement.

\par WOB acknowledges that the Johns Hopkins University occupies the unceded land of the Piscataway People, and acknowledes the Piscataway community, their elders both past and present, as well as future generations. \href{https://web.archive.org/web/20220816104243/http://trujhu.org/index.php/about-us/land-acknowledgement/}{JHU was founded on and presides over the exclusions and erasures of many people}, a fact no less true for being contentious, and one that bears repeating even in long acknowledgement sections.
\par WOB graciously acknowledges their cat, Morgoth, for her ``encouragement."

\newpage

\bibliography{hd72946B_gravity.bib}{}

\begin{thebibliography}{}
\expandafter\ifx\csname natexlab\endcsname\relax\def\natexlab#1{#1}\fi
\providecommand{\url}[1]{\href{#1}{#1}}
\providecommand{\dodoi}[1]{doi:~\href{http://doi.org/#1}{\nolinkurl{#1}}}
\providecommand{\doeprint}[1]{\href{http://ascl.net/#1}{\nolinkurl{http://ascl.net/#1}}}
\providecommand{\doarXiv}[1]{\href{https://arxiv.org/abs/#1}{\nolinkurl{https://arxiv.org/abs/#1}}}

\bibitem[{{Ackerman} \& {Marley}(2001)}]{Ackerman2001}
{Ackerman}, A.~S., \& {Marley}, M.~S. 2001, \apj, 556, 872, \dodoi{10.1086/321540}

\bibitem[{{Aguilera-G{\'o}mez} {et~al.}(2018){Aguilera-G{\'o}mez}, {Ram{\'\i}rez}, \& {Chanam{\'e}}}]{Aguilera-Gomez2018}
{Aguilera-G{\'o}mez}, C., {Ram{\'\i}rez}, I., \& {Chanam{\'e}}, J. 2018, \aap, 614, A55, \dodoi{10.1051/0004-6361/201732209}

\bibitem[{{Allard}(2014)}]{Allard2014}
{Allard}, F. 2014, in Exploring the Formation and Evolution of Planetary Systems, ed. M.~{Booth}, B.~C. {Matthews}, \& J.~R. {Graham}, Vol. 299, 271--272, \dodoi{10.1017/S1743921313008545}

\bibitem[{{Allard} {et~al.}(2003){Allard}, {Guillot}, {Ludwig}, {Hauschildt}, {Schweitzer}, {Alexander}, \& {Ferguson}}]{Allard2003}
{Allard}, F., {Guillot}, T., {Ludwig}, H.-G., {et~al.} 2003, in Brown Dwarfs, ed. E.~{Mart{\'\i}n}, Vol. 211, 325

\bibitem[{{Allard} {et~al.}(2011){Allard}, {Homeier}, \& {Freytag}}]{Allard2011}
{Allard}, F., {Homeier}, D., \& {Freytag}, B. 2011, in Astronomical Society of the Pacific Conference Series, Vol. 448, 16th Cambridge Workshop on Cool Stars, Stellar Systems, and the Sun, ed. C.~{Johns-Krull}, M.~K. {Browning}, \& A.~A. {West}, 91, \dodoi{10.48550/arXiv.1011.5405}

\bibitem[{{Allard} {et~al.}(2012){Allard}, {Homeier}, \& {Freytag}}]{Allard2012}
{Allard}, F., {Homeier}, D., \& {Freytag}, B. 2012, Philosophical Transactions of the Royal Society of London Series A, 370, 2765, \dodoi{10.1098/rsta.2011.0269}

\bibitem[{{Allard} {et~al.}(2013){Allard}, {Homeier}, {Freytag}, {Schaffenberger}, \& {Rajpurohit}}]{Allard2013}
{Allard}, F., {Homeier}, D., {Freytag}, B., {Schaffenberger}, W., \& {Rajpurohit}, A.~S. 2013, Memorie della Societa Astronomica Italiana Supplementi, 24, 128.
\newblock \doarXiv{1302.6559}

\bibitem[{{Amarsi} {et~al.}(2016){Amarsi}, {Lind}, {Asplund}, {Barklem}, \& {Collet}}]{amarsi2016}
{Amarsi}, A.~M., {Lind}, K., {Asplund}, M., {Barklem}, P.~S., \& {Collet}, R. 2016, \mnras, 463, 1518, \dodoi{10.1093/mnras/stw2077}

\bibitem[{{Amarsi} {et~al.}(2019){Amarsi}, {Nissen}, \& {Sk{\'u}lad{\'o}ttir}}]{amarsi2019}
{Amarsi}, A.~M., {Nissen}, P.~E., \& {Sk{\'u}lad{\'o}ttir}, {\'A}. 2019, \aap, 630, A104, \dodoi{10.1051/0004-6361/201936265}

\bibitem[{{Amarsi} {et~al.}(2020){Amarsi}, {Lind}, {Osorio}, {Nordlander}, {Bergemann}, {Reggiani}, {Wang}, {Buder}, {Asplund}, {Barklem}, {Wehrhahn}, {Sk{\'u}lad{\'o}ttir}, {Kobayashi}, {Karakas}, {Gao}, {Bland-Hawthorn}, {de Silva}, {Kos}, {Lewis}, {Martell}, {Sharma}, {Simpson}, {Zucker}, {{\v{C}}otar}, {Horner}, \& {Galah Collaboration}}]{amarsi2020}
{Amarsi}, A.~M., {Lind}, K., {Osorio}, Y., {et~al.} 2020, \aap, 642, A62, \dodoi{10.1051/0004-6361/202038650}

\bibitem[{{Asplund} {et~al.}(2021){Asplund}, {Amarsi}, \& {Grevesse}}]{asplund2021}
{Asplund}, M., {Amarsi}, A.~M., \& {Grevesse}, N. 2021, \aap, 653, A141, \dodoi{10.1051/0004-6361/202140445}

\bibitem[{{Baraffe} {et~al.}(2015){Baraffe}, {Homeier}, {Allard}, \& {Chabrier}}]{Baraffe2015}
{Baraffe}, I., {Homeier}, D., {Allard}, F., \& {Chabrier}, G. 2015, \aap, 577, A42, \dodoi{10.1051/0004-6361/201425481}

\bibitem[{{Bate}(2009)}]{Bate2009}
{Bate}, M.~R. 2009, \mnras, 392, 590, \dodoi{10.1111/j.1365-2966.2008.14106.x}

\bibitem[{{Blunt} {et~al.}(2020){Blunt}, {Wang}, {Angelo}, {Ngo}, {Cody}, {De Rosa}, {Graham}, {Hirsch}, {Nagpal}, {Nielsen}, {Pearce}, {Rice}, \& {Tejada}}]{Blunt2020}
{Blunt}, S., {Wang}, J.~J., {Angelo}, I., {et~al.} 2020, \aj, 159, 89, \dodoi{10.3847/1538-3881/ab6663}

\bibitem[{{Bonavita} {et~al.}(2022){Bonavita}, {Fontanive}, {Gratton}, {Mu{\v{z}}i{\'c}}, {Desidera}, {Mesa}, {Biller}, {Scholz}, {Sozzetti}, \& {Squicciarini}}]{Bonavita2022}
{Bonavita}, M., {Fontanive}, C., {Gratton}, R., {et~al.} 2022, \mnras, 513, 5588, \dodoi{10.1093/mnras/stac1250}

\bibitem[{{Bonnefoy} {et~al.}(2010){Bonnefoy}, {Chauvin}, {Rojo}, {Allard}, {Lagrange}, {Homeier}, {Dumas}, \& {Beuzit}}]{Bonnefoy2010}
{Bonnefoy}, M., {Chauvin}, G., {Rojo}, P., {et~al.} 2010, \aap, 512, A52, \dodoi{10.1051/0004-6361/200912688}

\bibitem[{{Borysow} {et~al.}(1989){Borysow}, {Frommhold}, \& {Moraldi}}]{Borysow1989}
{Borysow}, A., {Frommhold}, L., \& {Moraldi}, M. 1989, \apj, 336, 495, \dodoi{10.1086/167027}

\bibitem[{{Borysow} {et~al.}(2001){Borysow}, {Jorgensen}, \& {Fu}}]{Borysow2001}
{Borysow}, A., {Jorgensen}, U.~G., \& {Fu}, Y. 2001, \jqsrt, 68, 235, \dodoi{10.1016/S0022-4073(00)00023-6}

\bibitem[{{Borysow} {et~al.}(1988){Borysow}, {Frommhold}, \& {Birnbaum}}]{Borysow1988}
{Borysow}, J., {Frommhold}, L., \& {Birnbaum}, G. 1988, \apj, 326, 509, \dodoi{10.1086/166112}

\bibitem[{{Boss}(1997)}]{Boss1997}
{Boss}, A.~P. 1997, Science, 276, 1836, \dodoi{10.1126/science.276.5320.1836}

\bibitem[{{Bouchy} {et~al.}(2016){Bouchy}, {S{\'e}gransan}, {D{\'\i}az}, {Forveille}, {Boisse}, {Arnold}, {Astudillo-Defru}, {Beuzit}, {Bonfils}, {Borgniet}, {Bourrier}, {Courcol}, {Delfosse}, {Demangeon}, {Delorme}, {Ehrenreich}, {H{\'e}brard}, {Lagrange}, {Mayor}, {Montagnier}, {Moutou}, {Naef}, {Pepe}, {Perrier}, {Queloz}, {Rey}, {Sahlmann}, {Santerne}, {Santos}, {Sivan}, {Udry}, \& {Wilson}}]{Bouchy2016}
{Bouchy}, F., {S{\'e}gransan}, D., {D{\'\i}az}, R.~F., {et~al.} 2016, \aap, 585, A46, \dodoi{10.1051/0004-6361/201526347}

\bibitem[{{Bowler}(2016)}]{Bowler2016}
{Bowler}, B.~P. 2016, \pasp, 128, 102001, \dodoi{10.1088/1538-3873/128/968/102001}

\bibitem[{{Brandt} {et~al.}(2021){Brandt}, {Dupuy}, {Li}, {Chen}, {Brandt}, {Wong}, {Currie}, {Bowler}, {Liu}, {Best}, \& {Phillips}}]{mBrandt2021}
{Brandt}, G.~M., {Dupuy}, T.~J., {Li}, Y., {et~al.} 2021, \aj, 162, 301, \dodoi{10.3847/1538-3881/ac273e}

\bibitem[{{Brandt}(2018)}]{Brandt2018}
{Brandt}, T.~D. 2018, \apjs, 239, 31, \dodoi{10.3847/1538-4365/aaec06}

\bibitem[{{Brandt}(2021)}]{tBrandt2021}
---. 2021, \apjs, 254, 42, \dodoi{10.3847/1538-4365/abf93c}

\bibitem[{{Brandt} {et~al.}(2019){Brandt}, {Dupuy}, \& {Bowler}}]{Brandt2019}
{Brandt}, T.~D., {Dupuy}, T.~J., \& {Bowler}, B.~P. 2019, \aj, 158, 140, \dodoi{10.3847/1538-3881/ab04a8}

\bibitem[{{Brown-Sevilla} {et~al.}(2022){Brown-Sevilla}, {Maire}, {Molli{\`e}re}, {Samland}, {Feldt}, {Brandner}, {Henning}, {Gratton}, {Janson}, {Stolker}, {Hagelberg}, {Zurlo}, {Cantalloube}, {Boccaletti}, {Bonnefoy}, {Chauvin}, {Desidera}, {D'Orazi}, {Lagrange}, {Langlois}, {Menard}, {Mesa}, {Meyer}, {Pavlov}, {Petit}, {Rochat}, {Rouan}, {Schmidt}, {Vigan}, \& {Weber}}]{Brown-Sevilla2022}
{Brown-Sevilla}, S.~B., {Maire}, A.~L., {Molli{\`e}re}, P., {et~al.} 2022, arXiv e-prints, arXiv:2211.14330, \dodoi{10.48550/arXiv.2211.14330}

\bibitem[{{Buchner}(2017)}]{Buchner2017}
{Buchner}, J. 2017, arXiv e-prints, arXiv:1707.04476.
\newblock \doarXiv{1707.04476}

\bibitem[{{Buchner}(2021)}]{Buchner2021}
---. 2021, The Journal of Open Source Software, 6, 3001, \dodoi{10.21105/joss.03001}

\bibitem[{{Buchner} {et~al.}(2014){Buchner}, {Georgakakis}, {Nandra}, {Hsu}, {Rangel}, {Brightman}, {Merloni}, {Salvato}, {Donley}, \& {Kocevski}}]{Buchner2014}
{Buchner}, J., {Georgakakis}, A., {Nandra}, K., {et~al.} 2014, \aap, 564, A125, \dodoi{10.1051/0004-6361/201322971}

\bibitem[{{Burningham} {et~al.}(2017){Burningham}, {Marley}, {Line}, {Lupu}, {Visscher}, {Morley}, {Saumon}, \& {Freedman}}]{Burningham2017}
{Burningham}, B., {Marley}, M.~S., {Line}, M.~R., {et~al.} 2017, \mnras, 470, 1177, \dodoi{10.1093/mnras/stx1246}

\bibitem[{{Burningham} {et~al.}(2021){Burningham}, {Faherty}, {Gonzales}, {Marley}, {Visscher}, {Lupu}, {Gaarn}, {Fabienne Bieger}, {Freedman}, \& {Saumon}}]{Burningham2021}
{Burningham}, B., {Faherty}, J.~K., {Gonzales}, E.~C., {et~al.} 2021, \mnras, 506, 1944, \dodoi{10.1093/mnras/stab1361}

\bibitem[{{Burrows} {et~al.}(2001){Burrows}, {Hubbard}, {Lunine}, \& {Liebert}}]{Burrows2001}
{Burrows}, A., {Hubbard}, W.~B., {Lunine}, J.~I., \& {Liebert}, J. 2001, Reviews of Modern Physics, 73, 719, \dodoi{10.1103/RevModPhys.73.719}

\bibitem[{{Burrows} {et~al.}(1997){Burrows}, {Marley}, {Hubbard}, {Lunine}, {Guillot}, {Saumon}, {Freedman}, {Sudarsky}, \& {Sharp}}]{Burrows1997}
{Burrows}, A., {Marley}, M., {Hubbard}, W.~B., {et~al.} 1997, \apj, 491, 856, \dodoi{10.1086/305002}

\bibitem[{{Carnall}(2017)}]{Carnall2017}
{Carnall}, A.~C. 2017, arXiv e-prints, arXiv:1705.05165, \dodoi{10.48550/arXiv.1705.05165}

\bibitem[{{Casagrande} {et~al.}(2011){Casagrande}, {Sch{\"o}nrich}, {Asplund}, {Cassisi}, {Ram{\'\i}rez}, {Mel{\'e}ndez}, {Bensby}, \& {Feltzing}}]{Casagrande2011}
{Casagrande}, L., {Sch{\"o}nrich}, R., {Asplund}, M., {et~al.} 2011, \aap, 530, A138, \dodoi{10.1051/0004-6361/201016276}

\bibitem[{{Casagrande} {et~al.}(2021){Casagrande}, {Lin}, {Rains}, {Liu}, {Buder}, {Horner}, {Asplund}, {Lewis}, {Martell}, {Nordlander}, {Stello}, {Ting}, {Wittenmyer}, {Bland-Hawthorn}, {Casey}, {De Silva}, {D'Orazi}, {Freeman}, {Hayden}, {Kos}, {Lind}, {Schlesinger}, {Sharma}, {Simpson}, {Zucker}, \& {Zwitter}}]{casagrande2021}
{Casagrande}, L., {Lin}, J., {Rains}, A.~D., {et~al.} 2021, \mnras, 507, 2684, \dodoi{10.1093/mnras/stab2304}

\bibitem[{{Chabrier} \& {Baraffe}(2000)}]{Chabrier2000}
{Chabrier}, G., \& {Baraffe}, I. 2000, \araa, 38, 337, \dodoi{10.1146/annurev.astro.38.1.337}

\bibitem[{{Choi} {et~al.}(2016){Choi}, {Dotter}, {Conroy}, {Cantiello}, {Paxton}, \& {Johnson}}]{choi2016}
{Choi}, J., {Dotter}, A., {Conroy}, C., {et~al.} 2016, \apj, 823, 102, \dodoi{10.3847/0004-637X/823/2/102}

\bibitem[{{Chubb} {et~al.}(2021){Chubb}, {Rocchetto}, {Yurchenko}, {Min}, {Waldmann}, {Barstow}, {Molli{\`e}re}, {Al-Refaie}, {Phillips}, \& {Tennyson}}]{Chubb2021}
{Chubb}, K.~L., {Rocchetto}, M., {Yurchenko}, S.~N., {et~al.} 2021, \aap, 646, A21, \dodoi{10.1051/0004-6361/202038350}

\bibitem[{{Cushing} {et~al.}(2005){Cushing}, {Rayner}, \& {Vacca}}]{Cushing2005}
{Cushing}, M.~C., {Rayner}, J.~T., \& {Vacca}, W.~D. 2005, \apj, 623, 1115, \dodoi{10.1086/428040}

\bibitem[{{Cushing} {et~al.}(2006){Cushing}, {Roellig}, {Marley}, {Saumon}, {Leggett}, {Kirkpatrick}, {Wilson}, {Sloan}, {Mainzer}, {Van Cleve}, \& {Houck}}]{Cushing2006}
{Cushing}, M.~C., {Roellig}, T.~L., {Marley}, M.~S., {et~al.} 2006, \apj, 648, 614, \dodoi{10.1086/505637}

\bibitem[{{Czekala} {et~al.}(2015){Czekala}, {Andrews}, {Mandel}, {Hogg}, \& {Green}}]{Czekala2015}
{Czekala}, I., {Andrews}, S.~M., {Mandel}, K.~S., {Hogg}, D.~W., \& {Green}, G.~M. 2015, \apj, 812, 128, \dodoi{10.1088/0004-637X/812/2/128}

\bibitem[{{Dotter}(2016)}]{dotter2016}
{Dotter}, A. 2016, \apjs, 222, 8, \dodoi{10.3847/0067-0049/222/1/8}

\bibitem[{{Duch{\^e}ne} {et~al.}(2023){Duch{\^e}ne}, {Oon}, {De Rosa}, {Kantorski}, {Coy}, {Wang}, {Thomas}, {Patience}, {Pueyo}, {Nielsen}, \& {Konopacky}}]{Duchene2023}
{Duch{\^e}ne}, G., {Oon}, J.~T., {De Rosa}, R.~J., {et~al.} 2023, \mnras, 519, 778, \dodoi{10.1093/mnras/stac3527}

\bibitem[{{Dupuy} \& {Liu}(2012)}]{Dupuy2012}
{Dupuy}, T.~J., \& {Liu}, M.~C. 2012, \apjs, 201, 19, \dodoi{10.1088/0067-0049/201/2/19}

\bibitem[{{Dupuy} \& {Liu}(2017)}]{Dupuy2017}
---. 2017, \apjs, 231, 15, \dodoi{10.3847/1538-4365/aa5e4c}

\bibitem[{{Fernandes} {et~al.}(2019){Fernandes}, {Van Grootel}, {Salmon}, {Aringer}, {Burgasser}, {Scuflaire}, {Brassard}, \& {Fontaine}}]{Fernandes2019}
{Fernandes}, C.~S., {Van Grootel}, V., {Salmon}, S. J.~A.~J., {et~al.} 2019, \apj, 879, 94, \dodoi{10.3847/1538-4357/ab2333}

\bibitem[{{Feroz} \& {Hobson}(2008)}]{Feroz2008}
{Feroz}, F., \& {Hobson}, M.~P. 2008, \mnras, 384, 449, \dodoi{10.1111/j.1365-2966.2007.12353.x}

\bibitem[{{Feroz} {et~al.}(2009){Feroz}, {Hobson}, \& {Bridges}}]{Feroz2009}
{Feroz}, F., {Hobson}, M.~P., \& {Bridges}, M. 2009, \mnras, 398, 1601, \dodoi{10.1111/j.1365-2966.2009.14548.x}

\bibitem[{{Feroz} {et~al.}(2019){Feroz}, {Hobson}, {Cameron}, \& {Pettitt}}]{Feroz2019}
{Feroz}, F., {Hobson}, M.~P., {Cameron}, E., \& {Pettitt}, A.~N. 2019, The Open Journal of Astrophysics, 2, 10, \dodoi{10.21105/astro.1306.2144}

\bibitem[{{Filippazzo} {et~al.}(2015){Filippazzo}, {Rice}, {Faherty}, {Cruz}, {Van Gordon}, \& {Looper}}]{Filippazzo2015}
{Filippazzo}, J.~C., {Rice}, E.~L., {Faherty}, J., {et~al.} 2015, \apj, 810, 158, \dodoi{10.1088/0004-637X/810/2/158}

\bibitem[{{Fontanive} {et~al.}(2018){Fontanive}, {Biller}, {Bonavita}, \& {Allers}}]{2018MNRAS.479.2702F}
{Fontanive}, C., {Biller}, B., {Bonavita}, M., \& {Allers}, K. 2018, \mnras, 479, 2702, \dodoi{10.1093/mnras/sty1682}

\bibitem[{{Fontanive} {et~al.}(2019){Fontanive}, {Mu{\v{z}}i{\'c}}, {}, {Bonavita}, \& {Biller}}]{Fontanive2019}
{Fontanive}, C., {Mu{\v{z}}i{\'c}}, {}, K., {Bonavita}, M., \& {Biller}, B. 2019, \mnras, 490, 1120, \dodoi{10.1093/mnras/stz2587}

\bibitem[{{Foreman-Mackey} {et~al.}(2013){Foreman-Mackey}, {Hogg}, {Lang}, \& {Goodman}}]{Foreman-Mackey2013}
{Foreman-Mackey}, D., {Hogg}, D.~W., {Lang}, D., \& {Goodman}, J. 2013, \pasp, 125, 306, \dodoi{10.1086/670067}

\bibitem[{{Fouesneau} {et~al.}(2022){Fouesneau}, {Fr{\'e}mat}, {Andrae}, {Korn}, {Soubiran}, {Kordopatis}, {Vallenari}, {Heiter}, {Creevey}, {Sarro}, {de Laverny}, {Lanzafame}, {Lobel}, {Sordo}, {Rybizki}, {Slezak}, {{\'A}lvarez}, {Drimmel}, {Garabato}, {Delchambre}, {Bailer-Jones}, {Hatzidimitriou}, {Lorca}, {Le Fustec}, {Pailler}, {Mary}, {Robin}, {Utrilla}, {Abreu Aramburu}, {Bakker}, {Bellas-Velidis}, {Bijaoui}, {Blomme}, {Bouret}, {Brouillet}, {Brugaletta}, {Burlacu}, {Carballo}, {Casamiquela}, {Chaoul}, {Chiavassa}, {Contursi}, {Cooper}, {Dafonte}, {Demouchy}, {Dharmawardena}, {Garc{\'\i}a-Lario}, {Garc{\'\i}a-Torres}, {Gomez}, {Gonz{\'a}lez-Santamar{\'\i}a}, {Jean-Antoine Piccolo}, {Kontizas}, {Lebreton}, {Licata}, {Lindstr{\o}m}, {Livanou}, {Magdaleno Romeo}, {Manteiga}, {Marocco}, {Martayan}, {Marshall}, {Nicolas}, {Ordenovic}, {Palicio}, {Pallas-Quintela}, {Pichon}, {Poggio}, {Recio-Blanco}, {Riclet}, {Santove{\~n}a}, {Schultheis}, {Segol}, {Silvelo}, {Smart}, {S{\"u}veges}, {Th{\'e}venin},
  {Torralba Elipe}, {Ulla}, {van Dillen}, {Zhao}, \& {Zorec}}]{fouesneau2022}
{Fouesneau}, M., {Fr{\'e}mat}, Y., {Andrae}, R., {et~al.} 2022, arXiv e-prints, arXiv:2206.05992.
\newblock \doarXiv{2206.05992}

\bibitem[{{Franson} {et~al.}(2022){Franson}, {Bowler}, {Brandt}, {Dupuy}, {Tran}, {Brandt}, {Li}, \& {Kraus}}]{Franson2022}
{Franson}, K., {Bowler}, B.~P., {Brandt}, T.~D., {et~al.} 2022, \aj, 163, 50, \dodoi{10.3847/1538-3881/ac35e8}

\bibitem[{{Franson} {et~al.}(2023){Franson}, {Bowler}, {Bonavita}, {Brandt}, {Chen}, {Samland}, {Zhang}, {Lueber}, {Heng}, {Kitzmann}, {Wolf}, {Jones}, {Tran}, {Bardalez Gagliuffi}, {Biller}, {Chilcote}, {Crepp}, {Dupuy}, {Faherty}, {Fontanive}, {Groff}, {Gratton}, {Guyon}, {Jensen-Clem}, {Jovanovic}, {Kasdin}, {Lozi}, {Magnier}, {Mu{\v{z}}i{\'c}}, {Sanghi}, \& {Theissen}}]{Franson2023}
{Franson}, K., {Bowler}, B.~P., {Bonavita}, M., {et~al.} 2023, \aj, 165, 39, \dodoi{10.3847/1538-3881/aca408}

\bibitem[{{Gaia Collaboration} {et~al.}(2016){Gaia Collaboration}, {Prusti}, {de Bruijne}, {Brown}, {Vallenari}, {Babusiaux}, {Bailer-Jones}, {Bastian}, {Biermann}, {Evans}, {Eyer}, {Jansen}, {Jordi}, {Klioner}, {Lammers}, {Lindegren}, {Luri}, {Mignard}, {Milligan}, {Panem}, {Poinsignon}, {Pourbaix}, {Randich}, {Sarri}, {Sartoretti}, {Siddiqui}, {Soubiran}, {Valette}, {van Leeuwen}, {Walton}, {Aerts}, {Arenou}, {Cropper}, {Drimmel}, {H{\o}g}, {Katz}, {Lattanzi}, {O'Mullane}, {Grebel}, {Holland}, {Huc}, {Passot}, {Bramante}, {Cacciari}, {Casta{\~n}eda}, {Chaoul}, {Cheek}, {De Angeli}, {Fabricius}, {Guerra}, {Hern{\'a}ndez}, {Jean-Antoine-Piccolo}, {Masana}, {Messineo}, {Mowlavi}, {Nienartowicz}, {Ord{\'o}{\~n}ez-Blanco}, {Panuzzo}, {Portell}, {Richards}, {Riello}, {Seabroke}, {Tanga}, {Th{\'e}venin}, {Torra}, {Els}, {Gracia-Abril}, {Comoretto}, {Garcia-Reinaldos}, {Lock}, {Mercier}, {Altmann}, {Andrae}, {Astraatmadja}, {Bellas-Velidis}, {Benson}, {Berthier}, {Blomme}, {Busso}, {Carry}, {Cellino}, {Clementini},
  {Cowell}, {Creevey}, {Cuypers}, {Davidson}, {De Ridder}, {de Torres}, {Delchambre}, {Dell'Oro}, {Ducourant}, {Fr{\'e}mat}, {Garc{\'\i}a-Torres}, {Gosset}, {Halbwachs}, {Hambly}, {Harrison}, {Hauser}, {Hestroffer}, {Hodgkin}, {Huckle}, {Hutton}, {Jasniewicz}, {Jordan}, {Kontizas}, {Korn}, {Lanzafame}, {Manteiga}, {Moitinho}, {Muinonen}, {Osinde}, {Pancino}, {Pauwels}, {Petit}, {Recio-Blanco}, {Robin}, {Sarro}, {Siopis}, {Smith}, {Smith}, {Sozzetti}, {Thuillot}, {van Reeven}, {Viala}, {Abbas}, {Abreu Aramburu}, {Accart}, {Aguado}, {Allan}, {Allasia}, {Altavilla}, {{\'A}lvarez}, {Alves}, {Anderson}, {Andrei}, {Anglada Varela}, {Antiche}, {Antoja}, {Ant{\'o}n}, {Arcay}, {Atzei}, {Ayache}, {Bach}, {Baker}, {Balaguer-N{\'u}{\~n}ez}, {Barache}, {Barata}, {Barbier}, {Barblan}, {Baroni}, {Barrado y Navascu{\'e}s}, {Barros}, {Barstow}, {Becciani}, {Bellazzini}, {Bellei}, {Bello Garc{\'\i}a}, {Belokurov}, {Bendjoya}, {Berihuete}, {Bianchi}, {Bienaym{\'e}}, {Billebaud}, {Blagorodnova}, {Blanco-Cuaresma}, {Boch},
  {Bombrun}, {Borrachero}, {Bouquillon}, {Bourda}, {Bouy}, {Bragaglia}, {Breddels}, {Brouillet}, {Br{\"u}semeister}, {Bucciarelli}, {Budnik}, {Burgess}, {Burgon}, {Burlacu}, {Busonero}, {Buzzi}, {Caffau}, {Cambras}, {Campbell}, {Cancelliere}, {Cantat-Gaudin}, {Carlucci}, {Carrasco}, {Castellani}, {Charlot}, {Charnas}, {Charvet}, {Chassat}, {Chiavassa}, {Clotet}, {Cocozza}, {Collins}, {Collins}, {Costigan}, {Crifo}, {Cross}, {Crosta}, {Crowley}, {Dafonte}, {Damerdji}, {Dapergolas}, {David}, {David}, {De Cat}, {de Felice}, {de Laverny}, {De Luise}, {De March}, {de Martino}, {de Souza}, {Debosscher}, {del Pozo}, {Delbo}, {Delgado}, {Delgado}, {di Marco}, {Di Matteo}, {Diakite}, {Distefano}, {Dolding}, {Dos Anjos}, {Drazinos}, {Dur{\'a}n}, {Dzigan}, {Ecale}, {Edvardsson}, {Enke}, {Erdmann}, {Escolar}, {Espina}, {Evans}, {Eynard Bontemps}, {Fabre}, {Fabrizio}, {Faigler}, {Falc{\~a}o}, {Farr{\`a}s Casas}, {Faye}, {Federici}, {Fedorets}, {Fern{\'a}ndez-Hern{\'a}ndez}, {Fernique}, {Fienga}, {Figueras}, {Filippi},
  {Findeisen}, {Fonti}, {Fouesneau}, {Fraile}, {Fraser}, {Fuchs}, {Furnell}, {Gai}, {Galleti}, {Galluccio}, {Garabato}, {Garc{\'\i}a-Sedano}, {Gar{\'e}}, {Garofalo}, {Garralda}, {Gavras}, {Gerssen}, {Geyer}, {Gilmore}, {Girona}, {Giuffrida}, {Gomes}, {Gonz{\'a}lez-Marcos}, {Gonz{\'a}lez-N{\'u}{\~n}ez}, {Gonz{\'a}lez-Vidal}, {Granvik}, {Guerrier}, {Guillout}, {Guiraud}, {G{\'u}rpide}, {Guti{\'e}rrez-S{\'a}nchez}, {Guy}, {Haigron}, {Hatzidimitriou}, {Haywood}, {Heiter}, {Helmi}, {Hobbs}, {Hofmann}, {Holl}, {Holland}, {Hunt}, {Hypki}, {Icardi}, {Irwin}, {Jevardat de Fombelle}, {Jofr{\'e}}, {Jonker}, {Jorissen}, {Julbe}, {Karampelas}, {Kochoska}, {Kohley}, {Kolenberg}, {Kontizas}, {Koposov}, {Kordopatis}, {Koubsky}, {Kowalczyk}, {Krone-Martins}, {Kudryashova}, {Kull}, {Bachchan}, {Lacoste-Seris}, {Lanza}, {Lavigne}, {Le Poncin-Lafitte}, {Lebreton}, {Lebzelter}, {Leccia}, {Leclerc}, {Lecoeur-Taibi}, {Lemaitre}, {Lenhardt}, {Leroux}, {Liao}, {Licata}, {Lindstr{\o}m}, {Lister}, {Livanou}, {Lobel}, {L{\"o}ffler},
  {L{\'o}pez}, {Lopez-Lozano}, {Lorenz}, {Loureiro}, {MacDonald}, {Magalh{\~a}es Fernandes}, {Managau}, {Mann}, {Mantelet}, {Marchal}, {Marchant}, {Marconi}, {Marie}, {Marinoni}, {Marrese}, {Marschalk{\'o}}, {Marshall}, {Mart{\'\i}n-Fleitas}, {Martino}, {Mary}, {Matijevi{\v{c}}}, {Mazeh}, {McMillan}, {Messina}, {Mestre}, {Michalik}, {Millar}, {Miranda}, {Molina}, {Molinaro}, {Molinaro}, {Moln{\'a}r}, {Moniez}, {Montegriffo}, {Monteiro}, {Mor}, {Mora}, {Morbidelli}, {Morel}, {Morgenthaler}, {Morley}, {Morris}, {Mulone}, {Muraveva}, {Musella}, {Narbonne}, {Nelemans}, {Nicastro}, {Noval}, {Ord{\'e}novic}, {Ordieres-Mer{\'e}}, {Osborne}, {Pagani}, {Pagano}, {Pailler}, {Palacin}, {Palaversa}, {Parsons}, {Paulsen}, {Pecoraro}, {Pedrosa}, {Pentik{\"a}inen}, {Pereira}, {Pichon}, {Piersimoni}, {Pineau}, {Plachy}, {Plum}, {Poujoulet}, {Pr{\v{s}}a}, {Pulone}, {Ragaini}, {Rago}, {Rambaux}, {Ramos-Lerate}, {Ranalli}, {Rauw}, {Read}, {Regibo}, {Renk}, {Reyl{\'e}}, {Ribeiro}, {Rimoldini}, {Ripepi}, {Riva}, {Rixon},
  {Roelens}, {Romero-G{\'o}mez}, {Rowell}, {Royer}, {Rudolph}, {Ruiz-Dern}, {Sadowski}, {Sagrist{\`a} Sell{\'e}s}, {Sahlmann}, {Salgado}, {Salguero}, {Sarasso}, {Savietto}, {Schnorhk}, {Schultheis}, {Sciacca}, {Segol}, {Segovia}, {Segransan}, {Serpell}, {Shih}, {Smareglia}, {Smart}, {Smith}, {Solano}, {Solitro}, {Sordo}, {Soria Nieto}, {Souchay}, {Spagna}, {Spoto}, {Stampa}, {Steele}, {Steidelm{\"u}ller}, {Stephenson}, {Stoev}, {Suess}, {S{\"u}veges}, {Surdej}, {Szabados}, {Szegedi-Elek}, {Tapiador}, {Taris}, {Tauran}, {Taylor}, {Teixeira}, {Terrett}, {Tingley}, {Trager}, {Turon}, {Ulla}, {Utrilla}, {Valentini}, {van Elteren}, {Van Hemelryck}, {van Leeuwen}, {Varadi}, {Vecchiato}, {Veljanoski}, {Via}, {Vicente}, {Vogt}, {Voss}, {Votruba}, {Voutsinas}, {Walmsley}, {Weiler}, {Weingrill}, {Werner}, {Wevers}, {Whitehead}, {Wyrzykowski}, {Yoldas}, {{\v{Z}}erjal}, {Zucker}, {Zurbach}, {Zwitter}, {Alecu}, {Allen}, {Allende Prieto}, {Amorim}, {Anglada-Escud{\'e}}, {Arsenijevic}, {Azaz}, {Balm}, {Beck}, {Bernstein},
  {Bigot}, {Bijaoui}, {Blasco}, {Bonfigli}, {Bono}, {Boudreault}, {Bressan}, {Brown}, {Brunet}, {Bunclark}, {Buonanno}, {Butkevich}, {Carret}, {Carrion}, {Chemin}, {Ch{\'e}reau}, {Corcione}, {Darmigny}, {de Boer}, {de Teodoro}, {de Zeeuw}, {Delle Luche}, {Domingues}, {Dubath}, {Fodor}, {Fr{\'e}zouls}, {Fries}, {Fustes}, {Fyfe}, {Gallardo}, {Gallegos}, {Gardiol}, {Gebran}, {Gomboc}, {G{\'o}mez}, {Grux}, {Gueguen}, {Heyrovsky}, {Hoar}, {Iannicola}, {Isasi Parache}, {Janotto}, {Joliet}, {Jonckheere}, {Keil}, {Kim}, {Klagyivik}, {Klar}, {Knude}, {Kochukhov}, {Kolka}, {Kos}, {Kutka}, {Lainey}, {LeBouquin}, {Liu}, {Loreggia}, {Makarov}, {Marseille}, {Martayan}, {Martinez-Rubi}, {Massart}, {Meynadier}, {Mignot}, {Munari}, {Nguyen}, {Nordlander}, {Ocvirk}, {O'Flaherty}, {Olias Sanz}, {Ortiz}, {Osorio}, {Oszkiewicz}, {Ouzounis}, {Palmer}, {Park}, {Pasquato}, {Peltzer}, {Peralta}, {P{\'e}turaud}, {Pieniluoma}, {Pigozzi}, {Poels}, {Prat}, {Prod'homme}, {Raison}, {Rebordao}, {Risquez}, {Rocca-Volmerange}, {Rosen},
  {Ruiz-Fuertes}, {Russo}, {Sembay}, {Serraller Vizcaino}, {Short}, {Siebert}, {Silva}, {Sinachopoulos}, {Slezak}, {Soffel}, {Sosnowska}, {Strai{\v{z}}ys}, {ter Linden}, {Terrell}, {Theil}, {Tiede}, {Troisi}, {Tsalmantza}, {Tur}, {Vaccari}, {Vachier}, {Valles}, {Van Hamme}, {Veltz}, {Virtanen}, {Wallut}, {Wichmann}, {Wilkinson}, {Ziaeepour}, \& {Zschocke}}]{gaia2016}
{Gaia Collaboration}, {Prusti}, T., {de Bruijne}, J.~H.~J., {et~al.} 2016, \aap, 595, A1, \dodoi{10.1051/0004-6361/201629272}

\bibitem[{{Gaia Collaboration} {et~al.}(2018){Gaia Collaboration}, {Brown}, {Vallenari}, {Prusti}, {de Bruijne}, {Babusiaux}, {Bailer-Jones}, {Biermann}, {Evans}, {Eyer}, {Jansen}, {Jordi}, {Klioner}, {Lammers}, {Lindegren}, {Luri}, {Mignard}, {Panem}, {Pourbaix}, {Randich}, {Sartoretti}, {Siddiqui}, {Soubiran}, {van Leeuwen}, {Walton}, {Arenou}, {Bastian}, {Cropper}, {Drimmel}, {Katz}, {Lattanzi}, {Bakker}, {Cacciari}, {Casta{\~n}eda}, {Chaoul}, {Cheek}, {De Angeli}, {Fabricius}, {Guerra}, {Holl}, {Masana}, {Messineo}, {Mowlavi}, {Nienartowicz}, {Panuzzo}, {Portell}, {Riello}, {Seabroke}, {Tanga}, {Th{\'e}venin}, {Gracia-Abril}, {Comoretto}, {Garcia-Reinaldos}, {Teyssier}, {Altmann}, {Andrae}, {Audard}, {Bellas-Velidis}, {Benson}, {Berthier}, {Blomme}, {Burgess}, {Busso}, {Carry}, {Cellino}, {Clementini}, {Clotet}, {Creevey}, {Davidson}, {De Ridder}, {Delchambre}, {Dell'Oro}, {Ducourant}, {Fern{\'a}ndez-Hern{\'a}ndez}, {Fouesneau}, {Fr{\'e}mat}, {Galluccio}, {Garc{\'\i}a-Torres},
  {Gonz{\'a}lez-N{\'u}{\~n}ez}, {Gonz{\'a}lez-Vidal}, {Gosset}, {Guy}, {Halbwachs}, {Hambly}, {Harrison}, {Hern{\'a}ndez}, {Hestroffer}, {Hodgkin}, {Hutton}, {Jasniewicz}, {Jean-Antoine-Piccolo}, {Jordan}, {Korn}, {Krone-Martins}, {Lanzafame}, {Lebzelter}, {L{\"o}ffler}, {Manteiga}, {Marrese}, {Mart{\'\i}n-Fleitas}, {Moitinho}, {Mora}, {Muinonen}, {Osinde}, {Pancino}, {Pauwels}, {Petit}, {Recio-Blanco}, {Richards}, {Rimoldini}, {Robin}, {Sarro}, {Siopis}, {Smith}, {Sozzetti}, {S{\"u}veges}, {Torra}, {van Reeven}, {Abbas}, {Abreu Aramburu}, {Accart}, {Aerts}, {Altavilla}, {{\'A}lvarez}, {Alvarez}, {Alves}, {Anderson}, {Andrei}, {Anglada Varela}, {Antiche}, {Antoja}, {Arcay}, {Astraatmadja}, {Bach}, {Baker}, {Balaguer-N{\'u}{\~n}ez}, {Balm}, {Barache}, {Barata}, {Barbato}, {Barblan}, {Barklem}, {Barrado}, {Barros}, {Barstow}, {Bartholom{\'e} Mu{\~n}oz}, {Bassilana}, {Becciani}, {Bellazzini}, {Berihuete}, {Bertone}, {Bianchi}, {Bienaym{\'e}}, {Blanco-Cuaresma}, {Boch}, {Boeche}, {Bombrun}, {Borrachero},
  {Bossini}, {Bouquillon}, {Bourda}, {Bragaglia}, {Bramante}, {Breddels}, {Bressan}, {Brouillet}, {Br{\"u}semeister}, {Brugaletta}, {Bucciarelli}, {Burlacu}, {Busonero}, {Butkevich}, {Buzzi}, {Caffau}, {Cancelliere}, {Cannizzaro}, {Cantat-Gaudin}, {Carballo}, {Carlucci}, {Carrasco}, {Casamiquela}, {Castellani}, {Castro-Ginard}, {Charlot}, {Chemin}, {Chiavassa}, {Cocozza}, {Costigan}, {Cowell}, {Crifo}, {Crosta}, {Crowley}, {Cuypers}, {Dafonte}, {Damerdji}, {Dapergolas}, {David}, {David}, {de Laverny}, {De Luise}, {De March}, {de Martino}, {de Souza}, {de Torres}, {Debosscher}, {del Pozo}, {Delbo}, {Delgado}, {Delgado}, {Di Matteo}, {Diakite}, {Diener}, {Distefano}, {Dolding}, {Drazinos}, {Dur{\'a}n}, {Edvardsson}, {Enke}, {Eriksson}, {Esquej}, {Eynard Bontemps}, {Fabre}, {Fabrizio}, {Faigler}, {Falc{\~a}o}, {Farr{\`a}s Casas}, {Federici}, {Fedorets}, {Fernique}, {Figueras}, {Filippi}, {Findeisen}, {Fonti}, {Fraile}, {Fraser}, {Fr{\'e}zouls}, {Gai}, {Galleti}, {Garabato}, {Garc{\'\i}a-Sedano}, {Garofalo},
  {Garralda}, {Gavel}, {Gavras}, {Gerssen}, {Geyer}, {Giacobbe}, {Gilmore}, {Girona}, {Giuffrida}, {Glass}, {Gomes}, {Granvik}, {Gueguen}, {Guerrier}, {Guiraud}, {Guti{\'e}rrez-S{\'a}nchez}, {Haigron}, {Hatzidimitriou}, {Hauser}, {Haywood}, {Heiter}, {Helmi}, {Heu}, {Hilger}, {Hobbs}, {Hofmann}, {Holland}, {Huckle}, {Hypki}, {Icardi}, {Jan{\ss}en}, {Jevardat de Fombelle}, {Jonker}, {Juh{\'a}sz}, {Julbe}, {Karampelas}, {Kewley}, {Klar}, {Kochoska}, {Kohley}, {Kolenberg}, {Kontizas}, {Kontizas}, {Koposov}, {Kordopatis}, {Kostrzewa-Rutkowska}, {Koubsky}, {Lambert}, {Lanza}, {Lasne}, {Lavigne}, {Le Fustec}, {Le Poncin-Lafitte}, {Lebreton}, {Leccia}, {Leclerc}, {Lecoeur-Taibi}, {Lenhardt}, {Leroux}, {Liao}, {Licata}, {Lindstr{\o}m}, {Lister}, {Livanou}, {Lobel}, {L{\'o}pez}, {Managau}, {Mann}, {Mantelet}, {Marchal}, {Marchant}, {Marconi}, {Marinoni}, {Marschalk{\'o}}, {Marshall}, {Martino}, {Marton}, {Mary}, {Massari}, {Matijevi{\v{c}}}, {Mazeh}, {McMillan}, {Messina}, {Michalik}, {Millar}, {Molina}, {Molinaro},
  {Moln{\'a}r}, {Montegriffo}, {Mor}, {Morbidelli}, {Morel}, {Morris}, {Mulone}, {Muraveva}, {Musella}, {Nelemans}, {Nicastro}, {Noval}, {O'Mullane}, {Ord{\'e}novic}, {Ord{\'o}{\~n}ez-Blanco}, {Osborne}, {Pagani}, {Pagano}, {Pailler}, {Palacin}, {Palaversa}, {Panahi}, {Pawlak}, {Piersimoni}, {Pineau}, {Plachy}, {Plum}, {Poggio}, {Poujoulet}, {Pr{\v{s}}a}, {Pulone}, {Racero}, {Ragaini}, {Rambaux}, {Ramos-Lerate}, {Regibo}, {Reyl{\'e}}, {Riclet}, {Ripepi}, {Riva}, {Rivard}, {Rixon}, {Roegiers}, {Roelens}, {Romero-G{\'o}mez}, {Rowell}, {Royer}, {Ruiz-Dern}, {Sadowski}, {Sagrist{\`a} Sell{\'e}s}, {Sahlmann}, {Salgado}, {Salguero}, {Sanna}, {Santana-Ros}, {Sarasso}, {Savietto}, {Schultheis}, {Sciacca}, {Segol}, {Segovia}, {S{\'e}gransan}, {Shih}, {Siltala}, {Silva}, {Smart}, {Smith}, {Solano}, {Solitro}, {Sordo}, {Soria Nieto}, {Souchay}, {Spagna}, {Spoto}, {Stampa}, {Steele}, {Steidelm{\"u}ller}, {Stephenson}, {Stoev}, {Suess}, {Surdej}, {Szabados}, {Szegedi-Elek}, {Tapiador}, {Taris}, {Tauran}, {Taylor},
  {Teixeira}, {Terrett}, {Teyssandier}, {Thuillot}, {Titarenko}, {Torra Clotet}, {Turon}, {Ulla}, {Utrilla}, {Uzzi}, {Vaillant}, {Valentini}, {Valette}, {van Elteren}, {Van Hemelryck}, {van Leeuwen}, {Vaschetto}, {Vecchiato}, {Veljanoski}, {Viala}, {Vicente}, {Vogt}, {von Essen}, {Voss}, {Votruba}, {Voutsinas}, {Walmsley}, {Weiler}, {Wertz}, {Wevers}, {Wyrzykowski}, {Yoldas}, {{\v{Z}}erjal}, {Ziaeepour}, {Zorec}, {Zschocke}, {Zucker}, {Zurbach}, \& {Zwitter}}]{gaia2018}
{Gaia Collaboration}, {Brown}, A.~G.~A., {Vallenari}, A., {et~al.} 2018, \aap, 616, A1, \dodoi{10.1051/0004-6361/201833051}

\bibitem[{{Gaia Collaboration} {et~al.}(2021){Gaia Collaboration}, {Brown}, {Vallenari}, {Prusti}, {de Bruijne}, {Babusiaux}, {Biermann}, {Creevey}, {Evans}, {Eyer}, \& et~al.}]{GaiaCollaboration2021}
---. 2021, \aap, 649, A1, \dodoi{10.1051/0004-6361/202039657}

\bibitem[{{Gao} {et~al.}(2018){Gao}, {Marley}, \& {Ackerman}}]{Gao2018}
{Gao}, P., {Marley}, M.~S., \& {Ackerman}, A.~S. 2018, \apj, 855, 86, \dodoi{10.3847/1538-4357/aab0a1}

\bibitem[{{Gonzales} {et~al.}(2020){Gonzales}, {Burningham}, {Faherty}, {Cleary}, {Visscher}, {Marley}, {Lupu}, \& {Freedman}}]{Gonzales2020}
{Gonzales}, E.~C., {Burningham}, B., {Faherty}, J.~K., {et~al.} 2020, \apj, 905, 46, \dodoi{10.3847/1538-4357/abbee2}

\bibitem[{{Gravity Collaboration} {et~al.}(2017){Gravity Collaboration}, {Abuter}, {Accardo}, {Amorim}, {Anugu}, {{\'A}vila}, {Azouaoui}, {Benisty}, {Berger}, {Blind}, {Bonnet}, {Bourget}, {Brandner}, {Brast}, {Buron}, {Burtscher}, {Cassaing}, {Chapron}, {Choquet}, {Cl{\'e}net}, {Collin}, {Coud{\'e} Du Foresto}, {de Wit}, {de Zeeuw}, {Deen}, {Delplancke-Str{\"o}bele}, {Dembet}, {Derie}, {Dexter}, {Duvert}, {Ebert}, {Eckart}, {Eisenhauer}, {Esselborn}, {F{\'e}dou}, {Finger}, {Garcia}, {Garcia Dabo}, {Garcia Lopez}, {Gendron}, {Genzel}, {Gillessen}, {Gonte}, {Gordo}, {Grould}, {Gr{\"o}zinger}, {Guieu}, {Haguenauer}, {Hans}, {Haubois}, {Haug}, {Haussmann}, {Henning}, {Hippler}, {Horrobin}, {Huber}, {Hubert}, {Hubin}, {Hummel}, {Jakob}, {Janssen}, {Jochum}, {Jocou}, {Kaufer}, {Kellner}, {Kendrew}, {Kern}, {Kervella}, {Kiekebusch}, {Klein}, {Kok}, {Kolb}, {Kulas}, {Lacour}, {Lapeyr{\`e}re}, {Lazareff}, {Le Bouquin}, {L{\`e}na}, {Lenzen}, {L{\'e}v{\^e}que}, {Lippa}, {Magnard}, {Mehrgan}, {Mellein}, {M{\'e}rand},
  {Moreno-Ventas}, {Moulin}, {M{\"u}ller}, {M{\"u}ller}, {Neumann}, {Oberti}, {Ott}, {Pallanca}, {Panduro}, {Pasquini}, {Paumard}, {Percheron}, {Perraut}, {Perrin}, {Pfl{\"u}ger}, {Pfuhl}, {Phan Duc}, {Plewa}, {Popovic}, {Rabien}, {Ram{\'\i}rez}, {Ramos}, {Rau}, {Riquelme}, {Rohloff}, {Rousset}, {Sanchez-Bermudez}, {Scheithauer}, {Sch{\"o}ller}, {Schuhler}, {Spyromilio}, {Straubmeier}, {Sturm}, {Suarez}, {Tristram}, {Ventura}, {Vincent}, {Waisberg}, {Wank}, {Weber}, {Wieprecht}, {Wiest}, {Wiezorrek}, {Wittkowski}, {Woillez}, {Wolff}, {Yazici}, {Ziegler}, \& {Zins}}]{GravityCollaboration2017}
{Gravity Collaboration}, {Abuter}, R., {Accardo}, M., {et~al.} 2017, \aap, 602, A94, \dodoi{10.1051/0004-6361/201730838}

\bibitem[{{Gravity Collaboration} {et~al.}(2020){Gravity Collaboration}, {Nowak}, {Lacour}, {Molli{\`e}re}, {Wang}, {Charnay}, {van Dishoeck}, {Abuter}, {Amorim}, {Berger}, {Beust}, {Bonnefoy}, {Bonnet}, {Brandner}, {Buron}, {Cantalloube}, {Collin}, {Chapron}, {Cl{\'e}net}, {Coud{\'e} Du Foresto}, {de Zeeuw}, {Dembet}, {Dexter}, {Duvert}, {Eckart}, {Eisenhauer}, {F{\"o}rster Schreiber}, {F{\'e}dou}, {Garcia Lopez}, {Gao}, {Gendron}, {Genzel}, {Gillessen}, {Hau{\ss}mann}, {Henning}, {Hippler}, {Hubert}, {Jocou}, {Kervella}, {Lagrange}, {Lapeyr{\`e}re}, {Le Bouquin}, {L{\'e}na}, {Maire}, {Ott}, {Paumard}, {Paladini}, {Perraut}, {Perrin}, {Pueyo}, {Pfuhl}, {Rabien}, {Rau}, {Rodr{\'\i}guez-Coira}, {Rousset}, {Scheithauer}, {Shangguan}, {Straub}, {Straubmeier}, {Sturm}, {Tacconi}, {Vincent}, {Widmann}, {Wieprecht}, {Wiezorrek}, {Woillez}, {Yazici}, \& {Ziegler}}]{GravityCollaboration2020}
{Gravity Collaboration}, {Nowak}, M., {Lacour}, S., {et~al.} 2020, \aap, 633, A110, \dodoi{10.1051/0004-6361/201936898}

\bibitem[{{Green}(1985)}]{Green1985}
{Green}, R.~M. 1985, {Spherical Astronomy} (Cambridge University Press)

\bibitem[{{Grether} \& {Lineweaver}(2006)}]{Grether2006}
{Grether}, D., \& {Lineweaver}, C.~H. 2006, \apj, 640, 1051, \dodoi{10.1086/500161}

\bibitem[{{Henning} \& {Stognienko}(1996)}]{Henning1996}
{Henning}, T., \& {Stognienko}, R. 1996, \aap, 311, 291

\bibitem[{{H{\o}g} {et~al.}(2000){H{\o}g}, {Fabricius}, {Makarov}, {Urban}, {Corbin}, {Wycoff}, {Bastian}, {Schwekendiek}, \& {Wicenec}}]{hogg2000}
{H{\o}g}, E., {Fabricius}, C., {Makarov}, V.~V., {et~al.} 2000, \aap, 355, L27

\bibitem[{{Jaeger} {et~al.}(1998){Jaeger}, {Molster}, {Dorschner}, {Henning}, {Mutschke}, \& {Waters}}]{Jaeger1998}
{Jaeger}, C., {Molster}, F.~J., {Dorschner}, J., {et~al.} 1998, \aap, 339, 904

\bibitem[{{Johansson} {et~al.}(2003){Johansson}, {Litz{\'e}n}, {Lundberg}, \& {Zhang}}]{johansson2003}
{Johansson}, S., {Litz{\'e}n}, U., {Lundberg}, H., \& {Zhang}, Z. 2003, \apjl, 584, L107, \dodoi{10.1086/374037}

\bibitem[{{Kammerer} {et~al.}(2021){Kammerer}, {Lacour}, {Stolker}, {Molli{\`e}re}, {Sing}, {Nasedkin}, {Kervella}, {Wang}, {Ward-Duong}, {Nowak}, {Abuter}, {Amorim}, {Asensio-Torres}, {Baub{\"o}ck}, {Benisty}, {Berger}, {Beust}, {Blunt}, {Boccaletti}, {Bohn}, {Bolzer}, {Bonnefoy}, {Bonnet}, {Brandner}, {Cantalloube}, {Caselli}, {Charnay}, {Chauvin}, {Choquet}, {Christiaens}, {Cl{\'e}net}, {Coud{\'e} du Foresto}, {Cridland}, {Dembet}, {Dexter}, {de Zeeuw}, {Drescher}, {Duvert}, {Eckart}, {Eisenhauer}, {Gao}, {Garcia}, {Garcia Lopez}, {Gendron}, {Genzel}, {Gillessen}, {Girard}, {Haubois}, {Hei{\ss}el}, {Henning}, {Hinkley}, {Hippler}, {Horrobin}, {Houll{\'e}}, {Hubert}, {Jocou}, {Keppler}, {Kreidberg}, {Lagrange}, {Lapeyr{\`e}re}, {Le Bouquin}, {L{\'e}na}, {Lutz}, {Maire}, {M{\'e}rand}, {Monnier}, {Mouillet}, {M{\"u}ller}, {Ott}, {Otten}, {Paladini}, {Paumard}, {Perraut}, {Perrin}, {Pfuhl}, {Pueyo}, {Rameau}, {Rodet}, {Rousset}, {Rustamkulov}, {Shangguan}, {Shimizu}, {Stadler}, {Straub}, {Straubmeier},
  {Sturm}, {Tacconi}, {van Dishoeck}, {Vigan}, {Vincent}, {von Fellenberg}, {Widmann}, {Wieprecht}, {Wiezorrek}, {Woillez}, \& {Yazici}}]{Kammerer2021}
{Kammerer}, J., {Lacour}, S., {Stolker}, T., {et~al.} 2021, \aap, 652, A57, \dodoi{10.1051/0004-6361/202140749}

\bibitem[{{Knapp} {et~al.}(2004){Knapp}, {Leggett}, {Fan}, {Marley}, {Geballe}, {Golimowski}, {Finkbeiner}, {Gunn}, {Hennawi}, {Ivezi{\'c}}, {Lupton}, {Schlegel}, {Strauss}, {Tsvetanov}, {Chiu}, {Hoversten}, {Glazebrook}, {Zheng}, {Hendrickson}, {Williams}, {Uomoto}, {Vrba}, {Henden}, {Luginbuhl}, {Guetter}, {Munn}, {Canzian}, {Schneider}, \& {Brinkmann}}]{Knapp2004}
{Knapp}, G.~R., {Leggett}, S.~K., {Fan}, X., {et~al.} 2004, \aj, 127, 3553, \dodoi{10.1086/420707}

\bibitem[{{Kratter} {et~al.}(2010){Kratter}, {Murray-Clay}, \& {Youdin}}]{Kratter2010}
{Kratter}, K.~M., {Murray-Clay}, R.~A., \& {Youdin}, A.~N. 2010, \apj, 710, 1375, \dodoi{10.1088/0004-637X/710/2/1375}

\bibitem[{{Lacour} {et~al.}(2019){Lacour}, {Dembet}, {Abuter}, {F{\'e}dou}, {Perrin}, {Choquet}, {Pfuhl}, {Eisenhauer}, {Woillez}, {Cassaing}, {Wieprecht}, {Ott}, {Wiezorrek}, {Tristram}, {Wolff}, {Ram{\'\i}rez}, {Haubois}, {Perraut}, {Straubmeier}, {Brandner}, \& {Amorim}}]{Lacour2019}
{Lacour}, S., {Dembet}, R., {Abuter}, R., {et~al.} 2019, \aap, 624, A99, \dodoi{10.1051/0004-6361/201834981}

\bibitem[{{Lacour} {et~al.}(2020){Lacour}, {Wang}, {Nowak}, {Pueyo}, {Eisenhauer}, {Lagrange}, {Molli{\`e}re}, {Abuter}, {Amorin}, {Asensio-Torres}, {Baub{\"o}ck}, {Benisty}, {Berger}, {Beust}, {Blunt}, {Boccaletti}, {Bohn}, {Bonnefoy}, {Bonnet}, {Brandner}, {Cantalloube}, {Caselli}, {Charnay}, {Chauvin}, {Choquet}, {Christiaens}, {Cl{\'e}net}, {Cridland}, {de Zeeuw}, {Dembet}, {Dexter}, {Drescher}, {Duvert}, {Gao}, {Garcia}, {Garcia Lopez}, {Gardner}, {Gendron}, {Genzel}, {Gillessen}, {Girard}, {Haubois}, {Hei{\ss}el}, {Henning}, {Hinkley}, {Hippler}, {Horrobin}, {Houll{\'e}}, {Hubert}, {Jim{\'e}nez-Rosales}, {Jocou}, {Kammerer}, {Keppler}, {Kervella}, {Kreidberg}, {Lapeyr{\`e}re}, {Le Bouquin}, {L{\'e}na}, {Lutz}, {Maire}, {M{\'e}rand}, {Monnier}, {Mouillet}, {Muller}, {Nasedkin}, {Ott}, {Otten}, {Paladini}, {Paumard}, {Perraut}, {Perrin}, {Pfuhl}, {Rameau}, {Rodet}, {Rodriguez-Coira}, {Rousset}, {Shangguan}, {Shimizu}, {Stadler}, {Straub}, {Straubmeier}, {Sturm}, {Stolker}, {van Dishoeck}, {Vigan},
  {Vincent}, {von Fellenberg}, {Ward-Duong}, {Widmann}, {Wieprecht}, {Wiezorrek}, \& {Woillez}}]{Lacour2020}
{Lacour}, S., {Wang}, J.~J., {Nowak}, M., {et~al.} 2020, in Society of Photo-Optical Instrumentation Engineers (SPIE) Conference Series, Vol. 11446, Society of Photo-Optical Instrumentation Engineers (SPIE) Conference Series, 114460O, \dodoi{10.1117/12.2561667}

\bibitem[{{Lacour} {et~al.}(2021){Lacour}, {Wang}, {Rodet}, {Nowak}, {Shangguan}, {Beust}, {Lagrange}, {Abuter}, {Amorim}, {Asensio-Torres}, {Benisty}, {Berger}, {Blunt}, {Boccaletti}, {Bohn}, {Bolzer}, {Bonnefoy}, {Bonnet}, {Bourdarot}, {Brandner}, {Cantalloube}, {Caselli}, {Charnay}, {Chauvin}, {Choquet}, {Christiaens}, {Cl{\'e}net}, {Coud{\'e} Du Foresto}, {Cridland}, {Dembet}, {Dexter}, {de Zeeuw}, {Drescher}, {Duvert}, {Eckart}, {Eisenhauer}, {Gao}, {Garcia}, {Garcia Lopez}, {Gendron}, {Genzel}, {Gillessen}, {Girard}, {Haubois}, {Hei{\ss}el}, {Henning}, {Hinkley}, {Hippler}, {Horrobin}, {Houll{\'e}}, {Hubert}, {Jocou}, {Kammerer}, {Keppler}, {Kervella}, {Kreidberg}, {Lapeyr{\`e}re}, {Le Bouquin}, {L{\'e}na}, {Lutz}, {Maire}, {M{\'e}rand}, {Molli{\`e}re}, {Monnier}, {Mouillet}, {Nasedkin}, {Ott}, {Otten}, {Paladini}, {Paumard}, {Perraut}, {Perrin}, {Pfuhl}, {Rickman}, {Pueyo}, {Rameau}, {Rousset}, {Rustamkulov}, {Samland}, {Shimizu}, {Sing}, {Stadler}, {Stolker}, {Straub}, {Straubmeier}, {Sturm},
  {Tacconi}, {van Dishoeck}, {Vigan}, {Vincent}, {von Fellenberg}, {Ward-Duong}, {Widmann}, {Wieprecht}, {Wiezorrek}, {Woillez}, {Yazici}, {Young}, \& {Gravity Collaboration}}]{Lacour2021}
{Lacour}, S., {Wang}, J.~J., {Rodet}, L., {et~al.} 2021, \aap, 654, L2, \dodoi{10.1051/0004-6361/202141889}

\bibitem[{{Lapeyrere} {et~al.}(2014){Lapeyrere}, {Kervella}, {Lacour}, {Azouaoui}, {Garcia-Dabo}, {Perrin}, {Eisenhauer}, {Perraut}, {Straubmeier}, {Amorim}, \& {Brandner}}]{Lapeyrere2014}
{Lapeyrere}, V., {Kervella}, P., {Lacour}, S., {et~al.} 2014, in Society of Photo-Optical Instrumentation Engineers (SPIE) Conference Series, Vol. 9146, Optical and Infrared Interferometry IV, ed. J.~K. {Rajagopal}, M.~J. {Creech-Eakman}, \& F.~{Malbet}, 91462D, \dodoi{10.1117/12.2056850}

\bibitem[{{Li} {et~al.}(2016){Li}, {Kouwenhoven}, {Stamatellos}, \& {Goodwin}}]{Li2016}
{Li}, Y., {Kouwenhoven}, M.~B.~N., {Stamatellos}, D., \& {Goodwin}, S.~P. 2016, \apj, 831, 166, \dodoi{10.3847/0004-637X/831/2/166}

\bibitem[{{Li} {et~al.}(2023){Li}, {Brandt}, {Brandt}, {An}, {Franson}, {Dupuy}, {Chen}, {Bowens-Rubin}, {Lewis}, {Bowler}, {Gibbs}, {Kiman}, {Faherty}, {Currie}, {Jensen-Clem}, {Contreras-Martinez}, {Fitzgerald}, {Mazin}, \& {Millar-Blanchaer}}]{Li2023}
{Li}, Y., {Brandt}, T.~D., {Brandt}, G.~M., {et~al.} 2023, arXiv e-prints, arXiv:2301.10420, \dodoi{10.48550/arXiv.2301.10420}

\bibitem[{{Line} {et~al.}(2015){Line}, {Teske}, {Burningham}, {Fortney}, \& {Marley}}]{Line2015}
{Line}, M.~R., {Teske}, J., {Burningham}, B., {Fortney}, J.~J., \& {Marley}, M.~S. 2015, \apj, 807, 183, \dodoi{10.1088/0004-637X/807/2/183}

\bibitem[{{Luck}(2017)}]{Luck2017}
{Luck}, R.~E. 2017, \aj, 153, 21, \dodoi{10.3847/1538-3881/153/1/21}

\bibitem[{{Luck} \& {Heiter}(2006)}]{Luck2006}
{Luck}, R.~E., \& {Heiter}, U. 2006, \aj, 131, 3069, \dodoi{10.1086/504080}

\bibitem[{{Lueber} {et~al.}(2022){Lueber}, {Kitzmann}, {Bowler}, {Burgasser}, \& {Heng}}]{Lueber2022}
{Lueber}, A., {Kitzmann}, D., {Bowler}, B.~P., {Burgasser}, A.~J., \& {Heng}, K. 2022, \apj, 930, 136, \dodoi{10.3847/1538-4357/ac63b9}

\bibitem[{{Ma} \& {Ge}(2014)}]{Ma2014}
{Ma}, B., \& {Ge}, J. 2014, \mnras, 439, 2781, \dodoi{10.1093/mnras/stu134}

\bibitem[{{Magg} {et~al.}(2022){Magg}, {Bergemann}, {Serenelli}, {Bautista}, {Plez}, {Heiter}, {Gerber}, {Ludwig}, {Basu}, {Ferguson}, {Gallego}, {Gamrath}, {Palmeri}, \& {Quinet}}]{magg2022}
{Magg}, E., {Bergemann}, M., {Serenelli}, A., {et~al.} 2022, \aap, 661, A140, \dodoi{10.1051/0004-6361/202142971}

\bibitem[{{Maire} {et~al.}(2020){Maire}, {Baudino}, {Desidera}, {Messina}, {Brandner}, {Godoy}, {Cantalloube}, {Galicher}, {Bonnefoy}, {Hagelberg}, {Olofsson}, {Absil}, {Chauvin}, {Henning}, \& {Langlois}}]{Maire2020}
{Maire}, A.~L., {Baudino}, J.~L., {Desidera}, S., {et~al.} 2020, \aap, 633, L2, \dodoi{10.1051/0004-6361/201937134}

\bibitem[{{Manjavacas} {et~al.}(2014){Manjavacas}, {Bonnefoy}, {Schlieder}, {Allard}, {Rojo}, {Goldman}, {Chauvin}, {Homeier}, {Lodieu}, \& {Henning}}]{Manjavacas2014}
{Manjavacas}, E., {Bonnefoy}, M., {Schlieder}, J.~E., {et~al.} 2014, \aap, 564, A55, \dodoi{10.1051/0004-6361/201323016}

\bibitem[{{Marley} {et~al.}(2021){Marley}, {Saumon}, {Visscher}, {Lupu}, {Freedman}, {Morley}, {Fortney}, {Seay}, {Smith}, {Teal}, \& {Wang}}]{Marley2021}
{Marley}, M.~S., {Saumon}, D., {Visscher}, C., {et~al.} 2021, \apj, 920, 85, \dodoi{10.3847/1538-4357/ac141d}

\bibitem[{{Mel{\'e}ndez} {et~al.}(2014){Mel{\'e}ndez}, {Schirbel}, {Monroe}, {Yong}, {Ram{\'\i}rez}, \& {Asplund}}]{melendez2014}
{Mel{\'e}ndez}, J., {Schirbel}, L., {Monroe}, T.~R., {et~al.} 2014, \aap, 567, L3, \dodoi{10.1051/0004-6361/201424172}

\bibitem[{{Molli{\`e}re} {et~al.}(2019){Molli{\`e}re}, {Wardenier}, {van Boekel}, {Henning}, {Molaverdikhani}, \& {Snellen}}]{Molliere2019}
{Molli{\`e}re}, P., {Wardenier}, J.~P., {van Boekel}, R., {et~al.} 2019, \aap, 627, A67, \dodoi{10.1051/0004-6361/201935470}

\bibitem[{{Molli{\`e}re} {et~al.}(2020){Molli{\`e}re}, {Stolker}, {Lacour}, {Otten}, {Shangguan}, {Charnay}, {Molyarova}, {Nowak}, {Henning}, {Marleau}, {Semenov}, {van Dishoeck}, {Eisenhauer}, {Garcia}, {Garcia Lopez}, {Girard}, {Greenbaum}, {Hinkley}, {Kervella}, {Kreidberg}, {Maire}, {Nasedkin}, {Pueyo}, {Snellen}, {Vigan}, {Wang}, {de Zeeuw}, \& {Zurlo}}]{Molliere2020}
{Molli{\`e}re}, P., {Stolker}, T., {Lacour}, S., {et~al.} 2020, \aap, 640, A131, \dodoi{10.1051/0004-6361/202038325}

\bibitem[{{Molli{\`e}re} {et~al.}(2022){Molli{\`e}re}, {Molyarova}, {Bitsch}, {Henning}, {Schneider}, {Kreidberg}, {Eistrup}, {Burn}, {Nasedkin}, {Semenov}, {Mordasini}, {Schlecker}, {Schwarz}, {Lacour}, {Nowak}, \& {Schulik}}]{Molliere2022}
{Molli{\`e}re}, P., {Molyarova}, T., {Bitsch}, B., {et~al.} 2022, \apj, 934, 74, \dodoi{10.3847/1538-4357/ac6a56}

\bibitem[{{Morton}(2015)}]{morton2015}
{Morton}, T.~D. 2015, {isochrones: Stellar model grid package}.
\newblock \doeprint{1503.010}

\bibitem[{{Moultaka} {et~al.}(2004){Moultaka}, {Ilovaisky}, {Prugniel}, \& {Soubiran}}]{Moultaka2004}
{Moultaka}, J., {Ilovaisky}, S.~A., {Prugniel}, P., \& {Soubiran}, C. 2004, \pasp, 116, 693, \dodoi{10.1086/422177}

\bibitem[{{Mukherjee} {et~al.}(2022){Mukherjee}, {Fortney}, {Batalha}, {Karalidi}, {Marley}, {Visscher}, {Miles}, \& {Skemer}}]{Mukherjee2022}
{Mukherjee}, S., {Fortney}, J.~J., {Batalha}, N.~E., {et~al.} 2022, \apj, 938, 107, \dodoi{10.3847/1538-4357/ac8dfb}

\bibitem[{{Nelson} {et~al.}(2020){Nelson}, {Ford}, {Buchner}, {Cloutier}, {D{\'\i}az}, {Faria}, {Hara}, {Rajpaul}, \& {Rukdee}}]{Nelson2020}
{Nelson}, B.~E., {Ford}, E.~B., {Buchner}, J., {et~al.} 2020, \aj, 159, 73, \dodoi{10.3847/1538-3881/ab5190}

\bibitem[{{Nowak} {et~al.}(2020){Nowak}, {Lacour}, {Lagrange}, {Rubini}, {Wang}, {Stolker}, {Abuter}, {Amorim}, {Asensio-Torres}, {Baub{\"o}ck}, {Benisty}, {Berger}, {Beust}, {Blunt}, {Boccaletti}, {Bonnefoy}, {Bonnet}, {Brandner}, {Cantalloube}, {Charnay}, {Choquet}, {Christiaens}, {Cl{\'e}net}, {Coud{\'e} Du Foresto}, {Cridland}, {de Zeeuw}, {Dembet}, {Dexter}, {Drescher}, {Duvert}, {Eckart}, {Eisenhauer}, {Gao}, {Garcia}, {Garcia Lopez}, {Gardner}, {Gendron}, {Genzel}, {Gillessen}, {Girard}, {Grandjean}, {Haubois}, {Hei{\ss}el}, {Henning}, {Hinkley}, {Hippler}, {Horrobin}, {Houll{\'e}}, {Hubert}, {Jim{\'e}nez-Rosales}, {Jocou}, {Kammerer}, {Kervella}, {Keppler}, {Kreidberg}, {Kulikauskas}, {Lapeyr{\`e}re}, {Le Bouquin}, {L{\'e}na}, {M{\'e}rand}, {Maire}, {Molli{\`e}re}, {Monnier}, {Mouillet}, {M{\"u}ller}, {Nasedkin}, {Ott}, {Otten}, {Paumard}, {Paladini}, {Perraut}, {Perrin}, {Pueyo}, {Pfuhl}, {Rameau}, {Rodet}, {Rodr{\'\i}guez-Coira}, {Rousset}, {Scheithauer}, {Shangguan}, {Stadler}, {Straub},
  {Straubmeier}, {Sturm}, {Tacconi}, {van Dishoeck}, {Vigan}, {Vincent}, {von Fellenberg}, {Ward-Duong}, {Widmann}, {Wieprecht}, {Wiezorrek}, {Woillez}, \& {Gravity Collaboration}}]{Nowak2020}
{Nowak}, M., {Lacour}, S., {Lagrange}, A.~M., {et~al.} 2020, \aap, 642, L2, \dodoi{10.1051/0004-6361/202039039}

\bibitem[{{{\"O}berg} {et~al.}(2011){{\"O}berg}, {Murray-Clay}, \& {Bergin}}]{Oberg2011}
{{\"O}berg}, K.~I., {Murray-Clay}, R., \& {Bergin}, E.~A. 2011, \apjl, 743, L16, \dodoi{10.1088/2041-8205/743/1/L16}

\bibitem[{{Padoan} \& {Nordlund}(2004)}]{Padoan2004}
{Padoan}, P., \& {Nordlund}, {\r{A}}. 2004, \apj, 617, 559, \dodoi{10.1086/345413}

\bibitem[{{Patience} {et~al.}(2012){Patience}, {King}, {De Rosa}, {Vigan}, {Witte}, {Rice}, {Helling}, \& {Hauschildt}}]{Patience2012}
{Patience}, J., {King}, R.~R., {De Rosa}, R.~J., {et~al.} 2012, \aap, 540, A85, \dodoi{10.1051/0004-6361/201118058}

\bibitem[{{Paxton} {et~al.}(2011){Paxton}, {Bildsten}, {Dotter}, {Herwig}, {Lesaffre}, \& {Timmes}}]{paxton2011}
{Paxton}, B., {Bildsten}, L., {Dotter}, A., {et~al.} 2011, \apjs, 192, 3, \dodoi{10.1088/0067-0049/192/1/3}

\bibitem[{{Paxton} {et~al.}(2013){Paxton}, {Cantiello}, {Arras}, {Bildsten}, {Brown}, {Dotter}, {Mankovich}, {Montgomery}, {Stello}, {Timmes}, \& {Townsend}}]{paxton2013}
{Paxton}, B., {Cantiello}, M., {Arras}, P., {et~al.} 2013, \apjs, 208, 4, \dodoi{10.1088/0067-0049/208/1/4}

\bibitem[{{Paxton} {et~al.}(2015){Paxton}, {Marchant}, {Schwab}, {Bauer}, {Bildsten}, {Cantiello}, {Dessart}, {Farmer}, {Hu}, {Langer}, {Townsend}, {Townsley}, \& {Timmes}}]{paxton2015}
{Paxton}, B., {Marchant}, P., {Schwab}, J., {et~al.} 2015, \apjs, 220, 15, \dodoi{10.1088/0067-0049/220/1/15}

\bibitem[{{Paxton} {et~al.}(2018){Paxton}, {Schwab}, {Bauer}, {Bildsten}, {Blinnikov}, {Duffell}, {Farmer}, {Goldberg}, {Marchant}, {Sorokina}, {Thoul}, {Townsend}, \& {Timmes}}]{paxton2018}
{Paxton}, B., {Schwab}, J., {Bauer}, E.~B., {et~al.} 2018, \apjs, 234, 34, \dodoi{10.3847/1538-4365/aaa5a8}

\bibitem[{{Paxton} {et~al.}(2019){Paxton}, {Smolec}, {Schwab}, {Gautschy}, {Bildsten}, {Cantiello}, {Dotter}, {Farmer}, {Goldberg}, {Jermyn}, {Kanbur}, {Marchant}, {Thoul}, {Townsend}, {Wolf}, {Zhang}, \& {Timmes}}]{paxton2019}
{Paxton}, B., {Smolec}, R., {Schwab}, J., {et~al.} 2019, \apjs, 243, 10, \dodoi{10.3847/1538-4365/ab2241}

\bibitem[{{Peretti} {et~al.}(2019){Peretti}, {S{\'e}gransan}, {Lavie}, {Desidera}, {Maire}, {D'Orazi}, {Vigan}, {Baudino}, {Cheetham}, {Janson}, {Chauvin}, {Hagelberg}, {Menard}, {Heng}, {Udry}, {Boccaletti}, {Daemgen}, {Le Coroller}, {Mesa}, {Rouan}, {Samland}, {Schmidt}, {Zurlo}, {Bonnefoy}, {Feldt}, {Gratton}, {Lagrange}, {Langlois}, {Meyer}, {Carbillet}, {Carle}, {De Caprio}, {Gluck}, {Hugot}, {Magnard}, {Moulin}, {Pavlov}, {Pragt}, {Rabou}, {Ramos}, {Rousset}, {Sevin}, {Soenke}, {Stadler}, {Weber}, \& {Wildi}}]{Peretti2019}
{Peretti}, S., {S{\'e}gransan}, D., {Lavie}, B., {et~al.} 2019, \aap, 631, A107, \dodoi{10.1051/0004-6361/201732454}

\bibitem[{{Phillips} {et~al.}(2020){Phillips}, {Tremblin}, {Baraffe}, {Chabrier}, {Allard}, {Spiegelman}, {Goyal}, {Drummond}, \& {H{\'e}brard}}]{Phillips2020}
{Phillips}, M.~W., {Tremblin}, P., {Baraffe}, I., {et~al.} 2020, \aap, 637, A38, \dodoi{10.1051/0004-6361/201937381}

\bibitem[{{Placco} {et~al.}(2021){Placco}, {Sneden}, {Roederer}, {Lawler}, {Den Hartog}, {Hejazi}, {Maas}, \& {Bernath}}]{placco2021}
{Placco}, V.~M., {Sneden}, C., {Roederer}, I.~U., {et~al.} 2021, Research Notes of the American Astronomical Society, 5, 92, \dodoi{10.3847/2515-5172/abf651}

\bibitem[{{Reggiani} {et~al.}(2022){Reggiani}, {Schlaufman}, {Healy}, {Lothringer}, \& {Sing}}]{Reggiani2022}
{Reggiani}, H., {Schlaufman}, K.~C., {Healy}, B.~F., {Lothringer}, J.~D., \& {Sing}, D.~K. 2022, \aj, 163, 159, \dodoi{10.3847/1538-3881/ac4d9f}

\bibitem[{{Richard} {et~al.}(2012){Richard}, {Gordon}, {Rothman}, {Abel}, {Frommhold}, {Gustafsson}, {Hartmann}, {Hermans}, {Lafferty}, {Orton}, {Smith}, \& {Tran}}]{Richard2012}
{Richard}, C., {Gordon}, I.~E., {Rothman}, L.~S., {et~al.} 2012, \jqsrt, 113, 1276, \dodoi{10.1016/j.jqsrt.2011.11.004}

\bibitem[{{Rickman} {et~al.}(2020){Rickman}, {S{\'e}gransan}, {Hagelberg}, {Beuzit}, {Cheetham}, {Delisle}, {Forveille}, \& {Udry}}]{Rickman2020}
{Rickman}, E.~L., {S{\'e}gransan}, D., {Hagelberg}, J., {et~al.} 2020, \aap, 635, A203, \dodoi{10.1051/0004-6361/202037524}

\bibitem[{{Rickman} {et~al.}(2022){Rickman}, {Matthews}, {Ceva}, {S{\'e}gransan}, {Brandt}, {Zhang}, {Brandt}, {Forveille}, {Hagelberg}, \& {Udry}}]{Rickman2022}
{Rickman}, E.~L., {Matthews}, E., {Ceva}, W., {et~al.} 2022, \aap, 668, A140, \dodoi{10.1051/0004-6361/202244633}

\bibitem[{{Rowland} {et~al.}(2023){Rowland}, {Morley}, \& {Line}}]{Rowland2023}
{Rowland}, M.~J., {Morley}, C.~V., \& {Line}, M.~R. 2023, \apj, 947, 6, \dodoi{10.3847/1538-4357/acbb07}

\bibitem[{{Saumon} \& {Marley}(2008)}]{Saumon2008}
{Saumon}, D., \& {Marley}, M.~S. 2008, \apj, 689, 1327, \dodoi{10.1086/592734}

\bibitem[{{Scott} \& {Duley}(1996)}]{Scott1996}
{Scott}, A., \& {Duley}, W.~W. 1996, \apjs, 105, 401, \dodoi{10.1086/192321}

\bibitem[{{Skrutskie} {et~al.}(2006){Skrutskie}, {Cutri}, {Stiening}, {Weinberg}, {Schneider}, {Carpenter}, {Beichman}, {Capps}, {Chester}, {Elias}, {Huchra}, {Liebert}, {Lonsdale}, {Monet}, {Price}, {Seitzer}, {Jarrett}, {Kirkpatrick}, {Gizis}, {Howard}, {Evans}, {Fowler}, {Fullmer}, {Hurt}, {Light}, {Kopan}, {Marsh}, {McCallon}, {Tam}, {Van Dyk}, \& {Wheelock}}]{skrutskie2006}
{Skrutskie}, M.~F., {Cutri}, R.~M., {Stiening}, R., {et~al.} 2006, \aj, 131, 1163, \dodoi{10.1086/498708}

\bibitem[{{Sneden} {et~al.}(2012){Sneden}, {Bean}, {Ivans}, {Lucatello}, \& {Sobeck}}]{Sneden2012}
{Sneden}, C., {Bean}, J., {Ivans}, I., {Lucatello}, S., \& {Sobeck}, J. 2012, {MOOG: LTE line analysis and spectrum synthesis}, Astrophysics Source Code Library, record ascl:1202.009.
\newblock \doeprint{1202.009}

\bibitem[{{Sneden} {et~al.}(2009){Sneden}, {Lawler}, {Cowan}, {Ivans}, \& {Den Hartog}}]{sneden2009}
{Sneden}, C., {Lawler}, J.~E., {Cowan}, J.~J., {Ivans}, I.~I., \& {Den Hartog}, E.~A. 2009, \apjs, 182, 80, \dodoi{10.1088/0067-0049/182/1/80}

\bibitem[{{Sneden} {et~al.}(2016){Sneden}, {Lawler}, {den Hartog}, \& {Wood}}]{sneden2016}
{Sneden}, C., {Lawler}, J.~E., {den Hartog}, E.~A., \& {Wood}, M.~E. 2016, IAU Focus Meeting, 29A, 287, \dodoi{10.1017/S1743921316003069}

\bibitem[{{Sneden}(1973)}]{Sneden1973}
{Sneden}, C.~A. 1973, PhD thesis, University of Texas, Austin

\bibitem[{{Spina} {et~al.}(2016){Spina}, {Mel{\'e}ndez}, {Karakas}, {Ram{\'\i}rez}, {Monroe}, {Asplund}, \& {Yong}}]{Spina2016}
{Spina}, L., {Mel{\'e}ndez}, J., {Karakas}, A.~I., {et~al.} 2016, \aap, 593, A125, \dodoi{10.1051/0004-6361/201628557}

\bibitem[{{Squicciarini} {et~al.}(2022){Squicciarini}, {Gratton}, {Janson}, {Mamajek}, {Chauvin}, {Delorme}, {Langlois}, {Vigan}, {Ringqvist}, {Meeus}, {Reffert}, {Kenworthy}, {Meyer}, {Bonnefoy}, {Bonavita}, {Mesa}, {Samland}, {Desidera}, {D'Orazi}, {Engler}, {Alecian}, {Miglio}, {Henning}, {Quanz}, {Mayer}, {Flasseur}, \& {Marleau}}]{Squicciarini2022}
{Squicciarini}, V., {Gratton}, R., {Janson}, M., {et~al.} 2022, \aap, 664, A9, \dodoi{10.1051/0004-6361/202243675}

\bibitem[{{Stamatellos} {et~al.}(2007){Stamatellos}, {Hubber}, \& {Whitworth}}]{Stamatellos2007}
{Stamatellos}, D., {Hubber}, D.~A., \& {Whitworth}, A.~P. 2007, \mnras, 382, L30, \dodoi{10.1111/j.1745-3933.2007.00383.x}

\bibitem[{{Stamatellos} \& {Whitworth}(2009)}]{Stamatellos2009}
{Stamatellos}, D., \& {Whitworth}, A.~P. 2009, \mnras, 392, 413, \dodoi{10.1111/j.1365-2966.2008.14069.x}

\bibitem[{{Stephens} {et~al.}(2009){Stephens}, {Leggett}, {Cushing}, {Marley}, {Saumon}, {Geballe}, {Golimowski}, {Fan}, \& {Noll}}]{Stephens2009}
{Stephens}, D.~C., {Leggett}, S.~K., {Cushing}, M.~C., {et~al.} 2009, \apj, 702, 154, \dodoi{10.1088/0004-637X/702/1/154}

\bibitem[{{Stolker} {et~al.}(2020){Stolker}, {Quanz}, {Todorov}, {K{\"u}hn}, {Molli{\`e}re}, {Meyer}, {Currie}, {Daemgen}, \& {Lavie}}]{Stolker2020}
{Stolker}, T., {Quanz}, S.~P., {Todorov}, K.~O., {et~al.} 2020, \aap, 635, A182, \dodoi{10.1051/0004-6361/201937159}

\bibitem[{{STScI Development Team}(2018)}]{STScIDevelopmentTeam2018}
{STScI Development Team}. 2018, {synphot: Synthetic photometry using Astropy}, Astrophysics Source Code Library, record ascl:1811.001.
\newblock \doeprint{1811.001}

\bibitem[{{Su{\'a}rez} \& {Metchev}(2022)}]{Suarez2022}
{Su{\'a}rez}, G., \& {Metchev}, S. 2022, arXiv e-prints, arXiv:2205.00168.
\newblock \doarXiv{2205.00168}

\bibitem[{{Su{\'a}rez} \& {Metchev}(2023)}]{Suarez2023}
---. 2023, arXiv e-prints, arXiv:2306.01119, \dodoi{10.48550/arXiv.2306.01119}

\bibitem[{{Teske} {et~al.}(2014){Teske}, {Cunha}, {Smith}, {Schuler}, \& {Griffith}}]{teske2014}
{Teske}, J.~K., {Cunha}, K., {Smith}, V.~V., {Schuler}, S.~C., \& {Griffith}, C.~A. 2014, \apj, 788, 39, \dodoi{10.1088/0004-637X/788/1/39}

\bibitem[{{Thorngren} \& {Fortney}(2019)}]{Thorngren2019}
{Thorngren}, D., \& {Fortney}, J.~J. 2019, \apjl, 874, L31, \dodoi{10.3847/2041-8213/ab1137}

\bibitem[{{Tody}(1986)}]{Tody1986}
{Tody}, D. 1986, in Society of Photo-Optical Instrumentation Engineers (SPIE) Conference Series, Vol. 627, Instrumentation in astronomy VI, ed. D.~L. {Crawford}, 733, \dodoi{10.1117/12.968154}

\bibitem[{{Tody}(1993)}]{Tody1993}
{Tody}, D. 1993, in Astronomical Society of the Pacific Conference Series, Vol.~52, Astronomical Data Analysis Software and Systems II, ed. R.~J. {Hanisch}, R.~J.~V. {Brissenden}, \& J.~{Barnes}, 173

\bibitem[{{Tremblin} {et~al.}(2016){Tremblin}, {Amundsen}, {Chabrier}, {Baraffe}, {Drummond}, {Hinkley}, {Mourier}, \& {Venot}}]{Tremblin2016}
{Tremblin}, P., {Amundsen}, D.~S., {Chabrier}, G., {et~al.} 2016, \apjl, 817, L19, \dodoi{10.3847/2041-8205/817/2/L19}

\bibitem[{{Tremblin} {et~al.}(2015){Tremblin}, {Amundsen}, {Mourier}, {Baraffe}, {Chabrier}, {Drummond}, {Homeier}, \& {Venot}}]{Tremblin2015}
{Tremblin}, P., {Amundsen}, D.~S., {Mourier}, P., {et~al.} 2015, \apjl, 804, L17, \dodoi{10.1088/2041-8205/804/1/L17}

\bibitem[{{Tremblin} {et~al.}(2019){Tremblin}, {Padioleau}, {Phillips}, {Chabrier}, {Baraffe}, {Fromang}, {Audit}, {Bloch}, {Burgasser}, {Drummond}, {Gonz{\'a}lez}, {Kestener}, {Kokh}, {Lagage}, \& {Stauffert}}]{Tremblin2019}
{Tremblin}, P., {Padioleau}, T., {Phillips}, M.~W., {et~al.} 2019, \apj, 876, 144, \dodoi{10.3847/1538-4357/ab05db}

\bibitem[{{Umbreit} {et~al.}(2005){Umbreit}, {Burkert}, {Henning}, {Mikkola}, \& {Spurzem}}]{Umbreit2005}
{Umbreit}, S., {Burkert}, A., {Henning}, T., {Mikkola}, S., \& {Spurzem}, R. 2005, \apj, 623, 940, \dodoi{10.1086/428602}

\bibitem[{{Vacca} {et~al.}(2003){Vacca}, {Cushing}, \& {Rayner}}]{Vacca2003}
{Vacca}, W.~D., {Cushing}, M.~C., \& {Rayner}, J.~T. 2003, \pasp, 115, 389, \dodoi{10.1086/346193}

\bibitem[{{Vos} {et~al.}(2022){Vos}, {Faherty}, {Gagn{\'e}}, {Marley}, {Metchev}, {Gizis}, {Rice}, \& {Cruz}}]{Vos2022}
{Vos}, J.~M., {Faherty}, J.~K., {Gagn{\'e}}, J., {et~al.} 2022, \apj, 924, 68, \dodoi{10.3847/1538-4357/ac4502}

\bibitem[{{Vousden} {et~al.}(2016){Vousden}, {Farr}, \& {Mandel}}]{Vousden2016}
{Vousden}, W.~D., {Farr}, W.~M., \& {Mandel}, I. 2016, \mnras, 455, 1919, \dodoi{10.1093/mnras/stv2422}

\bibitem[{{Wang} {et~al.}(2022){Wang}, {Kolecki}, {Ruffio}, {Wang}, {Mawet}, {Baker}, {Bartos}, {Blake}, {Bond}, {Calvin}, {Cetre}, {Delorme}, {Doppmann}, {Echeverri}, {Finnerty}, {Fitzgerald}, {Jovanovic}, {Liu}, {Lopez}, {Morris}, {Pai Asnodkar}, {Pezzato}, {Ragland}, {Roy}, {Ruane}, {Sappey}, {Schofield}, {Skemer}, {Venenciano}, {Wallace}, {Wallack}, {Wizinowich}, \& {Xuan}}]{Wang2022}
{Wang}, J., {Kolecki}, J.~R., {Ruffio}, J.-B., {et~al.} 2022, arXiv e-prints, arXiv:2202.02477.
\newblock \doarXiv{2202.02477}

\bibitem[{{Wang} {et~al.}(2020){Wang}, {Ginzburg}, {Ren}, {Wallack}, {Gao}, {Mawet}, {Bond}, {Cetre}, {Wizinowich}, {De Rosa}, {Ruane}, {Liu}, {Absil}, {Alvarez}, {Baranec}, {Choquet}, {Chun}, {Defr{\`e}re}, {Delorme}, {Duch{\^e}ne}, {Forsberg}, {Ghez}, {Guyon}, {Hall}, {Huby}, {Jolivet}, {Jensen-Clem}, {Jovanovic}, {Karlsson}, {Lilley}, {Matthews}, {M{\'e}nard}, {Meshkat}, {Millar-Blanchaer}, {Ngo}, {Orban de Xivry}, {Pinte}, {Ragland}, {Serabyn}, {Catal{\'a}n}, {Wang}, {Wetherell}, {Williams}, {Ygouf}, \& {Zuckerman}}]{Wang2020}
{Wang}, J.~J., {Ginzburg}, S., {Ren}, B., {et~al.} 2020, \aj, 159, 263, \dodoi{10.3847/1538-3881/ab8aef}

\bibitem[{{Wang} {et~al.}(2021){Wang}, {Vigan}, {Lacour}, {Nowak}, {Stolker}, {De Rosa}, {Ginzburg}, {Gao}, {Abuter}, {Amorim}, {Asensio-Torres}, {Baub{\"o}ck}, {Benisty}, {Berger}, {Beust}, {Beuzit}, {Blunt}, {Boccaletti}, {Bohn}, {Bonnefoy}, {Bonnet}, {Brandner}, {Cantalloube}, {Caselli}, {Charnay}, {Chauvin}, {Choquet}, {Christiaens}, {Cl{\'e}net}, {Coud{\'e} Du Foresto}, {Cridland}, {de Zeeuw}, {Dembet}, {Dexter}, {Drescher}, {Duvert}, {Eckart}, {Eisenhauer}, {Facchini}, {Gao}, {Garcia}, {Garcia Lopez}, {Gardner}, {Gendron}, {Genzel}, {Gillessen}, {Girard}, {Haubois}, {Hei{\ss}el}, {Henning}, {Hinkley}, {Hippler}, {Horrobin}, {Houll{\'e}}, {Hubert}, {Jim{\'e}nez-Rosales}, {Jocou}, {Kammerer}, {Keppler}, {Kervella}, {Meyer}, {Kreidberg}, {Lagrange}, {Lapeyr{\`e}re}, {Le Bouquin}, {L{\'e}na}, {Lutz}, {Maire}, {M{\'e}nard}, {M{\'e}rand}, {Molli{\`e}re}, {Monnier}, {Mouillet}, {M{\"u}ller}, {Nasedkin}, {Ott}, {Otten}, {Paladini}, {Paumard}, {Perraut}, {Perrin}, {Pfuhl}, {Pueyo}, {Rameau}, {Rodet},
  {Rodr{\'\i}guez-Coira}, {Rousset}, {Scheithauer}, {Shangguan}, {Shimizu}, {Stadler}, {Straub}, {Straubmeier}, {Sturm}, {Tacconi}, {van Dishoeck}, {Vincent}, {von Fellenberg}, {Ward-Duong}, {Widmann}, {Wieprecht}, {Wiezorrek}, {Woillez}, \& {Gravity Collaboration}}]{Wang2021}
{Wang}, J.~J., {Vigan}, A., {Lacour}, S., {et~al.} 2021, \aj, 161, 148, \dodoi{10.3847/1538-3881/abdb2d}

\bibitem[{{Xuan} {et~al.}(2022){Xuan}, {Wang}, {Ruffio}, {Knutson}, {Mawet}, {Molli{\`e}re}, {Kolecki}, {Vigan}, {Mukherjee}, {Wallack}, {Wang}, {Baker}, {Bartos}, {Blake}, {Bond}, {Bryan}, {Calvin}, {Cetre}, {Chun}, {Delorme}, {Doppmann}, {Echeverri}, {Finnerty}, {Fitzgerald}, {Horstman}, {Inglis}, {Jovanovic}, {L{\'o}pez}, {Martin}, {Morris}, {Pezzato}, {Ragland}, {Ren}, {Ruane}, {Sappey}, {Schofield}, {Skemer}, {Venenciano}, {Wallace}, \& {Wizinowich}}]{Xuan2022}
{Xuan}, J.~W., {Wang}, J., {Ruffio}, J.-B., {et~al.} 2022, \apj, 937, 54, \dodoi{10.3847/1538-4357/ac8673}

\bibitem[{{Yana Galarza} {et~al.}(2019){Yana Galarza}, {Mel{\'e}ndez}, {Lorenzo-Oliveira}, {Valio}, {Reggiani}, {Carlos}, {Ponte}, {Spina}, {Haywood}, \& {Gandolfi}}]{galarza2019}
{Yana Galarza}, J., {Mel{\'e}ndez}, J., {Lorenzo-Oliveira}, D., {et~al.} 2019, \mnras, 490, L86, \dodoi{10.1093/mnrasl/slz153}

\bibitem[{{Zhang} {et~al.}(2019){Zhang}, {Burgasser}, {G{\'a}lvez-Ortiz}, {Lodieu}, {Zapatero Osorio}, {Pinfield}, \& {Allard}}]{Zhang2019}
{Zhang}, Z.~H., {Burgasser}, A.~J., {G{\'a}lvez-Ortiz}, M.~C., {et~al.} 2019, \mnras, 486, 1260, \dodoi{10.1093/mnras/stz777}

\end{thebibliography}
\bibliographystyle{aasjournal}

\clearpage
\appendix

\section{Posterior Distributions} \label{sec:posteriors}
\par
\begin{figure*}
    \centering
    \includegraphics[width=\textwidth]{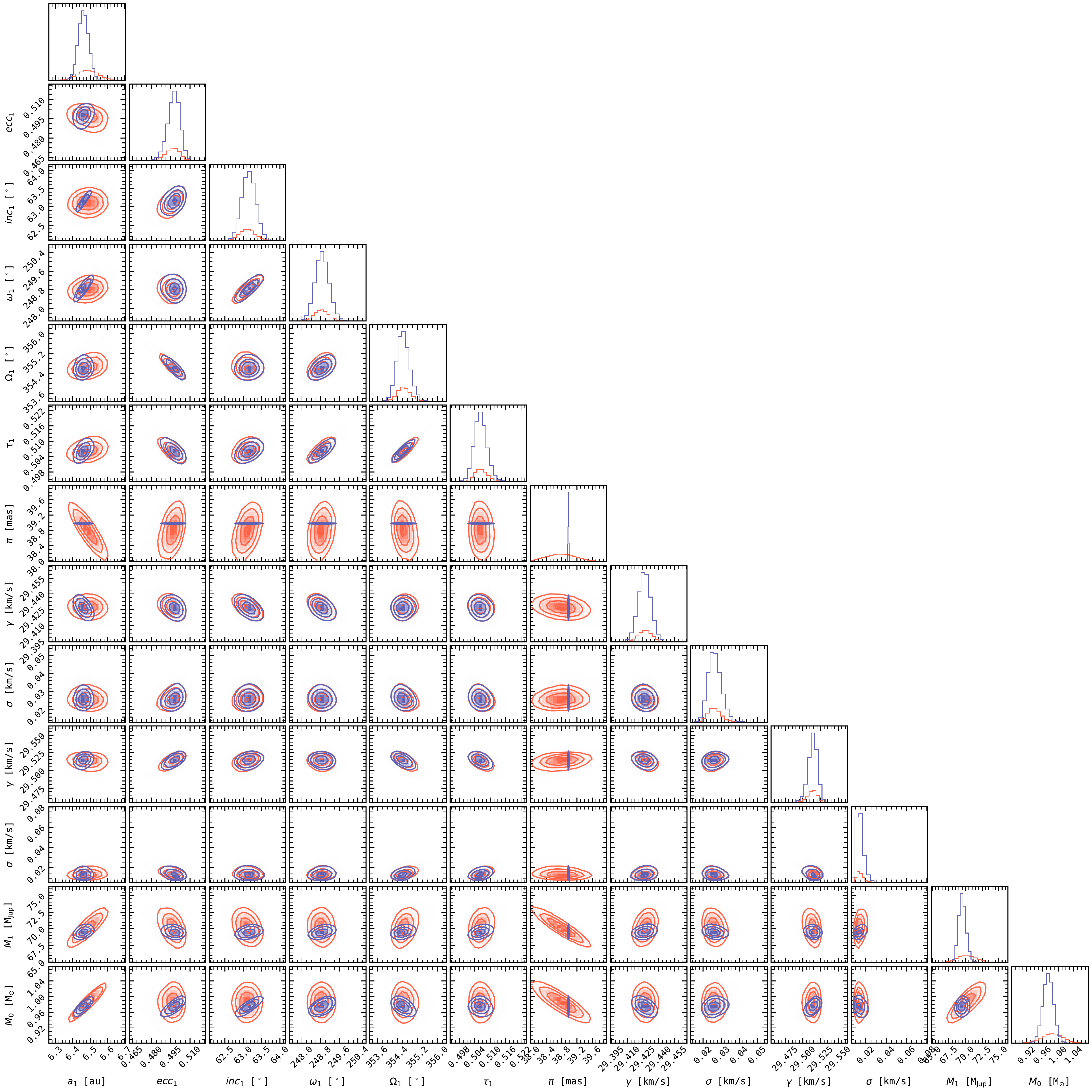}
    \caption{The posterior distributions of orbit fits to the HD~72946 system. The red contours and histograms plot the posterior for the no absolute astrometry fit, while the blue contours and histograms plot the posterior for the fit including absolute astrometry.}
    \label{fig:orbpost}
\end{figure*}

\begin{figure*}
    \centering
    \includegraphics[width=\textwidth]{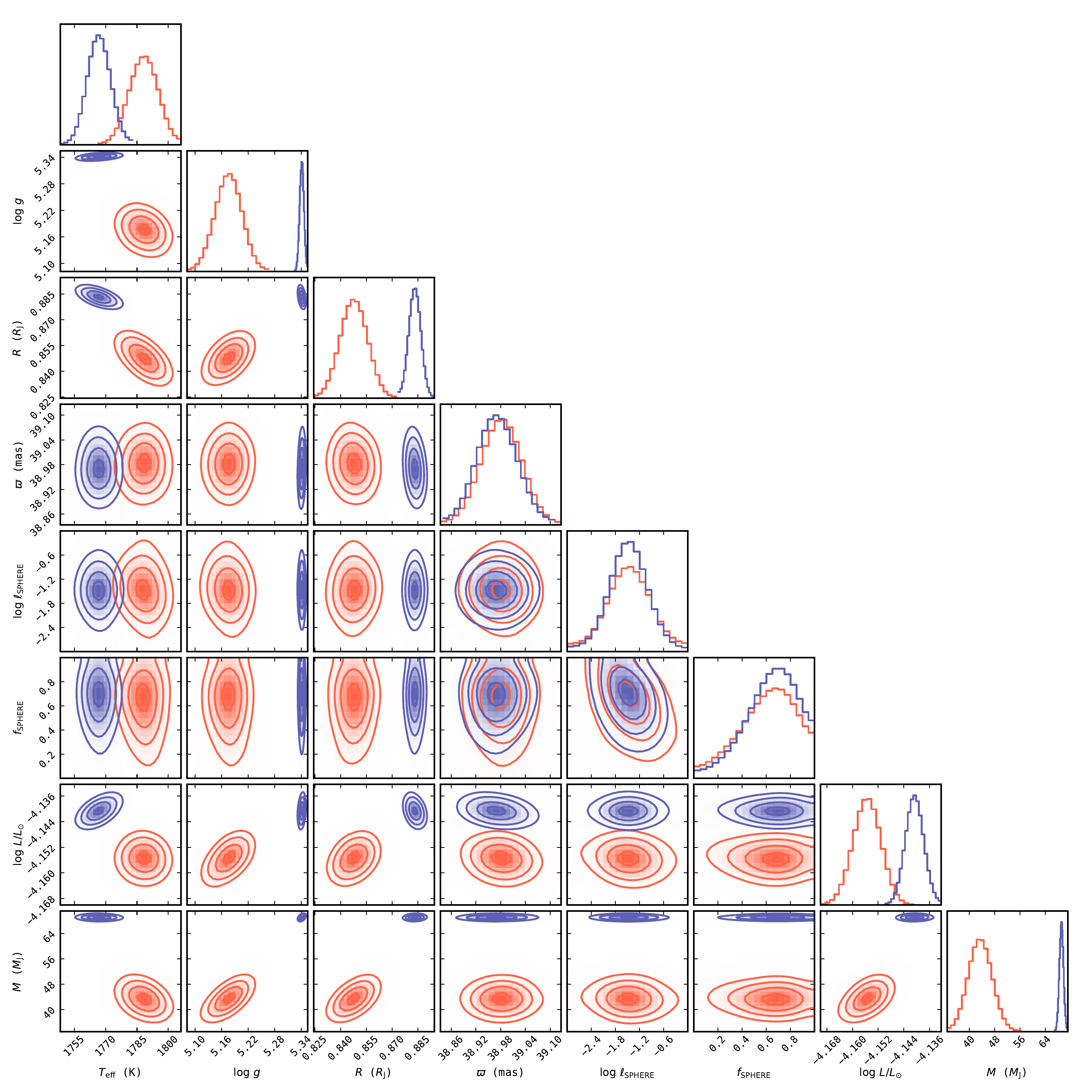}
    \caption{The posterior distribution of \texttt{BT-Settl} model spectra fit to observations of HD~72946~B. The dark blue, generally narrower, posteriors correspond to the fit with a prior on the mass of the object equivalent to the dynamical mass derived in \S\ref{sec:orbit}, while the red, generally wider, posterior corresponds to the free mass fit. Aside from the nuissance parameters (parallax, and the gaussian process ``correlation matrix" fit to the SPHERE/IFU data), the parameters are generally normally distributed and well constrained, but distinct between the two cases. Notably, the dynamical mass constrains the log(g) and radius to higher values, leading to a lower temperature.}
    \label{fig:btsettlpost}
\end{figure*}

\begin{figure*}
    \centering
    \includegraphics[width=\textwidth]{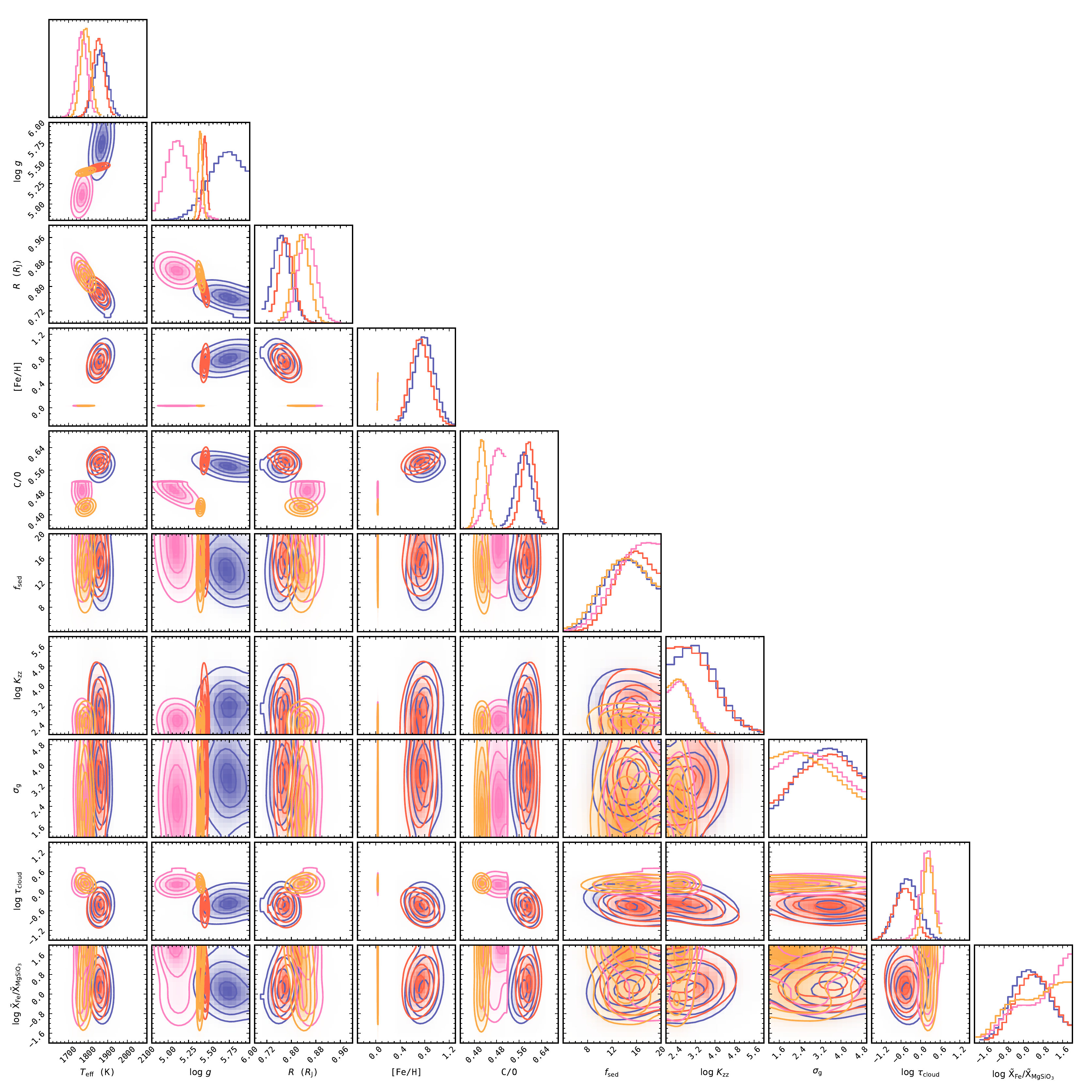}
    \caption{The posterior distributions of four \texttt{petitRADTRANS} retrievals with the EddySed cloud model. Each retrieval implements the 3-part P-T profile but the prior on the mass and on the abundances ([Fe/H] and C/O) are varied between retrievals. As in Figure \ref{fig:retrievalspec}, the red contours plot the posterior assuming a uniform mass and abundance prior, blue contours, assuming a prior on the dynamical mass but a uniform abundance prior, pink contours, assuming a uniform mass but Gaussian priors on the stellar abundances, and orange contours, assuming Gaussian priors on both the dynamical mass and stellar abundances.}
    \label{fig:retrievalpost}
\end{figure*}

 This appendix contains the posterior distributions of the various multi-dimensional models fit to data throughout this paper. Table \ref{tab:conservativeorbit} contains the median and 1$\,\sigma$ CI on the posterior distribution of orbits fit in excluding absolute astrometry, described in \ref{sec:orbit} and Table \ref{tab:optimisticorbit} for the orbit fit including absolute astrometry. Figure \ref{fig:orbpost} illustrates the comparative posterior distribution of orbital elements between the two orbit fits in Tables \ref{tab:conservativeorbit} and \ref{tab:optimisticorbit}. Figure \ref{fig:btsettlpost} plots the comparative posterior distribution of \texttt{BT-Settl-CIFIST} model spectra fit to HD~72946~B. Figure \ref{fig:retrievalpost} plots the comparative posterior distribution of \texttt{petitRADTRANS} retrievals using the EddySed cloud model recorded in Table \ref{tab:atmos}.

\begin{deluxetable*}{ccccc}
\tablewidth{\textwidth}
\tablecaption{Orbital parameters inferred for HD~72946~B excluding absolute astrometry. \label{tab:conservativeorbit}}
\tablehead{
\colhead{Parameter} & \colhead{Description} & \colhead{Median} & \colhead{Lower 1$\,\sigma$ CI} & \colhead{Upper 1$\,\sigma$ CI}
}
\startdata
a [au]                              & Semi-major axis                                & 6.487 & -0.059 & 0.058 \\
e                                   & Eccentricity                                   & 0.497 & -0.006 & 0.005 \\
i [rad]                        & Inclination                                    & 1.102 & -0.004 & 0.004 \\
$\omega$ [rad]                 & Argument of periastron                         & 4.343 & -0.005 & 0.005 \\
$\Omega$ [rad]                 & Longitude of Ascending Node                    & 6.190 & -0.004 & 0.005 \\
$\tau$ [dec. cal. yr] & Next periastron passage after $\tau_{\mathrm{ref}}^\mathrm{(a)}$ & 2028.164 & -0.126 & 0.129 \\
$\pi$ [mas]                         & Parallax                                       & 38.798 & -0.382 & 0.388 \\
$\gamma_{\mathrm{ELODIE}}$ [km/s]   & RV offset term                                 & 29.427 & -0.006 & 0.006 \\
$\sigma_{\mathrm{ELODIE}}$ [km/s]   & RV jitter term                                 & 0.026 & -0.003 & 0.004 \\
$\gamma_{\mathrm{SOPHIE}}$ [km/s]   & RV offset term                                 & 29.514 & -0.007 & 0.006 \\
$\sigma_{\mathrm{SOPHIE}}$ [km/s]   & RV jitter term                                 & 0.014 & -0.003 & 0.004 \\
$\mathrm{M_B}$ [$\mathrm{M_\odot}$] & Mass of B                                      & 0.067 & -0.001 & 0.001 \\
$\mathrm{M_A}$ [$\mathrm{M_\odot}$] & Mass of A                                      & 0.985 & -0.026 & 0.026
\enddata
\tablecomments{We report the median and 68\% confidence interval on each parameter derived from the posterior visualized in Figure \ref{fig:orbpost}. This orbit analysis did not include absolute astrometry. $^\mathrm{(a)}$We set $\tau_{\mathrm{ref}}=2020.0$}
\end{deluxetable*}

\begin{deluxetable*}{ccccc}
\tablewidth{\textwidth}
\tablecaption{Orbital parameters inferred for HD~72946~B including absolute astrometry. \label{tab:optimisticorbit}}
\tablehead{
\colhead{Parameter} & \colhead{Description} & \colhead{Median} & \colhead{Lower 1$\,\sigma$ CI} & \colhead{Upper 1$\,\sigma$ CI}
}
\startdata
a [au]                              & Semi-major axis                                & 6.462 & -0.029 & 0.030 \\
e                                   & Eccentricity                                   & 0.498 & -0.005 & 0.005 \\
i [rad]                        & Inclination                                    & 1.102 & -0.003 & 0.003 \\
$\omega$ [rad]                 & Argument of periastron                         & 4.343 & -0.005 & 0.005 \\
$\Omega$ [rad]                 & Longitude of Ascending Node                    & 6.189 & -0.004 & 0.005 \\
$\tau$ [dec. cal. yr]           & Next periastron passage after $\tau_{\mathrm{ref}}^\mathrm{(a)}$ & 2028.154 & -0.080 & 0.083 \\
$\pi$ [mas]                         & Parallax                                       & 38.981 & -0.010 & 0.010 \\
$\gamma_{\mathrm{ELODIE}}$ [km/s]   & RV offset term                                 & 29.427 & -0.006 & 0.006 \\
$\sigma_{\mathrm{ELODIE}}$ [km/s]   & RV jitter term                                 & 0.026 & -0.003 & 0.004 \\
$\gamma_{\mathrm{SOPHIE}}$ [km/s]   & RV offset term                                 & 29.514 & -0.007 & 0.006 \\
$\sigma_{\mathrm{SOPHIE}}$ [km/s]   & RV jitter term                                 & 0.014 & -0.003 & 0.004 \\
$\mathrm{M_B}$ [$\mathrm{M_\odot}$] & Mass of B                                      & 0.066 & -0.001 & 0.001 \\
$\mathrm{M_A}$ [$\mathrm{M_\odot}$] & Mass of A                                      & 0.975 & -0.013 & 0.013
\enddata
\tablecomments{We report the median and 68\% confidence interval on each parameter derived from the posterior visualized in Figure \ref{fig:orbpost}. This orbit analysis included absolute astrometry from the HGCA. $^\mathrm{(a)}$We set $\tau_{\mathrm{ref}}=2020.0$}
\end{deluxetable*}


\onecolumngrid
\section{Host Star} \label{hostappendix}

\par
\begin{figure*}
    \centering
    \includegraphics[width=0.95\textwidth]{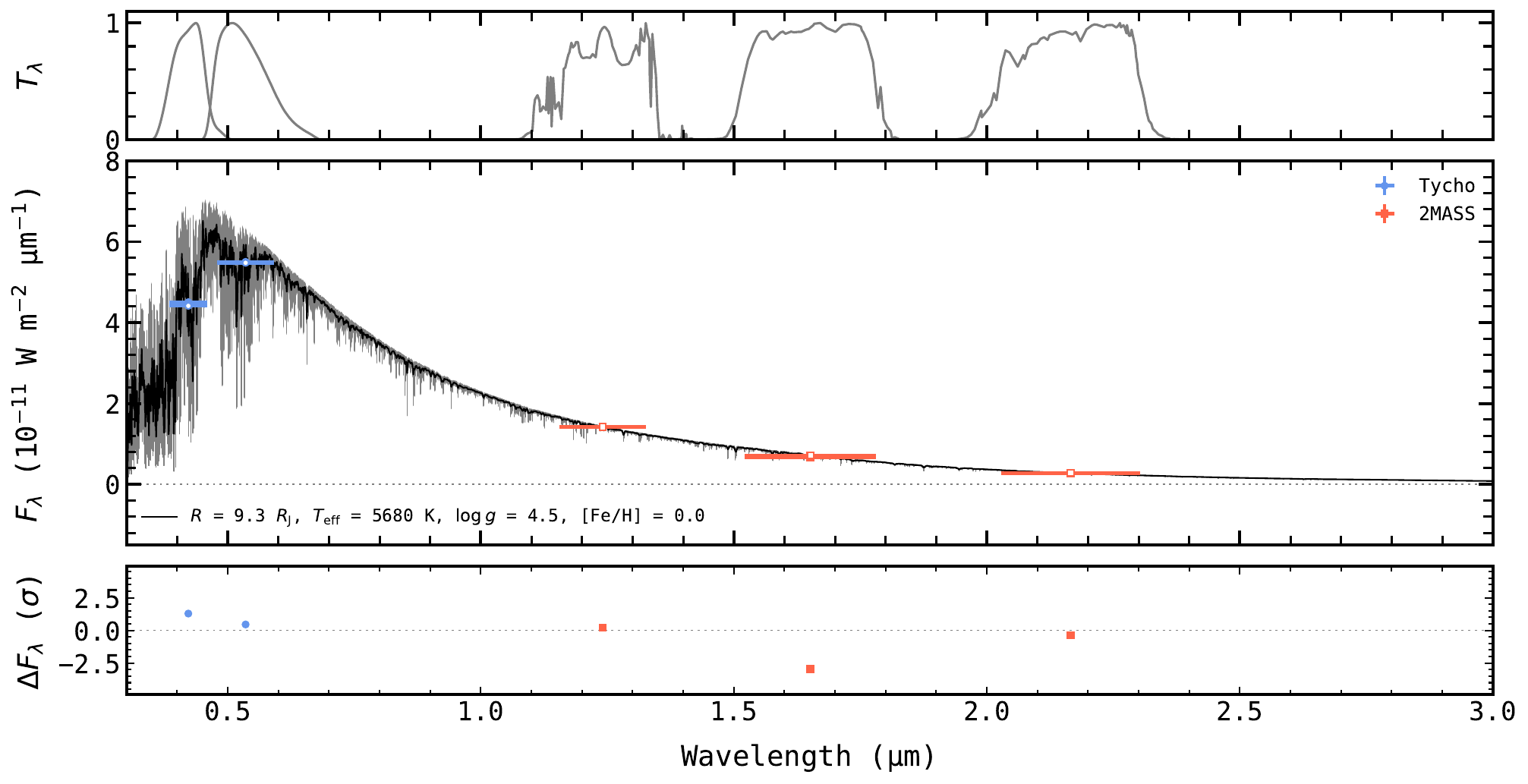}
    \caption{A model \texttt{BT-NextGen} spectrum of HD~72946~A with parameters based on our host star analysis scaled to the archival photometry of the star. Transmission profiles for the photometric filters are plotted along the upper panel. The flux density of the spectrum is plotted in the middle panel, with the best fit in black and 30 random samples in grey, and the measured photometry overlaid as colored squares. The residuals to the scaling fit are shown in the bottom panel.}
    \label{fig:hostspec}
\end{figure*}

 This appendix contains both the table of derived abundances for the host star HD~72946~A, Table \ref{host_chemical_abundances}, and Figure \ref{fig:hostspec} illustrating the scaled \texttt{BT-Nextgen} model stellar atmosphere fit to the literature photometry recorded in Table \ref{tab:host}.

\begin{deluxetable*}{lcccccc}
\tablecaption{Elemental Abundances for HD~72946~A}\label{host_chemical_abundances}
\tablewidth{0pt}
\tablehead{
\colhead{Species} &
\colhead{$A(\mathrm{X})$} &
\colhead{[X/H]} &
\colhead{$\sigma_{\mathrm{[X/H]}}$} & 
\colhead{[X/Fe]} &
\colhead{$\sigma_{[\mathrm{X/Fe}]}$} &
\colhead{$n$}}
\startdata
\multicolumn{6}{l}{\textbf{LTE abundances}} \\
\ion{C}{1} & $8.369$ & $-0.091$ & $0.038$ & $-0.127$ & $0.054$ & $2$\\ 
\ion{O}{1} & $8.660$ & $\cdots$ & $\cdots$ & $\cdots$ & $\cdots$ & $1$\\ 
\ion{Na}{1} & $6.191$ & $-0.029$ & $0.078$ & $-0.065$ & $0.048$ & $4$\\ 
\ion{Mg}{1} & $7.600$ & $0.050$ & $0.094$ & $0.014$ & $0.054$ & $5$\\ 
\ion{Al}{1} & $6.409$ & $-0.021$ & $0.015$ & $-0.057$ & $0.022$ & $2$\\ 
\ion{Si}{1} & $7.580$ & $0.070$ & $0.081$ & $0.034$ & $0.025$ & $14$\\ 
\ion{S}{1} & $7.107$ & $-0.013$ & $0.000$ & $-0.049$ & $0.037$ & $1$\\ 
\ion{Ca}{1} & $6.373$ & $0.073$ & $0.043$ & $0.037$ & $0.035$ & $11$\\ 
\ion{Sc}{1} & $3.121$ & $-0.019$ & $0.022$ & $-0.055$ & $0.034$ & $3$\\ 
\ion{Sc}{2} & $3.167$ & $0.027$ & $0.212$ & $-0.009$ & $0.078$ & $10$\\ 
\ion{Sc}{1} & $3.121$ & $-0.019$ & $0.022$ & $-0.055$ & $0.034$ & $3$\\ 
\ion{Sc}{2} & $3.167$ & $0.027$ & $0.212$ & $-0.009$ & $0.078$ & $10$\\ 
\ion{Ti}{1} & $4.951$ & $-0.019$ & $0.100$ & $-0.055$ & $0.046$ & $17$\\ 
\ion{Ti}{2} & $5.076$ & $0.106$ & $0.069$ & $0.070$ & $0.041$ & $13$\\ 
\ion{V}{1} & $3.993$ & $0.093$ & $0.061$ & $0.057$ & $0.044$ & $9$\\ 
\ion{Cr}{1} & $5.696$ & $0.076$ & $0.056$ & $0.040$ & $0.037$ & $14$\\ 
\ion{Cr}{2} & $5.741$ & $0.121$ & $0.155$ & $0.085$ & $0.074$ & $7$\\ 
\ion{Mn}{1} & $5.307$ & $-0.113$ & $0.105$ & $-0.149$ & $0.049$ & $8$\\ 
\ion{Fe}{1} & $7.491$ & $0.031$ & $0.069$ & $-0.005$ & $0.026$ & $63$\\ 
\ion{Fe}{2} & $7.521$ & $0.061$ & $0.072$ & $0.025$ & $0.047$ & $18$\\ 
\ion{Ni}{1} & $6.297$ & $0.097$ & $0.067$ & $0.061$ & $0.026$ & $18$\\ 
\ion{Cu}{1} & $4.173$ & $-0.007$ & $0.076$ & $-0.043$ & $0.058$ & $3$\\ 
\ion{Zn}{1} & $4.647$ & $0.087$ & $0.098$ & $0.051$ & $0.074$ & $3$\\ 
\ion{Sr}{1} & $2.724$ & $-0.106$ & $0.000$ & $-0.142$ & $0.055$ & $1$\\ 
\ion{Y}{2} & $2.360$ & $0.150$ & $0.058$ & $0.114$ & $0.056$ & $4$\\ 
\ion{Zr}{2} & $2.631$ & $0.041$ & $0.166$ & $0.005$ & $0.123$ & $3$\\ 
\ion{Ba}{2} & $2.478$ & $0.208$ & $0.062$ & $0.172$ & $0.050$ & $4$\\ 
\ion{La}{2} & $1.331$ & $0.221$ & $0.000$ & $0.185$ & $0.033$ & $1$\\ 
\ion{Ce}{2} & $1.776$ & $0.196$ & $0.148$ & $0.160$ & $0.110$ & $3$\\ 
\ion{Nd}{2} & $1.592$ & $0.172$ & $0.072$ & $0.136$ & $0.061$ & $3$\\ 
\ion{Dy}{2} & $1.132$ & $0.032$ & $0.000$ & $-0.004$ & $0.033$ & $1$\\ 
\hline
\multicolumn{7}{l}{\textbf{1D non-LTE abundances}}\\
\ion{Al}{1} & $6.368$ & $-0.062$ & $\cdots$ &$-0.131$ & $\cdots$ & $1$ \\
\ion{Ca}{1} & $6.034$ & $-0.266$ & $0.07$ & $-0.335$ & $\cdots$ & $2$ \\
\ion{Fe}{1} & $7.524$ & $0.064$ & $0.072$ & $\cdots$ & $\cdots$ & $63$ \\
\ion{Fe}{2} & $7.547$ & $0.087$ & $0.063$ & $\cdots$ & $\cdots$ & $18$ \\
\hline
\multicolumn{6}{l}{\textbf{3D non-LTE abundances}} \\
\ion{C}{1} & $8.352$ & $-0.108$ & $0.077$ & $-0.177$ & $\cdots$ & $5$ \\
\hline
\multicolumn{7}{l}{\textbf{Additional abundance ratios of interest}} \\
\multicolumn{7}{l}{$[\mathrm{Fe/H}]_{\mathrm{1D non-LTE}} = 0.069 \pm 0.080$} \\
\multicolumn{7}{l}{$[\mathrm{(C+O)/H}] = -0.121\pm0.038 $} \\
\multicolumn{7}{l}{$[\mathrm{C/O}] = -0.061\pm 0.038$} \\
\multicolumn{7}{l}{$\mathrm{C/O}_{\mathrm{1D LTE}} = 0.512^{+0.047}_{-0.043}$} \\
\multicolumn{7} {l}{$\mathrm{Mg/Si}_{\mathrm{1D LTE}} = 1.047\pm0.124$}
\enddata
\end{deluxetable*}

\allauthors


\end{document}